\documentclass[10pt]{article}
\usepackage{hyperref}
\usepackage{amssymb}
\usepackage{amsmath}
\usepackage{stmaryrd}

\usepackage{color}

\usepackage{amsmath}
\usepackage{latexsym}
\usepackage{amssymb}

\usepackage{amsthm}

\font\tengoth=eufm10 at 10pt
\font\sevengoth=eufm7 at 6pt
\newfam\gothfam
\textfont\gothfam=\tengoth
\scriptfont\gothfam=\sevengoth

\newcommand{\mlabel}[1]{\marginpar{#1}\label{#1}}

\newcommand{\g}{{\mathfrak g}}

\newcommand{\fa}{{\mathfrak a}}
\newcommand{\fb}{{\mathfrak b}}
\newcommand{\fc}{{\mathfrak c}}

\newcommand{\fe}{{\mathfrak e}}

\newcommand{\fg}{{\mathfrak g}}
\newcommand{\fh}{{\mathfrak h}}

\newcommand{\fj}{{\mathfrak j}}
\newcommand{\fk}{{\mathfrak k}}
\newcommand{\fl}{{\mathfrak l}}

\newcommand{\fq}{{\mathfrak q}}
\newcommand{\fp}{{\mathfrak p}}

\newcommand{\fz}{{\mathfrak z}}

\renewcommand\sp{\mathfrak {sp}}

\newcommand\heis{\mathfrak {heis}}

\renewcommand{\:}{\colon}
\newcommand{\1}{\mathbf{1}}

\newcommand{\cD}{\mathcal{D}}

\newcommand{\cF}{\mathcal{F}}
\newcommand{\cG}{\mathcal{G}}
\newcommand{\cH}{\mathcal{H}}
\newcommand{\cK}{\mathcal{K}}
\newcommand{\cL}{\mathcal{L}}
\newcommand{\cM}{\mathcal{M}}

\newcommand{\cO}{\mathcal{O}}

\newcommand{\cQ}{\mathcal{Q}}
\newcommand{\cR}{\mathcal{R}}
\newcommand{\cS}{\mathcal{S}}
\newcommand{\cT}{\mathcal{T}}
\newcommand{\cU}{\mathcal{U}}

\newcommand{\cW}{\mathcal{W}}

\newcommand{\bO}{\mathbf{O}}

\newcommand{\eset}{\emptyset}

\newcommand{\dd}{{\tt d}}

\newcommand{\trile}{\trianglelefteq}
\newcommand{\subeq}{\subseteq}
\newcommand{\supeq}{\supseteq}

\newcommand{\into}{\hookrightarrow}
\newcommand{\eps}{\varepsilon}

\newcommand{\shalf}{{\textstyle{\frac{1}{2}}}}

\newcommand{\N}{{\mathbb N}}

\newcommand{\R}{{\mathbb R}}
\newcommand{\C}{{\mathbb C}}

\renewcommand{\H}{{\mathbb H}}

\newcommand{\bS}{{\mathbb S}}

\renewcommand{\hat}{\widehat}

\renewcommand{\tilde}{\widetilde}

\renewcommand{\L}{\mathop{\bf L{}}\nolimits}

\newcommand{\Aff}{\mathop{{\rm Aff}}\nolimits}

\newcommand{\GL}{\mathop{{\rm GL}}\nolimits}
\newcommand{\SL}{\mathop{{\rm SL}}\nolimits}

\newcommand{\AU}{\mathop{{\rm AU}}\nolimits}

\newcommand{\PSL}{\mathop{{\rm PSL}}\nolimits}
\newcommand{\SO}{\mathop{{\rm SO}}\nolimits}
\newcommand{\SU}{\mathop{{\rm SU}}\nolimits}

\newcommand{\U}{\mathop{\rm U{}}\nolimits}

\newcommand{\Sp}{\mathop{{\rm Sp}}\nolimits}
\newcommand{\Sym}{\mathop{{\rm Sym}}\nolimits}

\newcommand{\gl}  {\mathop{{\mathfrak{gl} }}\nolimits}

\newcommand{\fsl} {\mathop{{\mathfrak{sl} }}\nolimits}

\newcommand{\su}  {\mathop{{\mathfrak{su} }}\nolimits}
\newcommand{\so}  {\mathop{{\mathfrak{so} }}\nolimits}

\newcommand{\Exp}{\mathop{{\rm Exp}}\nolimits}
\newcommand{\Fix}{\mathop{{\rm Fix}}\nolimits}

\newcommand{\ad}{\mathop{{\rm ad}}\nolimits}
\newcommand{\Ad}{\mathop{{\rm Ad}}\nolimits}

\renewcommand{\Re}{\mathop{{\rm Re}}\nolimits}
\renewcommand{\Im}{\mathop{{\rm Im}}\nolimits}

\newcommand{\Hol}{\mathop{{\rm Hol}}\nolimits}

\newcommand{\Herm}{\mathop{{\rm Herm}}\nolimits}

\newcommand{\Aut}{\mathop{{\rm Aut}}\nolimits}
\newcommand{\Conf}{\mathop{\rm Conf{}}\nolimits}

\newcommand{\Diff}{\mathop{{\rm Diff}}\nolimits}

\newcommand{\End}{\mathop{{\rm End}}\nolimits}
\newcommand{\id}{\mathop{{\rm id}}\nolimits}

\renewcommand{\dim}{\mathop{{\rm dim}}\nolimits}

\newcommand{\im}{\mathop{{\rm im}}\nolimits}

\newcommand{\supp}{\mathop{{\rm supp}}\nolimits}

\newcommand{\Inn}{\mathop{{\rm Inn}}\nolimits}

\newcommand{\Spann}{\mathop{{\rm span}}\nolimits}
\newcommand{\ev}{\mathop{{\rm ev}}\nolimits}

\newcommand{\Str}{\mathop{{\rm Str}}\nolimits}
\newcommand{\dS}{\mathop{{\rm dS}}\nolimits}

\renewcommand{\phi}{\varphi}

\newcommand{\Rarrow}{\Rightarrow}
\newcommand{\nin}{\noindent} 
\newcommand{\oline}{\overline}

\newcommand{\la}{\langle}
\newcommand{\ra}{\rangle}

\newcommand{\res}{\vert}

\newcommand{\spann}{{\rm span}}

\newcommand{\Spec}{{\rm Spec}}
\newcommand{\Spin}{{\rm Spin}}

\newcommand{\ssssarr}{\hbox to 15pt{\rightarrowfill}}
\newcommand{\sssarr}{\hbox to 20pt{\rightarrowfill}}
\newcommand{\ssarr}{\hbox to 30pt{\rightarrowfill}}
\newcommand{\sarr}{\hbox to 40pt{\rightarrowfill}}
\newcommand{\arr}{\hbox to 60pt{\rightarrowfill}}
\newcommand{\larr}{\hbox to 60pt{\leftarrowfill}}
\newcommand{\Arr}{\hbox to 80pt{\rightarrowfill}}

\newcommand{\mapright}[1]{\smash{\mathop{\arr}\limits^{#1}}}

\def\theoremname{Theorem}
\def\propositionname{Proposition}
\def\corollaryname{Corollary}
\def\lemmaname{Lemma}
\def\remarkname{Remark}
\def\conjecturename{Conjecture} 

\def\definitionname{Definition}
\def\exercisename{Exercise}
\def\examplename{Example}
\def\examplesname{Examples}
\def\problemname{Problem}
\def\problemsname{Problems}

\def\satzname{Satz} 
\def\koroname{Korollar}
\def\folgname{Folgerung}
\def\bemerkname{Bemerkung}
\def\aufgname{Aufgabe}

\def\beisname{Beispiel}
\def\beissname{Beispiele}
\def\bewname{Beweis}

\def\@thmcounter#1{\noexpand\arabic{#1}}
\def\@thmcountersep{}
\def\@begintheorem#1#2{\it \trivlist \item[\hskip 
\labelsep{\bf #1\ #2.\quad}]}
\def\@opargbegintheorem#1#2#3{\it \trivlist
      \item[\hskip \labelsep{\bf #1\ #2.\quad{\rm #3}}]}
\makeatother
\newtheorem{theor}{\theoremname}[section]
\newtheorem{propo}[theor]{\propositionname}
\newtheorem{coro}[theor]{\corollaryname}
\newtheorem{lemm}[theor]{\lemmaname}

\newenvironment{thm}{\begin{theor}\it}{\end{theor}}

\newenvironment{prop}{\begin{propo}\it}{\end{propo}}

\newenvironment{cor}{\begin{coro}\it}{\end{coro}}

\newenvironment{lem}{\begin{lemm}\it}{\end{lemm}}

\newtheorem{rema}[theor]{\remarkname}

\newenvironment{rem}{\begin{rema}\rm}{\end{rema}}

\newtheorem{stepnow}[theor]{}

\newtheorem{defin}[theor]{\definitionname} 

\newenvironment{defn}{\begin{defin}\rm}{\end{defin}}

\newtheorem{exerc}{\exercisename}[section]

\newtheorem{exa}[theor]{\examplename}

\newenvironment{ex}{\begin{exa}\rm}{\end{exa}}

\newtheorem{exas}[theor]{\examplesname}

\newtheorem{conj}[theor]{\conjecturename}

\newtheorem{pro}[theor]{\problemname}

\newtheorem{prs}[theor]{\problemsname}

\newtheorem{aufg}{\aufgname}[section]

\newenvironment{prf}{\begin{proof}}{\end{proof}}
 
{\hfill\qed\end{trivlist}}

\newenvironment{beweis*}{\begin{trivlist}\item[\hskip%
\labelsep{\bf\bewname.\quad}]}%
{\end{trivlist}}

\newtheorem{satzn}[theor]{\satzname}

\newtheorem{koro}[theor]{\koroname}

\newtheorem{folg}[theor]{\folgname}

\newtheorem{bem}[theor]{\bemerkname}

\newtheorem{aufgn}[theor]{\aufgname}

\newtheorem{beis}[theor]{\beisname}

\newtheorem{beiss}[theor]{\beissname}

\newcommand\bo{{\rm{bd}}}

\newcommand{\sE}{{\sf E}}
\newcommand{\sF}{{\sf F}}
\newcommand{\sH}{{\sf H}}

\newcommand{\sV}{{\tt V}}

\newcommand{\bd}{\mathbf d} 
\renewcommand{\bO}{\mathbb O} 
\newcommand{\csp}{{\mathfrak{csp}}} 
\newcommand{\hcsp}{{\mathfrak{hcsp}}}

\newcommand{\Mob}{{\rm\textsf{M\"ob}}}

\renewcommand{\phi}{\varphi} 

\newcommand{\AdS}{\mathop{{\rm AdS}}\nolimits}

\renewcommand\mlabel{\label}

\begin{document}

\title{Nets of standard subspaces on Lie groups}
\author{Karl-Hermann Neeb, Gestur \'Olafsson} 

\maketitle

\abstract{Let $G$ be a Lie group with Lie algebra $\g$, 
$h \in \g$ an element for which the derivation $\ad h$ defines a 
$3$-grading of $\g$ and $\tau_G$ an involutive automorphism of $G$ 
inducing on $\g$ the involution $e^{\pi i \ad h}$. 
We consider antiunitary representations 
$(U,\cH)$ of the Lie group $G_\tau = G \rtimes \{\1,\tau_G\}$ 
for which the positive cone 
$C_U = \{ x \in \g \: -i \partial U(x) \geq 0\}$ and 
$h$ span $\g$. To a real subspace $\sE \subeq \cH^{-\infty}$ 
of distribution vectors invariant under $\exp(\R h)$ 
and an open subset $\cO \subeq G$, we 
associate the real subspace $\sH_\sE(\cO) \subeq \cH$,  
generated by the subspaces $U^{-\infty}(\phi)\sE$, where 
$\phi \in C^\infty_c(\cO,\R)$ is a real-valued test function on $\cO$. 
Then $\sH_\sE(\cO)$ generates the complex Hilbert space 
$\cH_E(G) := \oline{\sH_\sE(G) + i \sH_\sE(G)}$ for every non-empty open subset 
$\cO \subeq G$ (Reeh--Schlieder property). 

For the real standard subspace 
$\sV \subeq \cH$, for which $J_\sV = U(\tau_G)$ is the modular conjugation 
and $\Delta_\sV^{-it/2\pi} = U(\exp th)$ is the modular group, 
we obtain sufficient conditions to be of the form 
$\sH_\sE(S)$ for an open subsemigroup $S \subeq G$. 
If $\g$ is semisimple with simple hermitian ideals 
of tube type, we verify these criteria and obtain nets 
of cyclic subspaces $\sH_\sE(\cO)$, $\cO \subeq G$, satisfying the 
Bisognano--Wichmann property in a suitable sense. Our construction also yields such 
nets on simple Jordan space-times and compactly causal symmetric 
spaces of Cayley type. By 
second quantization, these nets lead to 
free quantum fields in the sense of Haag--Kastler on causal 
homogeneous spaces whose groups are generated by modular groups 
and conjugations. }

\tableofcontents 

\section{Introduction} 
\mlabel{sec:1}

Let $\cH$ be a complex Hilbert space and 
$\cM \subeq B(\cH)$ be a von Neumann algebra. 
A unit vector $\Omega \in \cH$ is called 
{\it cyclic} for $\cM$ if $\cM\Omega$ is dense in $\cH$, and 
{\it separating} if the map $\cM \to \cH,M \mapsto M\Omega$ is injective. 
If $\Omega$ is both, cyclic  and separating for $\cM$, 
then the Tomita--Takesaki Theorem (\cite[Thm.~2.5.14]{BR87}) 
asserts in particular that the closed real subspace 
\[ \sV := \oline{\{ M\Omega \: M = M^* \in \cM\}} \] 
is {\it standard}, i.e., 
\begin{equation}
  \label{eq:stansub}
  \sV \cap i \sV = \{0\}\ \ (\sV \ \mbox{is separating}) 
\quad \mbox{ and } \quad \cH = \oline{\sV + i \sV}
\ \ \ \ (\sV \ \mbox{is cyclic}) 
\end{equation} 
(cf.\ \cite{Lo08} for the basic theory of standard subspaces). 
To the standard subspace $\sV$, we associate a {\it pair of modular 
objects} $(\Delta_\sV, J_\sV)$: the {\it modular operator} $\Delta_\sV$ is a positive 
selfadjoint operator, $J_\sV$ is a {\it conjugation} (an antiunitary 
involution), and these two operators satisfy the modular relation 
$J_\sV \Delta_\sV J_\sV = \Delta_\sV^{-1}$. The pair $(\Delta_\sV, J_\sV)$ is obtained
by the polar decomposition $\sigma_\sV = J_\sV \Delta_\sV^{1/2}$ of the closed operator 
\[ \sigma_\sV \: \sV + i \sV \to \cH, \quad 
x + i y \mapsto x- iy \] 
with $\sV = \Fix(\sigma_\sV)$. The main assertion of the 
Tomita--Takesaki Theorem 
is  that 
\[ J_\sV \cM J_\sV = \cM' \quad \mbox{ and } \quad 
\Delta_\sV^{it} \cM \Delta_\sV^{-it}  = \cM \quad \mbox{ for }  \quad t \in \R.\] 
So we obtain a one-parameter group of automorphisms of $\cM$ 
(the modular group) 
and a symmetry between $\cM$ and its commutant $\cM'$, implemented by $J_\sV$. 

In Algebraic Quantum Field Theory (AQFT) 
in the sense of Haag--Kastler, one considers 
{\it nets} of von Neumann algebras $\cM(\cO) \subeq B(\cH)$, 
associated to  regions $\cO$ in some space-time manifold~$M$ 
(\cite{Ha96}). 
The hermitian elements of the algebra $\cM(\cO)$ are interpreted as observables  
that can be measured in the ``laboratory'' $\cO$. 
Accordingly, one requires {\it isotony}, i.e., that $\cO_1 \subeq \cO_2$ implies 
$\cM(\cO_1) \subeq \cM(\cO_2)$. 
Causality enters by the {\it locality} assumption that 
$\cM(\cO_1)$ and $\cM(\cO_2)$ commute if 
$\cO_1$ and $\cO_2$ are space-like separated, i.e., cannot correspond 
with each other. One further assumes a unitary representation 
$U \: G \to \U(\cH)$  of a Lie group~$G$, acting as a space-time symmetry 
group on $M$, such that $U(g) \cM(\cO) U(g)^* = \cM(g\cO)$ for 
$g \in G$. 
In addition, one assumes a $U(G)$-fixed 
unit vector $\Omega \in \cH$, representing typically 
a vacuum state 
of a quantum field. 
The domains $\cO \subeq M$ for which 
$\Omega$ is cyclic and separating for $\cM(\cO)$ are of particular 
relevance. For these domains $\cO$, the von Neumann algebra 
$\cM(\cO)$ specifies a standard subspace $\sV(\cO)\subeq \cH$ 
which determines a pair $(\Delta_\cO,J_\cO)$ of modular objects and in particular 
a modular automorphism group 
$\alpha_t(M) = \Delta_\cO^{-it/2\pi} M \Delta_\cO^{it/2\pi}$. 
It is now an interesting question if this modular group is ``geometric'' 
in the sense that it is implemented by a one-parameter subgroup of~$G$, hence 
corresponds to a one-parameter group of symmetries of~$M$. For the modular 
conjugation $J_\cO$, we may likewise ask for the existence of an involutive 
automorphism $\tau_G$ of $G$ and an involution $\tau_M$ on $M$ 
reversing the causal structure, such that 
\[ J_\cO \cM(\tilde\cO) J_\cO = \cM(\tau_M(\tilde \cO)), \qquad 
J_\cO U(g) J_\cO = U(\tau_G(g))\quad 
 \mbox{ for } \quad g \in G, \tilde \cO \subeq M,\] 
and that $\tau_M$ and $\tau_G$ are compatible in the sense that 
\[ \tau_M \circ g = \tau_G(g) \circ \tau_M  \quad \mbox{ for } \quad g \in G. \] 
We are particularly interested in the geometric realizations of modular 
groups and involutions in the sense explained above. 
The present paper contributes to this project 
by exhibiting large classes of nets of standard subspaces 
where the action of the modular group is geometric. 

To study this question systematically, we 
first observe that, passing from operator algebras 
to the corresponding standard subspaces 
is a tremendous reduction of information, but 
the net of standard subspaces $\sV(\cO)$ 
still encodes the geometric features of the original theory 
and in particular it still reflects the action of the symmetry group 
$G$ and its extensions by involutions.  

Conversely, one can use the functorial process provided by Second Quantization 
(\cite{Si74}) to associate to each standard subspace $\sV \subeq \cH$   
a pair $(\cR_\pm(\sV),\Omega)$, where $\cR_\pm(\sV)$ is a von Neumann algebra 
on the bosonic/fermionic Fock space $\cF_\pm(\cH)$, for which the 
vacuum vector is cyclic and separating. 
This method has been developed by Araki and Woods  in the context of 
free bosonic quantum fields (\cite{Ar63, Ar64, AW63, AW68}); 
some of the corresponding fermionic results are more recent 
(cf.\ \cite{EO73}, \cite{BJL02}). Other statistics (anyons) 
are discussed in \cite{Schr97} and more recent deformations of 
this procedure are discussed in \cite[\S 3]{Le15}. 
Throughout this paper we only deal with nets of standard subspaces, 
but it is important to keep in mind that there are functorial 
constructions that associate to such nets various types of 
free quantum fields on homogeneous spaces. 

The current interest in standard subspaces arose 
in the 1990s from the work of 
Borchers 
and 
Wiesbrock (\cite{Bo92, Wi93}). 
This in turn led to the concept of modular localization 
in Quantum Field Theory introduced by  
Brunetti, Guido and Longo in \cite{BGL02, BGL93}; see also \cite{BDFS00} 
and \cite{Le15, LL15} for important applications of this technique. 

We start with  a unitary representation $(U,\cH)$ of $G$ which extends  
to an {\it antiunitary representation} of the  extended group~$G_\tau 
 = G \rtimes \{\1,\tau_G\}$, containing $G$ as a subgroup of index~$2$. 
Then $J := U(\tau_G)$ is a conjugation satisfying 
$J U(g) J = U(\tau_G(g))$ for $g \in G$. 
We then obtain for each pair 
$(h,\tau_G)$ for which $h$ is fixed by the Lie algebra 
involution $\tau_\g := \L(\tau_G)$, a standard subspace 
$\sV := \sV_{(h,\tau_G,U)}$, specified by 
\begin{equation}
  \label{eq:bgl}
J_\sV = U(\tau_G) \quad \mbox{ and } \quad 
\Delta_{\sV}^{-it/2\pi} = U(\exp th) \quad \mbox{ for } \quad t \in \R.
\end{equation}
This assignment is called the {\it Brunetti--Guido--Longo 
(BGL) construction} (see \cite{BGL02}). 
As a consequence, standard subspaces can be associated to antiunitary 
representations in abundance, but only a few of them 
carry interesting geometric 
information. In particular, we would like to understand 
when a standard subspace of the form $\sV_{(h,\tau_G,U)}$ arises from a 
natural family $\sV(\cO)$ of real subspaces associated to open 
subsets of a homogeneous space $M = G/P$ and which domains $\cO \subeq M$ 
(so-called {\it generalized wedge domains}) correspond to such standard subspaces.  
The geometric investigation of such domains in causal symmetric spaces 
will be pursued further in \cite{NO20}. 

In the present paper we develop a method for the construction of  
nets of standard subspaces for antiunitary representations of Lie groups 
in spaces of distributions which are boundary values of holomorphic functions. 
This construction provides for a large class of triples
 $(h,\tau_G, U)$ a realization of the corresponding standard subspace 
$\sV_{(h,\tau_G,U)}$ as some $\sV(\cW)$, associated to an open subset 
$\cW$ of the group $G$ or of a homogeneous space. We obtain such a 
realization on three levels: 
\begin{itemize}
\item[(GL)] the group level, where $\cW \subeq G$ is an open subsemigroup. 
\item[(SL)] the level of symmetric spaces, where $\cW$ is  an open 
domain in the symmetric space $G/G^{\tau_G}$. 
\item[(CL)] the conformal level, where $\cW$ is an open domain in a {\it Jordan space-time}, 
i.e., the universal cover of the conformal completion 
of a euclidean Jordan algebra~$E$. These are the Jordan space-times in the sense of 
G\"unaydin \cite{Gu93}, resp., the simple space-time manifolds in the sense of 
Mack--de Riese (\cite{MdR07}). 
\end{itemize}
In Physics, these three levels arise for one-dimensional conformal 
field theory, where $G$ is the $3$-dimensional M\"obius group 
$\Mob \cong \PSL_2(\R)$, the corresponding Jordan space-time 
is the real line, considered as the simply connected covering of the circle, and 
the corresponding causal symmetric space is the $2$-dimensional 
Anti-de Sitter space $\AdS^2$. For the universal cover 
$G = \tilde\SO_{2,d}(\R)_0$ of the conformal group of 
$d$-dimensional Minkowski space, 
the corresponding Jordan space-time is the simply connected cover of the 
conformal completion of Minkowski space. \\

The following table contains some information on the 
simple hermitian Lie algebras of tube type $\g$, the corresponding 
simple euclidean Jordan algebra $E \cong \g^1(h) = \ker(\ad h-\1)$, the rank of 
$E$, which is the real rank of the simple 
real Lie algebra $\g$, the subalgebra $\fh = \g^0(h) = 
\ker(\ad h)$, 
and the topology of the simple Jordan space-time~$M$, 
considered as a product of $\R$ with a Riemannian symmetric space. Here 
$E_6^c$ stands for the simply connected compact Lie group of type $E_6$ and 
$F_4^c \subeq E_6^c$ for a connected subgroup of type $F_4$.\\

\hspace{-8mm}
\begin{tabular}{||l|l|l|l|l||}\hline
{} $\g$ \mbox{(conf. Lie alg)}  &\ $E$\ \mbox{(Jordan alg.)} & ${\rm rk}(E)$ &\ \ $\fh = \g^0(h)$ \phantom{\Big(}& $M$ \\ 
\hline\hline 
 $\so_{2,n}(\R), n > 2$ & \phantom{\Big(} $\R^{1,n-1}$ & $2$ & $\R \oplus \so_{1,n-1}(\R)$  
& $\R \times \bS^{n-1}$ \\ 
\hline
 $\sp_{2n}(\R)$ &  $\Sym_n(\R)$ & $n$ & $\R \oplus \fsl_n(\R) \cong \gl_n(\R)$  
& $\R \times \SU_n(\R)/\SO_n(\R)$ \\ 
\hline
 $\su_{n,n}(\C)$ &  $\Herm_n(\C)$ & $n$ & $\R \oplus \fsl_n(\C)$  
& $\R \times \SU_n(\C)$  \\ 
\hline
 $\so^*(4n)$ &  $\Herm_n(\H)$ & $n$ & $\R \oplus \fsl_n(\H) \cong \gl_n(\H)$ 
& $\R \times \SU_{2n}(\C)/\U_n(\H)$  \\ 
\hline
 $\fe_{7(-25)}$ &  $\Herm_3(\bO)$ & $3$ & $\R \oplus \fe_{6(-26)}$  
& $\R \times E_6^c/F_4^c$  \\ 
\hline
\end{tabular}

\vspace{2mm}

Our construction of nets of standard subspaces is based on 
distribution vectors of unitary representations. 
To introduce this concept, we first observe that the subspace 
$\cH^\infty \subeq \cH$ of vectors $v \in \cH$ for which the orbit map 
$U^v \: G \to \cH, g \mapsto U(g)v$, is smooth 
(the {\it smooth vectors}) is dense 
(we assume throughout that $\dim G < \infty$). 
It carries a natural Fr\'echet topology for which the action of 
$G$ on this space is smooth (\cite{Ne10}). 
The space $\cH^{-\infty}$ of continuous antilinear functionals $\eta \: \cH^\infty \to \C$ 
(the {\it distribution vectors}) 
contains in particular Dirac's kets 
$\la \cdot, v \ra$, $v \in \cH$, so that 
we obtain a {\it rigged Hilbert space} 
\[ \cH^\infty \into \cH \into \cH^{-\infty},\] 
where $G$ acts on all three spaces 
by representations denoted $U^\infty, U$ and $U^{-\infty}$, respectively, 
and its Lie algebra $\g$ acts on $\cH^{\infty}$ and $\cH^{-\infty}$ 
by the derived representation 
(see Appendix~\ref{app:a} for details). 

To any real subspace $\sE \subeq \cH^{-\infty}$ and every open subset 
$\cO \subeq G$, we associate the real subspace 
\begin{equation}
  \label{eq:HE}
 \sH_\sE(\cO) := \oline{\spann_\R \{ U^{-\infty}(\phi) \eta \: 
\phi \in C^\infty_c(\cO,\R), \eta \in \sE\}},
\end{equation}
where 
\[ U^{-\infty}(\phi) = \int_G \phi(g)U^{-\infty}(g)\, dg, \qquad 
\phi \in C^\infty_c(G) \] 
denotes the integrated representation of the convolution algebra 
$C^\infty_c(G)$ of test functions on $G$ on the space~$\cH^{-\infty}$.  
This assignment has already 
two obvious properties: 
\begin{itemize}
\item[\rm(Iso)] Isotony: $\cO_1 \subeq \cO_2$ implies 
$\sH_\sE(\cO_1) \subeq \sH_\sE(\cO_2)$. 
\item[\rm(Cov)] Covariance: $U(g)\sH_\sE(\cO) = \sH_\sE(g\cO)$. 
\end{itemize}

One of our main results consists in 
specifying sufficient conditions for 
\eqref{eq:HE} to produce 
standard subspaces including some $\sV_{(h,\tau_G,U)}$ 
obtained from the Brunetti--Guido--Longo construction. 
Important questions in this context are: 
\begin{itemize}
\item[\rm(RS)] Does $\sH_\sE(\cO)$ generate $\cH$ 
for each non-empty open subset $\cO \subeq G$? 
({\it Reeh--Schlieder property}) 
\item[\rm(BW)] If $\sH_\sE(\cO)$ is standard, when is its modular group implemented by a 
one-parameter subgroup of $G$ as in \eqref{eq:bgl}? 
If this is the case for the one-parameter group 
generated by $h \in \g$, we say that $\sH_\sE(\cO)$ has the 
{\it Bisognano--Wichmann property} with respect to~$h$.
\end{itemize}

A structural property with strong impact in this context is the {\it spectrum condition} 
on the infinitesimal generators $\partial U(x)$ of the one-parameter groups 
$(U(\exp tx))_{t \in \R}$. This is the requirement that the closed convex cone 
\[ C_U := \{ x \in \g \: -i \partial U(x) \geq 0 \} \] 
(the {\it positive cone of $U$}) 
is ``large'' in the sense that the ideal $\g_{C_U} := C_U - C_U$ 
satisfies \break $\g = \g_{C_U} + \R  h$. 
We shall see in Theorem~\ref{thm:sw64-gen-abs} 
that this already implies the Reeh--Schlieder property~(RS).
Under more specific assumptions 
on $\sE$ and $\cO$, we show that $\sH_\sE(\cO)$ is actually standard. 
We refer to Section~\ref{sec:3} for details. 

\nin {\bf Content of this paper:} 
Section~\ref{sec:2} is devoted to properties of smooth and distribution 
vectors for representations for which $C_U$ is pointed and generating, i.e., 
$\g_{C_U} = \g$. Then $U$ extends by 
\[ U(g \exp(ix)) = U(g) e^{i\partial U(x)}\]  
to a representation of the complex Olshanski semigroup $S_{C_U} = G \exp(iC_U)$ 
by contractions. 
In Subsection~\ref{subsec:2.1} we study the action of $S_{C_U}$ 
on the space of distribution vectors and establish continuity 
and holomorphy of orbit maps in the weak-$*$-topology. 
This is used in Subsection~\ref{subsec:2.2} to verify the Reeh--Schlieder property 
(Theorem~\ref{thm:sw64-gen-abs}). 

In Section~\ref{sec:3} we turn to a more specific situation, where 
$\sV = \sV_{(h,\tau_G,U)}$ is a standard subspace obtained from the BGL 
construction, and the semigroup 
\begin{equation} \label{eq:sv}
  S_\sV = \{ g \in G \: U(g)\sV \subeq \sV\} 
\end{equation}
of endomorphisms of $\sV$ is large in the sense that its Lie wedge 
\[ \L(S_\sV) = \{ x \in \g \: \exp(\R_+ x) \subeq S_\sV \} \] 
(the set of infinitesimal generators of one-parameter subsemigroups of $S_\sV$) 
spans the Lie algebra~$\g$. Here we build on the previous work 
 \cite{Ne19, Ne19b} of the first author on these semigroups. 
The main results of these two papers are easy to describe. 
In \cite{Ne19} the Lie wedge $\L(S_\sV)$ is calculated. 
As a consequence of its explicit description, 
the assumption that it spans $\g$ implies that 
$\ad h$ defines a $3$-grading in the sense that 
\begin{equation}
  \label{eq:3grad}
\g = \g^1 \oplus \g^0 \oplus \g^{-1} \quad \mbox{ for } \quad 
\g^j := \ker(\ad h - j\id_\g).
\end{equation}
Assuming \eqref{eq:3grad}, the semigroup $S_\sV$ has been completely 
determined in \cite{Ne19b}. To describe the structure of $S_\sV$, 
let 
\[ C_\pm := \pm C_U \cap \g_{\pm 1},\quad \mbox{ and write } \quad 
G_\sV = \{ g \in G \: U(g) \sV = \sV \}\]
for the stabilizer group of $\sV$ in~$G$. Then  
\begin{equation}
  \label{eq:SE}
S_\sV = \exp(C_+) G_\sV \exp(C_-) = G_\sV \exp(C_+ + C_-).
\end{equation}
In the setting of \eqref{eq:3grad}, we define in Subsection~\ref{subsec:3.1} 
the space $\sV^{-\infty}$ of distribution vectors associated to the 
standard subspace $\sV$. 
This requires an extension of 
the symplectic form on $\cH$ to $\cH^\infty \times \cH^{-\infty}$ 
and the symplectic orthogonal space $\sV'$ of~$\sV$, so that we can define 
$\sV^{-\infty}$ as the annihilator of $\sV' \cap \cH^\infty$. 
Here the difficulty  is to identify elements of this space. 
This is achieved in Subsection~\ref{subsec:3.2}, where we extend the following 
characterization of elements in a standard subspaces from \cite[Prop.~2.1]{NOO20} 
to distribution vectors: An element $\xi \in \cH$ is contained in $\sV$ 
if and only if the orbit map $\alpha^\xi \: \R \to \cH, 
\alpha^\xi(t) :=  \Delta_\sV^{-it/2\pi}\xi$ extends to a 
continuous map on the closure of the strip 
\[ \cS_\pi = \{ z \in \C \: 0 <\Im z < \pi\}, \] 
which is holomorphic on $\cS_\pi$ and satisfies the boundary 
relation $\alpha^\xi(\pi i) = J_\sV \xi$. 
A suitable extension of this requirement to distribution vectors specifies 
a linear subspace $\cH^{-\infty}_{{\rm ext},J}\subeq \cH^{-\infty}$ 
which is invariant under 
$U^{-\infty}(S_\sV^0)$ and $U^{-\infty}(C^\infty(S_\sV^0))$. 
This leads to the inclusion 
$\cH^{-\infty}_{{\rm ext},J}\subeq \sV^{-\infty}$ 
(Lemma~\ref{lem:key}) which is one of our key tools. 
We conclude Section~\ref{sec:3} with the observation that 
$\cH^{-\infty}_{\rm ext}$ is invariant under the Lie algebra and 
discuss the irreducible antiunitary representation of the $ax+b$-group 
on $L^2(\R_+)$ in some detail. 

The results in Section~\ref{sec:3} are valid 
for general Lie groups whose Lie algebra contains an invariant cone 
$C$ and an element $h$ defining a $3$-grading for which 
the cones $C_\pm = \pm C \cap \g^{\pm 1}$ are generating. 
To construct representations to which the theory developed in 
Section~\ref{sec:3} applies, we specialize in Section~\ref{sec:4} 
to semisimple Lie groups. Assuming, in addition, that $\g^0$ contains 
no non-zero ideals, $\g$ is a direct sum of simple hermitian ideals of tube type 
and $E := \g^1$ thus carries the structure of a unital euclidean Jordan algebra 
such that $C_+$ is the closure of the open positive cone $E_+$ of invertible 
squares in~$E$ (see \cite{FK94} and the table above). 

This structure is used in Section~\ref{sec:4} to show that every 
irreducible antiunitary representation $(U,\cH)$ of $G_\tau$, which is 
{\it $C$-positive} in the sense that $C \subeq C_U$, can be realized on 
a Hilbert space $\cH_\rho$ of holomorphic vector-valued functions 
on the tube domain $\cT := E + i E_+$ on which the group $G$ acts as a 
group of biholomorphic automorphisms,  and $\tau_G$ corresponds to the 
antiholomorphic involution $\tau_E(z) = -\oline z$. The input for this construction 
is a real irreducible representation $(\rho, \cK_\R)$ of the simply 
connected Lie group $\tilde H$ with Lie algebra $\g^0$ which satisfies 
a certain positivity condition. Then $\cH_\rho \subeq \Hol(\cT,\cK)$, 
the space of holomorphic functions with values in the 
complexification $\cK$ of $\cK_\R$. 
The positivity condition for $(\rho,\cK_\R)$ is hard to evaluate explicitly 
for non-scalar representations, an explicit classification 
can be derived by combining the classification 
of unitary highest modules of $\g$,  
due to Enright, Howe and Wallach (\cite{EHW83}), 
with the translation into the present context, carried out in \cite{HN01}. 
The main point of this realization is that evaluation in $0 \in \partial \cT$ 
defines a continuous linear map $\cH_\rho^\infty \to \cK$, which embeds 
$\cK$ into $\cH^{-\infty}$. 
In Section~\ref{sec:5} we show that this leads to a 
real subspace $\sE \subeq \cK$ contained in $\sV^{-\infty}$, 
so that we can apply Section~\ref{sec:3}. We then have the identity 
$\sV = \sH_\sE(S_\sV^0)$ and obtain a net $\sV(\cO) := \sH_\sE(\cO)$ 
of cyclic subspaces associated to non-empty open subsets $\cO \subeq G$. 
The set of generalized wedge domains in $G$ in the sense of (GL) is the set 
\[ \cW = \{ g S \: g \in G \}\]
of left translates of the connected component $S$ of the 
open semigroup $S_\sV^0$ containing $e$ in its closure. If $\cO$ is contained 
in a left translate of the semigroup $S_\sV^0$, then 
the cyclic subspace $\sV(\cO)$ is standard. 

As the real subspace $\sE$ is invariant under the subgroups 
$H := \la \exp_G \g^0 \ra$ and $P^- := \exp(\g^{-1})H$ of $G$, 
it easily follows that $\sV(\cO) = \sV(\cO \cdot H) = \sV(\cO \cdot P^-)$, 
so that these nets of cyclic subspaces also define nets on the homogeneous spaces 
$G/H$ and $G/P^-$. Here $G/H$ is a compactly causal symmetric spaces 
of Cayley type (\cite{HO97}) und $G/P^-$ is a  Jordan space-time in the sense of 
G\"unaydin \cite{Gu93} and a simple space-time manifold in the sense of 
Mack--de Riese (\cite{MdR07}).

\nin {\bf Perspectives:} In the present paper we construct nets 
of standard subspaces on homogeneous spaces which are associated 
to unitary representations $(U,\cH)$ for which the positive cone $C_U$ 
is non-trivial. If this cone is trivial, one needs other 
techniques to specify suitable subspaces $\sE \subeq \sV^{-\infty}$. 
A prototypical example is  the action of the Lorentz 
group $G = \SO_{1,d}(\R)_0$ on de Sitter space $\dS^d$, which is a 
non-compactly causal symmetric space. 
In the physics literature one finds several 
replacements for the spectrum condition that are used to deal 
with quantum fields on curved spaces such as de Sitter space. 
Of particular relevance is the 
``microlocal spectrum condition''; 
see \cite{TS16} and \cite{PEGW19}.

We expect that the  reflection positivity condition and the 
KMS condition that characterize the modular objects 
(see the appendix of \cite{NOO20} for details) 
can be used to express a condition on the distribution vectors in $\sE$ 
to ensure that the subspace 
$\sH_\sE(\cW)$ associated to a ``wedge domain''  $\cW \subeq M$  is 
the standard subspace corresponding to the pair 
$(h,\tau)$. 

In the forthcoming paper \cite{NO20} we develop a theory of 
generalized wedge domains in causal symmetric spaces that 
will shed additional light on the class of domains $\cW \subeq M$ 
for which one may expect a relation like $\sH_\sE(\cW) = \sV$. 

In all these constructions, a good understanding of the class of 
generalized wedge domains is of crucial importance. 
This motivated the abstract approach to these domains 
in \cite{MN20}, where the set 
\[ \cG(G) := \{ (x,\sigma) \in \g \times (G \times \{\tau_G\}) 
\subeq \g \times G_\tau\:
\Ad(\sigma) x = x, \sigma^2 = e \} \] 
is studied as a candidate of an index set for nets of standard subspaces. 
In this context the {\it Euler couples}, i.e., those for which 
$\ad x$ is diagonalizable with eigenvalues $\{-1,0,1\}$ and 
$\Ad(\sigma) = e^{\pi i \ad x}$ are of particular relevance. 
For instance, the assumptions in Section~\ref{sec:3} imply 
that $(h,\tau)$ is an Euler couple. 

\nin {\bf Acknowledgment:} We thank Roberto Longo for asking us on several 
occasions to construct quantum field theories on Lie groups. The present paper 
is a first step in this direction. We are most 
grateful to Daniel Oeh for several valuable 
comments on an earlier version of this manuscript and for pointing 
out an error in Theorem~\ref{thm:5.4}.

\subsection*{Notation} 

\begin{itemize}
\item For a Lie group $G$ with neutral element $e$, the 
identity component is denoted $G_0$. 
We write $\g$ for its Lie algebra, 
$\Ad \: G \to \Aut(\g)$ for the adjoint action of $G$ on $\g$, induced by the 
conjugation action of $G$ on itself, and $\ad x(y) = [x,y]$ for the adjoint 
action of $\g$ on itself. 
\item For a group action $G \times M \to M$ and $m \in M$, we write 
$G^m := \{ g \in G \: g.m = m\}$ for the stabilizer subgroup of $m$. 
\item For a connected Lie group $G$, we write $q_G \: \tilde G \to G$ 
for the universal covering morphism and 
$\eta_G \: G \to G_\C$ for the universal complexification. 
\item If $M$ is a smooth manifold, we write $C^\infty_c(M)$ for the space 
of complex-valued test functions on $M$, endowed with the natural LF topology, 
and $C^{-\infty}_c(M)$ for the space of {\bf antilinear} continuous 
linear functionals  on this space, i.e., the space of 
distributions on $M$. 
\item We likewise consider tempered distributions 
$D \in \cS'(E)$ on a real finite dimensional vector space $E$ 
as antilinear functionals on the Schwartz space $\cS(E)$. 
The {\it Fourier transform of an $L^1$-function $f$} on $E$ is defined by 
\begin{equation}
  \label{eq:ftrafo}
\hat f(\lambda) := \int_E e^{-i\lambda(x)} f(x) \,  d\mu_E(x), \quad 
\lambda \in E^*, 
\end{equation}
where $\mu_E$ is a Haar measure on $E$. 
For tempered distributions $D \in \cS'(E)$, we define the Fourier transform by 
\begin{equation}
  \label{eq:2}
\hat D(\phi) := D(\tilde \phi), \quad \mbox{ where} 
\quad 
\tilde\phi(\lambda) 
:= \hat\phi(-\lambda) 
= \int_E e^{i\lambda(x)} \phi(x) \,  d\mu_E(x).
\end{equation}
The distribution 
$D_f(\phi) := \int_E \oline{\phi(x)} f(x)\, d\mu_E(x)$, defined by 
an $L^1$-function, then satisfies  $\hat{D_f} = D_{\hat f}$. 
\end{itemize}

\section{Distribution vectors of $C$-positive representations} 
\mlabel{sec:2}

Let $G$ be a Lie group with Lie algebra $\g$ 
and $C \subeq \g$ a closed convex pointed generating 
$\Ad(G)$-invariant cone (see \cite{Ne99} for a structure theory of these 
configurations). 
In this section we study distribution vectors of 
unitary representations $(U,\cH)$ of $G$ which are 
$C$-positive in the sense that $C \subeq C_U$.\begin{footnote}{In Section~\ref{sec:3}, this 
section will be applied to the 
ideal $\g_C := C - C$ generated by the invariant cone~$C$.} \end{footnote} 
Then the representation $U \: G \to \U(\cH)$ extends analytically 
to a representation of an Olshanski semigroup $S_C= G \exp(iC)$. 

In Subsection~\ref{subsec:2.1} we start with some results concerning 
the action of the semigroup $S_C$ on the spaces of smooth and 
distribution vectors. In particular, we show that 
the closed semigroup $S_C$ acts on the space $\cH^{-\infty}$ of distribution vectors  with weak-$*$-continuous orbit maps that are holomorphic on the 
interior of~$S_C$. From this we derive 
the interesting result that every distribution vector 
$\eta \in \cH^{-\infty}$ defines a $G$-right equivariant map 
$\cH^{-\infty} \to \Hol(S_C^0)$ whose range consists of holomorphic 
functions with distributional boundary values 
(Theorem~\ref{thm:2.5}). 

In Subsection~\ref{subsec:2.2} we turn to real subspaces $\sH_\sE(\cO)$ 
generated by $U^{-\infty}(C^\infty_c(\cO,\R))\sE$, where 
$\cO \subeq G$ is open and $\sE \subeq \cH^{-\infty}$ 
is a real subspace. Here the main result is the 
Reeh--Schlieder property (Theorem~\ref{thm:sw64-gen-abs}), asserting 
that $\sH_\sE(\cO)$ is total in $\cH_\sE(G)$ for every 
non-empty open subset $\cO \subeq G$ and every 
real subspace $\sE \subeq \cH^{-\infty}$. 

\subsection{The semigroup action on distribution vectors} 
\mlabel{subsec:2.1} 

We start by recalling the construction of the complex Olshanski semigroup 
$S_C$ associated to the invariant cone~$C \subeq \g$. 

\begin{defn} \mlabel{def:olshanski} (Olshanski semigroups) 
Let $G$ be a connected Lie group with Lie algebra $\g$ 
 and $C \subeq \g$ be a pointed generating 
$\Ad(G)$-invariant closed convex cone. 
Then $C$ is {\it elliptic} 
in the sense that, for every element $x \in C^0$, 
the derivation $\ad x$ is semisimple with purely imaginary spectrum 
(\cite[Prop.~VII.3.4(b)]{Ne99}). 
The corresponding {\it Olshanski semigroup} 
$S_C$ is defined as follows.

Let $q_G \:  \tilde G \to G$ be the universal covering group 
of $G$ and $\tilde G_\C$ be the $1$-connected Lie group with Lie algebra $\g_\C$, 
so that the universal complexification 
$\eta_{\tilde G} \: \tilde G \to \tilde G_\C$ simply is the 
canonical morphism of Lie groups 
for which $\L(\eta_{\tilde G}) \: \g \into \g_\C$ is the inclusion. 
\begin{footnote}{In general the map $\eta_G$ is not injective, as the example $G = \tilde\SL_2(\R)$ with
$G_\C = \SL_2(\C)$ shows.}   
\end{footnote}
As $\tilde G_\C$ is simply connected, the complex conjugation 
on $\g_\C$ integrates to an antiholomorphic 
involution $\sigma \: \tilde G_\C \to \tilde G_\C$ with 
$\sigma \circ \eta_{\tilde G} = \eta_{\tilde G}$, and this implies that
$\eta_{\tilde G}(\tilde G)$ coincides with the subgroup $(\tilde G_\C)^\sigma$ 
of $\sigma$-fixed points in $\tilde G_\C$.
\begin{footnote}{Since $\tilde G_\C$ is simply 
connected, this subgroup is connected by \cite[Thm.~IV.3.4]{Lo69}.}
\end{footnote}
As $C$ is elliptic, 
Lawson's Theorem (\cite[Thm.~XIII.5.6]{Ne99}) asserts that
\[S'_C := (\tilde G_\C)^\sigma \exp(iC) \subeq \tilde G_\C \]
is a closed subsemigroup of $\tilde G_\C$ stable under the antiholomorphic
involution $s^* := \sigma(s)^{-1}$ and the polar map
\[ (\tilde G_\C)^\sigma \times C \to S_C, \quad (g,x) \mapsto g \exp(ix) \]
is a homeomorphism. 
Next we observe that $\ker \eta_{\tilde G}$ is a discrete subgroup of 
$\tilde G$ and define 
$\tilde S_C$ as the simply connected covering space of $S_C'$ and 
\[ S_C := \tilde S_C/\ker(q_G) \] 
(\cite[Def.~XI.1.11]{Ne99}). The quotient map is denoted 
$q_{S_C} \: \tilde S_C \to S_C$. It is the universal covering of~$S_C$.
We write 
\[ \tilde\exp \: \g + i C \to \tilde S_C \] 
for the continuous lift of the exponential function 
$\exp \:  \g + i C = \L(S_C')\to S_C' \subeq \tilde G_\C$ 
and 
\[ \exp := q_{S_C} \circ \tilde\exp \: \g + i C \to S_C.\]
This function extends the exponential 
function of the group~$G$ and,  for every $x \in \g + i C$, 
the curve $\gamma_x(t) := \exp(tx)$ 
is a continuous one-parameter semigroup of $S_C$.  
Basic covering theory implies that $S_C$ inherits an involution 
given by 
\begin{equation}
  \label{eq:eq*}
 (g \exp(ix))^* = \exp(ix)g^{-1} = g^{-1} \exp(\Ad(g)ix)
\end{equation}
 and a homeomorphic  polar map $G \times C \to S_C,  (g,x) \mapsto g \exp(ix)$. 

If $C$ has interior points, then 
the polar map maps $(\tilde G_\C)^\sigma \times C^0$ 
diffeomorphically onto the interior $S_{C^0}'$ of $S_C'$. 
Hence $S_C^0 = S_{C^0}= G \exp(i C^0)$ is a complex manifold 
with a holomorphic multiplication, and the exponential function 
$\g + i C^0 \to S_{C^0}$ is holomorphic, 
whereas the involution $*$ is antiholomorphic 
(\cite[Thm.~XI.1.12]{Ne99}). 
\end{defn}

The following theorem ensures the existence of extensions of a 
unitary representation~$U$ of $G$ to the semigroup $S_C$, provided its positive cone $C_U$ contains~$C$.

\begin{thm} {\rm(Holomorphic Extension Theorem)} \mlabel{thm:extens}
If $(U,\cH)$ is a unitary representation of $G$ 
satisfying $C \subeq C_U$, then it extends by 
\[ U(g \exp(ix)) = U(g) e^{i\partial U(x)}, \quad g \in G, x \in C, \] 
to a $*$-representation 
of the closed complex Olshanski semigroup $S_C= G \exp(iC)$ by contractions on $\cH$ 
which defines a continuous action of $S_C$ on $\cH$. 
If $C$ has interior points, then $U$ defines a holomorphic map 
$S_{C^0} \to B(\cH)$. 
\end{thm}

\begin{prf} The existence of the holomorphic extension to $S_C$ follows from 
\cite[Thm.~XI.2.5]{Ne99} and its weak continuity from 
\cite[Prop.~XI.3.7]{Ne99}. As it is a representation by contractions, 
strong continuity follows from \cite[Cor.~IV.1.18]{Ne99}. 
Since the representation of $S_C$ on $\cH$ is strongly continuous 
and by contractions, it defines a continuous action of $S_C$ on $\cH$ 
because 
\[ \|U(s')\xi' - U(s)\xi \|
\leq  \|U(s')(\xi' - \xi) \| + \|(U(s') - U(s))\xi\| 
\leq \|\xi' - \xi\| +  \|(U(s') - U(s))\xi\| \] 
for $\xi,\xi' \in \cH$ and $s,s' \in S_C$.  
\end{prf}

The main purpose of this subsection is to analyze the representation 
of the semigroup $S_C$ on the spaces of smooth and distribution 
vectors. The first crucial step is to show that $U(S_C)$ leaves 
$\cH^\infty$ invariant (Proposition~\ref{prop:2.2}).

\begin{lem}
  \mlabel{lem:2.1} 
The operators $(U(s))_{s \in S_C}$ preserve the dense subspace 
$\cH^\infty$. For every smooth vector $\xi \in \cH^\infty$, the orbit map 
\[ U^\xi \: S_C \to \cH^\infty, \quad 
s \mapsto U(s)\xi \] 
is continuous on $S_C$ and holomorphic on the interior $S_C^0$.   
\end{lem}

\begin{prf} We consider $\cH$ as a subspace of $\cH^{-\infty}$ 
as in \eqref{eq:embindistr} in Appendix~\ref{app:a}. 
As the locally convex topology on $\cH^\infty$ is defined by the 
topological embedding 
\begin{equation}
  \label{eq:topsmo}
\cH^\infty \into \prod_{D \in \cU(\g)} \cH, \quad 
\xi \mapsto (\dd U(D)\xi)_{D \in \cU(\g)}
\end{equation}
(see \eqref{eq:tophinfty} in Appendix~\ref{app:a}), 
we have to show that: 
\begin{itemize}
\item[\rm(a)] For every $D \in \cU(\g)$ and $\xi \in \cH^\infty$, 
we have $f(s) := \dd U^{-\infty}(D) U(s) \xi \in \cH$ 
(Lemma~\ref{lem:charsmooth}). 
\item[\rm(b)] The function $f \: S_C \to \cH$ is continuous and holomorphic on~$S_C^0$. 
\end{itemize}

We first observe that the adjoint action of $G$ on $\g$ 
extends to a locally finite holomorphic representation of the universal 
complexification $G_\C$ (\cite[Thm.~15.1.4]{HN12}). As the semigroup 
$S_C$ has a natural continuous homomorphism 
\[ \eta_S \: S_C \to G_\C, \quad \eta_S(g \exp(ix)) 
= \eta_G(g) \exp_{G_\C}(ix) \] 
which is holomorphic on the interior 
(Definition~\ref{def:olshanski}), 
we thus obtain in particular a continuous representation  
\[ \Ad_\C \: S_C \to \Aut(\cU(\g))  \] 
of $S_C$ on the complex enveloping algebra 
$\cU(\g)$ which is holomorphic on $S_C^0$. 
For $g \in G$ and $D \in \cU(\g)$, we have the relation 
\begin{equation}
  \label{eq:adrel1}
\dd U(D) U(g) = U(g) \dd U(\Ad(g^{-1})D) 
\: \cH^\infty \to \cH.
\end{equation}
We claim that 
\begin{equation}
  \label{eq:s-id}
 \dd U^{-\infty}(D) U(s) = U(s) \dd U(\Ad_\C(s)^{-1}D) 
\: \cH^\infty \to \cH^{-\infty} 
\quad \mbox{ for }\quad 
s \in S_C.
\end{equation}
Note that both sides define linear maps $\cH^\infty \to \cH^{-\infty}$ 
and that the right hand side maps into~$\cH$. To verify 
\eqref{eq:s-id}, we have to show that 
\begin{equation}
  \label{eq:s-id2}
\la \xi, \dd U^{-\infty}(D) U(s)\eta \ra 
= \la \xi, U(s) \dd U(\Ad_\C(s)^{-1}D) \eta \ra
\quad \mbox{ for } \quad 
\xi,\eta \in \cH^\infty.
\end{equation}
The left hand side equals 
$\la \dd U(D^*) \xi,  U(s)\eta \ra,$
which is continuous on $S_C$ and holomorphic on the interior. 
For the right hand side we obtain 
\[ \la \xi, U(s) \dd U(\Ad_\C(s)^{-1}D) \eta \ra 
= \la U(s^*)\xi, \dd U(\Ad_\C(s)^{-1}D) \eta \ra.\] 
Here $U(s^*)\xi$ is continuous on $S_C$ and antiholomorphic 
on $S_C^0$ and $\dd U(\Ad_\C(s)^{-1}D) \eta$ is also 
continuous on $S_C$ and holomorphic on $S_C^0$. Therefore the 
right hand side of \eqref{eq:s-id2} 
is continuous on $S_C$ and holomorphic on the 
interior. As both sides of \eqref{eq:s-id2} coincide on $G$ by 
\eqref{eq:adrel1}, they are equal 
(\cite[Lemma~A.III.6]{Ne99}). 
This implies \eqref{eq:s-id}, and hence that 
$\dd U^{-\infty}(D) U(s) \cH^\infty \subeq \cH$. 
As 
\[ \cH^\infty = \{ \xi \in \cH \subeq \cH^{-\infty} \: (\forall D \in \cU(\g)) 
\ \dd U^{-\infty}(D)\xi \in \cH\} \] 
by Lemma~\ref{lem:charsmooth}, it follows that 
$U(S_C)$ preserves $\cH^\infty$. 
Since the right hand side of \eqref{eq:s-id} 
defines a continuous linear map $\cH^\infty \to \cH$, 
the definition of the topology on $\cH^\infty$ by the embedding 
\eqref{eq:topsmo} also shows that the restrictions 
\begin{equation}
  \label{eq:dag}
 U^\infty(s) := U(s)\res_{\cH^\infty} \: \cH^\infty \to \cH^\infty, \quad s \in S_C,  
\end{equation}
are continuous linear maps. 
In particular, we have 
$f(s) \in \cH^\infty$ for $s \in S_C$. To see that $f$ is continuous, 
we write it with \eqref{eq:s-id} as 
\[ f(s) = U(s) \dd U(\Ad_\C(s)^{-1}D) \xi.\] 
As the representation of $S_C$ defines a continuous 
action on $\cH$ (Theorem~\ref{thm:extens}),
the continuity of $f$ follows from the continuity of the map 
\begin{equation}
  \label{eq:admap}
S_C \to \cH, \quad s \mapsto \dd U(\Ad_\C(s)^{-1}D) \xi
\end{equation}
which actually ``extends'' to a holomorphic map on all of~$G_\C$. 

That $f$ is holomorphic on $S_C^0$ follows likewise from the holomorphy 
of the action of $S_C^0$ on the complex manifold $\cH$ and the 
holomorphy of \eqref{eq:admap} on $S_C^0$. 
\end{prf}

With Lemma~\ref{lem:2.1}, we immediately get: 

\begin{prop} \mlabel{prop:2.2}
The prescription 
\[ U^{-\infty}(s)\alpha := \alpha\circ U^\infty(s^*) 
 \quad \mbox{ for } \quad 
\alpha \in \cH^{-\infty}, s \in S_C, \] 
defines a representation of $S_C$ on $\cH^{-\infty}$ 
whose orbit maps are weak-$*$-continuous and weak-$*$-holomorphic on 
$S_C^0$. 
\end{prop} 

Although the closed semigroup $S_C$ acts on smooth and distribution 
vectors, its interior $S_{C^0}$ has particular regularizing properties 
which are described in the following two lemmas. 

\begin{lem} For $s \in S_C^0$, we  have $U(s)\cH\subeq \cH^\infty$, and the maps 
  \begin{equation}
    \label{eq:smoothop}
 U(s) \: \cH \to \cH^\infty, \qquad s \in S_C^0, 
  \end{equation}
are continuous.   
\end{lem}

\begin{prf} Let $\xi \in \cH$. 
Then the orbit map $U^\xi \:  S_C^0 \to \cH, U^\xi(s) := U(s)\xi$ 
is  holomorphic, so that its range lies in particular in 
the space of smooth vectors for $G$. 
For $D = x_1 \cdots x_n \in \cU(\g)$, $x_j \in \g$, we consider the 
differential operator defined by 
\[ D^R \: C^\infty(S_C^0) \to C^\infty(S_C^0), \quad 
(D^R\phi)(s) 
= \frac{\partial^n}{\partial t_1 \cdots \partial t_n}\Big|_{t_j = 0} 
\phi(\exp(t_n x_n) \cdots \exp(t_1 x_1)s).\] 
That the maps \eqref{eq:smoothop} 
are continuous follows with \eqref{eq:tophinfty} in Appendix~\ref{app:a} 
from the fact that, for every $D \in \cU(\g)$,  the composition 
$\dd U(D) \circ U^\xi$ is obtained by applying the right invariant 
differential operator $D^R$ 
to the smooth functions $g \mapsto U^\xi(gs)$, $s \in S_C^0$. 
\end{prf}

\begin{lem} \mlabel{lem:9.14} 
For $s \in S_C^0$, we have $U^{-\infty}(s)\cH^{-\infty} 
\subeq \cH^\infty$ and the orbit maps 
\begin{equation}
  \label{eq:orbitmapint}
U^{-\infty,\eta} \: S_C^0 \to \cH^\infty, \quad
s \mapsto U^{-\infty}(s) \eta, \qquad 
\eta \in \cH^{-\infty},  
\end{equation}
are holomorphic. 
\end{lem}

\begin{prf} For $D \in \cU(\g)$,  let $D^L$ denote the corresponding 
left invariant differential operator. 
For $x_1, \ldots, x_n \in \g$ and  $D = x_1 \cdots x_n$, it is defined by 
\[ (D^L\phi)(g) 
= \frac{\partial^n}{\partial t_1 \cdots \partial t_n}\Big|_{t_j = 0} 
 \phi(g \exp(t_1x_1) \cdots \exp(t_n x_n)). \] 
 
For $\xi \in \cH$, the orbit map $U^\xi \: S^0_C \to \cH$ is 
holomorphic  and $G$-equivariant with respect to left multiplications in the sense that 
\begin{equation}
  \label{eq:leftcov}
U^\xi \circ \lambda_g = U(g) \circ U^\xi \quad \mbox{ for } \quad g \in G,
\lambda_g(h) = gh, \end{equation}
in particular $U^\xi(S_C^0) \subeq \cH^\infty$.
It follows that, for $D \in \cU(\g)$, the function 
\[ D^L U^\xi \: S_C^0 \to \cH, \] 
obtained by applying a left invariant differential operator, 
is also holomorphic and $G$-equivariant in the sense of \eqref{eq:leftcov}, 
hence takes values in~$\cH^\infty$. 

Next we observe that, for $\eta \in \cH^\infty$ and 
$s \in S_C^0$,  the relation 
\[ \la U^\xi(s), \eta \ra = \la \xi, U(s^*) \eta \ra \] 
leads for $D := x_1 \cdots x_k$, $x_j \in \g$, to 
\begin{align*}
\la (D^L U^\xi)(s), \eta \ra 
&= \frac{\partial^k}{\partial t_1 \cdots \partial t_k}\Big|_{t_1= \cdots = t_k = 0} 
\la U^\xi(s \exp(t_1 x_1) \cdots \exp(t_k x_k)), \eta \ra  \\
&= \frac{\partial^k}{\partial t_1 \cdots \partial t_k} \Big|_{t_1= \cdots = t_k = 0}  
 \la U(\exp(t_1 x_1) \cdots \exp(t_k x_k))\xi, U(s^*) \eta \ra \\
&=  \la \dd U^{-\infty}(D)\xi, U(s^*) \eta \ra 
=  \la U^{-\infty}(s)\dd U^{-\infty}(D)\xi, \eta \ra.
\end{align*}
Here we use that $U(s^*)\eta \in \cH^\infty$ by Lemma~\ref{lem:2.1} and that 
\[ \frac{\partial^k}{\partial t_1 \cdots \partial t_k} 
\Big|_{t_1= \cdots = t_k = 0}  
  U(\exp(t_1 x_1) \cdots \exp(t_k x_k))\xi  
= \dd U^{-\infty}(D) \xi \in \cH^{-\infty} \] 
holds in the weak-$*$-topology on $\cH^{-\infty}$. 
For $\xi \in \cH$, we thus obtain 
\[ \cH^\infty \ni (D^L U^\xi)(s) = U^{-\infty}(s) \dd U^{-\infty}(D) \xi.\]
As $\cH^{-\infty}$ is spanned by $\dd U^{-\infty}(\g)\cH$ 
(see Lemma~\ref{lem:charsmooth}(b)), it follows that 
$U^{-\infty}(s)\cH^{-\infty} \subeq~\cH^\infty$. 

To see that the orbit maps \eqref{eq:orbitmapint} 
are holomorphic, we have to show that the maps 
\[  f\: S_C^0 \to \cH, \quad
s \mapsto \dd U(D) U^{-\infty}(s) \dd U^{-\infty}(D')\xi, \qquad  \xi \in \cH, D, D' \in 
\cU(\g),\] 
are holomorphic because $\cH^{-\infty}$ is spanned by $\dd U^{-\infty}(\cU(\g))\cH$ 
(Lemma~\ref{lem:charsmooth}(b)). 
As \[ f = D^R (D')^L U^\xi \: S_C^0 \to \cH \] and $U^\xi$ is holomorphic, 
$f$ is holomorphic as well. 
\end{prf}

The following theorem shows that any distribution vector generates a 
subrepresentation that can be realized in holomorphic functions on 
$S_C^0$ with distributional boundary values. 

\begin{thm} {\rm(Realization in holomorphic functions)} 
  \mlabel{thm:2.5} Let $\eta \in \cH^{-\infty}$. By 
\[ j_\eta^S \: \cH^{-\infty} \to \Hol(S_C^0), \quad 
j_\eta^S(\alpha)(s) := \alpha(U^{-\infty}(s^*)\eta), \] 
we obtain a map which intertwines $U^{-\infty}$ with the action of 
$S_C$ on $\Hol(S_C^0)$ by right translations. 
Every function $j_\eta^S(\alpha) \in \Hol(S_C^0)$ has the 
distributional boundary value $j_\eta^\vee(\alpha) 
:= j_\eta(\alpha)^\vee\in C^{-\infty}(G)$ 
in the sense that 
\begin{equation}
  \label{eq:ddag}
j_\eta^\vee(\alpha)(\phi) 
= \lim_{s \to e} \int_G \oline{\phi(g)} j_\eta^S(\alpha)(gs)\, dg 
\quad \mbox{ for } \quad 
\phi \in C^\infty_c(G).  
\end{equation}
\end{thm}

\begin{prf} (a) 
As $U^{-\infty}(s)\eta \in \cH^\infty$ for $s \in S_C^0$ by 
Lemma~\ref{lem:9.14}, the function $j_\eta^S(\alpha) \: S_C^0 \to \C$ 
is defined for each distribution vector $\alpha$. 
That it is holomorphic follows from the holomorphy of the orbit 
maps \eqref{eq:orbitmapint} in Lemma~\ref{lem:9.14} 
and the antilinearity of $\alpha$ on $\cH^\infty$. 

\nin (b) By definition (see \eqref{eq:dist-inc2}), 
the left hand side of \eqref{eq:ddag} equals 
\[ j_\eta^\vee(\alpha)(\phi)  = \alpha(U^{-\infty}(\phi^\vee)\eta)
\quad \mbox{ for } \quad 
\phi^\vee(g) := \Delta_G(g)^{-1} \phi(g^{-1}).\] 
Using that 
\[ \int_G \phi(g) U(g^{-1})\, dg 
=   \int_G \phi^\vee(g) U(g)\, dg = U(\phi^\vee),\] 
we obtain for the integral on the right hand side of \eqref{eq:ddag} 
\[ \int_G \oline{\phi(g)} \alpha(U^{-\infty}(s^*g^{-1})\eta)\, dg 
= \alpha\big(U^{-\infty}(s^*) U^{-\infty}(\phi^\vee)\eta\big) 
= (U^{-\infty}(s)\alpha)\big(U^{-\infty}(\phi^\vee)\eta\big).\] 
Evaluating on the smooth vector $U^{-\infty}(\phi^\vee)\eta$, the assertion 
now follows from the weak-$*$-continuity of the map 
$S_C \to \cH^{-\infty}, s \mapsto U^{-\infty}(s)\alpha$ 
(Proposition~\ref{prop:2.2}). 
\end{prf}

\subsection{Real subspaces generated by distribution vectors} 
\mlabel{subsec:2.2}

In this subsection we turn to real subspaces of $C$-positive 
unitary representations $(U,\cH)$ of $G$. Our main result 
is the Reeh--Schlieder property (Theorem~\ref{thm:sw64-gen-abs}). 
We shall use the following terminology concerning real subspaces of 
complex Hilbert spaces, which is inspired by cyclic and separating 
vectors for Neumann algebras, as defined in  the introduction.

\begin{defn} Let $\sH$ be a closed real subspace of the complex Hilbert space~$\cH$.

\nin (a) We write 
\begin{equation}
  \label{eq:sympdual}
\sH' := \{ \xi \in \cH \: 
(\forall \eta \in \sH) \Im \la \xi, \eta \ra = 0\} 
\end{equation}
for its {\it symplectic orthogonal space} and note that $\sH = \sH''$ follows 
from the closedness of $\sH$. 

\nin (b) $\cH$ is said to be 
\begin{itemize}
\item {\it cyclic} if $\sH + i \sH$ is dense in $\cH$. 
\item {\it separating} if $\sH \cap i \sH = \{0\}$. 
As $\sH \cap i \sH = (\sH' + i \sH')'$, this is equivalent to is $\sH'$ being cyclic. 
\item {\it standard} if it is cyclic and separating. 
\end{itemize}
\end{defn}

\begin{defn} \mlabel{def:HE}
Let $(U,\cH)$ be a unitary representation of $G$. 
If $\sE \subeq \cH^{-\infty}$ is a real linear subspace, then we consider 
for an open subset $\cO \subeq G$ the subspaces 
\begin{equation}
  \label{eq:he1}
\sH_\sE(\cO) := \oline{\spann_\R \big(U^{-\infty}(C^\infty_c(\cO,\R))\sE\big)} 
\subeq \cH_\sE(\cO) := \oline{\spann_\C \big(U^{-\infty}(C^\infty_c(\cO))\sE\big)}
\subeq \cH.
\end{equation}
For $\eta \in \cH^{-\infty}$, we also put 
\begin{equation}
  \label{eq:heta}
 \sH_\eta(\cO) := \sH_{\R \eta}(\cO) \quad \mbox{ and } \quad 
 \cH_\eta(\cO) := \cH_{\R \eta}(\cO).
\end{equation}
 By construction and \eqref{eq:covl1}  in Appendix~\ref{app:a}, we have
 \begin{equation}
   \label{eq:hecov}
U(g) \sH_\sE(\cO) = \sH_\sE(g\cO) \quad \mbox{ for } \quad g \in G, 
\cO \subeq G. 
 \end{equation}
\end{defn}

In this subsection we show that, if the cone $C_U$ has interior points, 
then $\cH_\eta(\cO) = \cH_\eta(G)$ for every non-empty open subset 
$\cO \subeq G$ (Reeh--Schlieder property). 
If $\eta$ is cyclic, i.e., $\cH_\eta(G) = \cH$, this means that 
$\sH_\eta(\cO)$ is a cyclic real subspace.

\begin{lem} \mlabel{lem:fragment} {\rm(Fragmentation Lemma)} 
Let $\eset\not=\cO \subeq G$ be open.
Then the following assertions hold: 
\begin{itemize}
\item[\rm(a)] If $P \subeq G$ is a closed subgroup, then 
  \begin{itemize}
  \item[\rm(i)] every test function $\phi \in C^\infty_c(\cO P,\R)$ is a finite sum of 
test functions of the form 
\[ \psi \circ \rho_p \: G \to \C, \quad g \mapsto \psi(gp),
\quad 
\psi \in C^\infty_c(\cO,\R), p \in P.\] 
  \item[\rm(ii)] every test function $\phi \in C^\infty_c(P\cO,\R)$ is a finite sum of 
test functions of the form 
\[ \psi \circ \lambda_p \: G \to \C, \quad g \mapsto \psi(pg),
\quad 
\psi \in C^\infty_c(\cO,\R), p \in P.\] 
  \end{itemize}
\item[\rm(b)] Every $\phi \in C^\infty_c(G,\R)$ 
is a finite sum $\sum_{j = 1}^n \phi_j \circ \lambda_{g_j}$ 
with $\phi_j \in C^\infty_c(\cO,\R)$ and $g_j \in G$. 
\end{itemize}
\end{lem}

\begin{prf} (a)(i) The family $(\cO p)_{p \in P}$ is an open cover of the 
compact subset $\supp(\phi)$, so that there exist 
$p_1, \ldots, p_n \in P$ with 
\[ \supp(\phi) \subeq \cO p_1 \cup \cdots \cup \cO p_n.\] 
Let $\chi_0, \ldots, \chi_n$ be a smooth partition of unity subordinated to the 
open cover 
\[ G \setminus \supp(\phi), \quad \cO p_1, \ldots, \cO p_n\] 
of $G$. Then 
$\phi = \sum_{j = 1}^n \phi_j$, where $\phi_j := \chi_j \phi$ 
satisfies $\supp(\phi_j) \subeq \cO p_j$. 
Then $\psi_j := \phi_j \circ \rho_{p_j} \in C^\infty_c(\cO,\R)$ and 
$\phi = \sum_{j= 1}^n \psi_j \circ \rho_{p_j^{-1}}$. 

\nin (a)(ii) and (b) are proved along the same lines. For (b), we use 
the open cover $(g\cO)_{g \in G}$ of the group $G$. 
\end{prf}

\begin{lem}  \mlabel{lem:sw64-genb} 
Let $(U, \cH)$ be a unitary representation of $G$, let 
$\sE \subeq \cH^{-\infty}$ be a real linear subspace, 
$P \subeq G$ a closed subgroup 
and $\eset\not=\cO \subeq G$. 
Then the following assertions hold: 
\begin{itemize}
\item[\rm(a)] $\sH_\sE(\cO P) = \sH_\sE(\cO)$ if 
$\sE$ is $P$-invariant. 
\item[\rm(b)] $\sH_\sE(P\cO)$ is the closed real span of $U(P)\sH_\sE(\cO)$. 
\item[\rm(c)] The real subspace 
spanned by $U(G)\sH_\sE(\cO)$ is dense in $\sH_\sE(G)$. 
\end{itemize}
\end{lem}

\begin{prf} (a) The inclusion 
$\sH_\sE(\cO) \subeq \sH_\sE(\cO P)$ is trivial. 
Conversely, 
for $\phi = \psi \circ \rho_p$, $\psi \in C^\infty_c(\cO)$ and $p \in P$, we 
obtain with \eqref{eq:rightrel} in Appendix~\ref{app:a}
\[ U^{-\infty}(\phi)\sE 
= U^{-\infty}(\psi \circ \rho_p) \sE
= \Delta_G(p)^{-1} U^{-\infty}(\psi) U^{-\infty}(p^{-1})\sE
=  U^{-\infty}(\psi)\sE\subeq \sE.\] 
Hence the assertion follows from Lemma~\ref{lem:fragment}(a). 

\nin (b)  From \eqref{eq:hecov} we know  that 
$U(p) \sH_\sE(\cO) = \sH_\sE(p\cO) \subeq \sH_\sE(P\cO)$ 
for $p \in P$. Now the assertion follows from Lemma~\ref{lem:fragment}(b). 

\nin (c) is an immediate consequence of (b), applied with $P = G$. 
\end{prf}

The following lemma is of particular interest for semidirect products 
such as the Poincar\'e group. 

\begin{lem} \mlabel{lem:semidir}
Let $G= L \ltimes_\sigma N$ be a semidirect product group 
and $(U,\cH)$ be a unitary representation of $G$. We write 
$U_N := U\res_N$ for the restriction to $N$. 
Suppose that $\sE \subeq \cH^{-\infty}(U_N) \subeq \cH^{-\infty}(U)$ 
is an $L$-invariant real subspace. Then we have for every non-empty open 
$\sigma(L)$-invariant subset $\cO \subeq N$ the relation  
\[ \sH_\sE^N(\cO) 
:= \oline{\spann_\R U^{-\infty}(C^\infty_c(\cO,\R)) \sE} 
= \sH_\sE(\cO L).\] 
\end{lem}

\begin{prf} First we show that 
$\sH_\sE^N(\cO) \subeq  \sH_\sE(\cO L).$
As subsets of $G$, we have $\cO L = L \cO$, and the multiplication map 
$L \times \cO \to L\cO$ is a diffeomorphism onto an open subset of $G$. 
On $L\cO$ we thus have test functions $\phi \in C^\infty_c(L\cO)$ of the 
form 
\[ \phi_n(l,x) = \delta_n(l)\psi(x), \quad \mbox{ where } \quad 
\psi \in C^\infty_c(\cO,\R), \] 
and $(\delta_n)_{n \in \N}$ is a 
$\delta$-sequence in $C^\infty_c(L,\R)$ 
(See Appendix~\ref{subsec:app1}). 
A left invariant Haar measure of $G$ 
is given by 
\[ d\mu_G(l,n) = d\mu_L(l) d\mu_N(n), \] 
where $\mu_L$ and $\mu_N$ are left invariant Haar measures on $L$ and 
$N$, respectively.  Then 
\begin{equation}
  \label{eq:intprodform}
U(\phi_n) 
=  \int_\cO \int_L \delta_n(l) U(l) \psi(x) U(x)\, d\mu_L(l)\, d\mu_N(x) 
=   U(\delta_n) U(\psi) 
\end{equation}
leads for $\eta \in \sE \subeq \cH^{-\infty}(U_N)$ to 
\[ U^{-\infty}(\phi_n)\eta = U(\delta_n) U^{-\infty}(\psi) \eta \to 
U^{-\infty}(\psi) \eta\] 
for $n \to \infty$. We conclude that 
$\sH_\sE^N(\cO) \subeq \sH_\sE(\cO L)$. 

To verify the converse inclusion, we recall 
that the functions 
\[ \phi(l,x) = \phi_L(l) \phi_N(x) \quad \mbox{ where } \quad 
\phi_L \in C^\infty_c(L,\R), 
\phi_N \in C^\infty_c(\cO,\R) \] 
span a dense subspace of $C^\infty_c(L\cO)$ 
(\cite[Thm.~5.16]{Tr67}). In view of \eqref{eq:intprodform}, 
it thus suffices to show that, 
\[ U(\phi_L) U^{-\infty}(\phi_N) \sE \subeq 
U(\phi_L) \sH_\sE^N(\cO) \subeq \sH_\sE^N(\cO).\] 
The first inclusion is trivial, and the second one 
follows from the $L$-invariance and the closedness of $\sH_\sE^N(\cO)$, 
which in turn follows from the $\sigma(L)$-invariance of $\cO$. 
\end{prf}

\subsection{The Reeh--Schlieder property} 

The following lemma is our key tool in the derivation of the Reeh--Schlieder 
property. 
\begin{lem}  \mlabel{lem:sw64-gen2} 
Let $(U, \cH)$ be a unitary representation of the connected Lie 
group $G$ for which the positive cone $C_U$ 
has interior points. If $\xi, \eta \in \cH$ are such that the matrix coefficient 
\[ U^{\xi,\eta} \:  G\to \C, \quad g \mapsto \la \xi, U(g) \eta \ra \] 
vanishes on an open subset of $G$, then $U^{\xi,\eta} =0$ on~$G$.
\end{lem}

\begin{prf} Passing to the quotient Lie group $G/\ker(U)$, we may 
w.l.o.g.\ assume that $U$ is injective. 
Then the 
closed convex cone $C_U\subeq \g$ is pointed, and by assumption 
it is also generating, so that it defines a closed complex Olshanski semigroup 
$S_{C_U} = G \exp(iC_U)$ (Definition~\ref{def:olshanski}), 
and the matrix coefficient 
$U^{\xi,\eta}$ extends to a continuous function on $S_{C_U}$ which is holomorphic on 
the interior~$S_{C_U}^0 = G \exp(i C_U^0)$, which is a complex manifold 
(Theorem~\ref{thm:extens}). 

Suppose that $U^{\xi,\eta}$ vanishes on the non-empty open subset~$\cO$. 
Replacing $\eta$ by $U(g_0)\eta$ for some $g_0 \in \cO$, we may assume that 
$e \in \cO$. 
The exponential function 
$\exp \: \g + i C_U \to S_{C_U}$ 
is continuous, and holomorphic on $\g + i C_U^0$. 
Therefore the function $U^{\xi,\eta} \circ \exp \: \g \to \C$ 
extends to a continuous function on the closed wedge 
$\g + i C_U$, holomorphic on the interior and vanishing on 
$\exp^{-1}(\cO) \not=\eset$. 
Now \cite[Lemma~A.III.6]{Ne99} implies that 
$U^{\xi,\eta} \circ \exp = 0$. 
Therefore the regularity of the exponential 
function $\g + i C_U^0 \to S_{C_U}^0$ near $0$ implies that 
$U^{\xi,\eta} = 0$ on $S_{C_U}$, and hence its restriction to $G$ vanishes. 
\end{prf}

The following theorem is the main result of this section. 
It is a key tool for geometric constructions of standard subspaces 
for unitary representations satisfying a spectrum condition, 
such as $C \subeq C_U$.

\begin{thm}  \mlabel{thm:sw64-gen-abs} 
{\rm(Reeh--Schlieder property)} 
Let $\eset\not=\cO\subeq G$  be an open subset and 
$(U, \cH)$ be a unitary representation of the connected Lie 
group $G$ for which the positive cone $C_U$ 
has interior points. Then $\sH_\sE(\cO)$ is total in $\cH_\sE(G)$ 
for every real subspace $\sE \subeq  \cH^{-\infty}$. 
\end{thm}

\begin{prf}  Suppose that $\xi \in \cH_\sE(G)$ is orthogonal to 
$\sH_\sE(\cO)$ and let $\phi \in C^\infty_c(\cO,\R)$. 
Then $\supp(\phi)$ is a compact subset of $\cO$, so that there exists an 
open $e$-neighborhood $V \subeq G$ with $V \supp(\phi) \subeq \cO$. 
For $g \in V$ and $\eta \in \sE$, we then have by 
\eqref{eq:covl1} in Appendix~\ref{app:a} 
\[ \xi \bot U(g)U^{-\infty}(\phi) \eta 
= U^{-\infty}(\lambda_g \phi) \eta \quad \mbox{ because } \quad 
\supp(\lambda_g\phi) =g\supp(\phi) \subeq \cO.\] 
If follows that, for $\eta' := U^{-\infty}(\phi) \eta\in \cH$, 
the continuous function 
\[ U^{\xi,\eta'} \: G \to \C, \quad g \mapsto \la \xi, U(g) \eta' \ra \] 
vanishes on the identity neighborhood $V$, and by 
Lemma~\ref{lem:sw64-gen2} it vanishes on~$G$.
We conclude that $\xi \bot U(G) \sH_\sE(\cO)$, so that 
Lemma~\ref{lem:sw64-genb} shows that $\xi = 0$. 
\end{prf}

\subsection{Support properties of corresponding distributions} 

Although we will not use them below, we record some immediate consequences 
of our discussion for supports of certain distributions. We expect these 
results to become relevant in subsequent work. 
For the abelian case, we refer to \cite{NOO20} for more details. 

From Theorem~\ref{thm:sw64-gen-abs} we immediately obtain 
the following criterion for $\sH_\sE(\cO)$ to be standard in 
terms of support properties of distributions. 

\begin{cor} Let $\cO, \cO' \subeq G$ be two non-empty open subsets, let 
$(U,\cH)$ be a unitary representation with $C_U^0\not=\eset$, 
let $\eta \in  \cH^{-\infty}$, and consider the distribution 
$D = j_\eta(\eta) \in C^{-\infty}(G)$ 
{\rm(see \eqref{eq:dist-inc} in Appendix~\ref{app:a})}. 
Then the following assertions hold: 
\begin{itemize}
\item[\rm(a)] $\sH_\eta(\cO') \subeq \sH_\eta(\cO)'$ is equivalent to 
\begin{equation}
  \label{eq:imdcond}
\Im D(\psi^** \phi) = 0 \quad \mbox{ for } \quad 
\phi \in C^\infty_c(\cO,\R), \psi \in C^\infty_c(\cO',\R). 
\end{equation}
If this is the case, then $\sH_\eta(\cO)$ is a standard subspace of $\cH_\eta(G)$. 

\item[\rm(b)] Suppose that $\cO \subeq G$ is an open subsemigroup with 
$e \in \oline{\cO}$. Then 
\begin{equation}
  \label{eq:supprest}
\sH_\eta(\cO^{-1}) \subeq \sH_\eta(\cO)' \quad \Leftrightarrow \quad 
 \supp(\Im D) \subeq G \setminus \cO.
\end{equation}
\end{itemize}
\end{cor}

\begin{prf} (a) From Theorem~\ref{thm:sw64-gen-abs}
we infer that $\sH_\eta(\cO)$ is total in $\cH_\eta(G)$. 
With the notation from \eqref{eq:cHD}  and \eqref{eq:scalproHd} 
in Appendix~\ref{app:a}, we see that 
\[ \Im D(\psi^* * \phi) 
= \Im \la \phi * D, \psi * D\ra_{\cH_D} 
= \Im \la U^{-\infty}(\phi)\eta, U^{-\infty}(\psi)\eta \ra \] 
vanishes  for all 
$\phi \in C^\infty_c(\cO,\R)$ and $\psi \in C^\infty_c(\cO',\R)$ 
if and only if $\sH_\eta(\cO') \subeq \sH_\eta(\cO)'.$ 
As the real subspace $\sH_\eta(\cO')$ is cyclic in $\cH_\eta(G)$ 
(Theorem~\ref{thm:sw64-gen-abs}), 
the above condition implies that 
the subspace $\sH_\eta(\cO)$ is also separating, hence standard. 

\nin (b) For $\phi \in C^\infty_c(\cO,\R)$ and $\psi \in C^\infty_c(\cO',\R)$, 
the convolution product 
$\psi^** \phi$ is a test function on~$\cO$. 
Using a $\delta$-sequence in $C^\infty_c(\cO,\R)$, it follows that 
\eqref{eq:imdcond} is equivalent to $\Im(D)$ vanishing on 
$C^\infty_c(\cO,\R)$, which means that $\supp(\Im D) \subeq G \setminus \cO$.
\end{prf}

\begin{cor} Let $G$ be a connected Lie group, 
$D\in C^{-\infty}(G)$ be a positive definite distribution, 
and $(U_D,\cH_D)$ be the corresponding unitary representation of~$G$ 
on the Hilbert subspace $\cH_D \subeq C^{-\infty}(G)$. 
If $C_U^0 \not=\eset$ then, for any non-empty open subset $\cO \subeq G$, we have: 
  \begin{itemize}
  \item[\rm(1)] The real subspace $\sH_D(\cO):= 
\oline{C^\infty_c(\cO,\R) * D}$ is cyclic in $\cH_D$. 
  \item[\rm(2)] If $\cO$ is an open subsemigroup with 
$e \in \oline{\cO}$ and $\Im D$ vanishes on $G \setminus \cO$, 
then $\sH_D(\cO)$ is a standard subspace of $\cH_D$ with 
$\sH_D(\cO^{-1}) \subeq \sH_D(\cO)'$. 
  \end{itemize}
\end{cor}

\begin{lem} \mlabel{lem:supp1} 
Let $\eta \in \cH^{-\infty}$ and $\cO \subeq G$ be an open subset for which 
$\sH_\eta(\cO)$ is total in $\cH$.
If $\alpha \in \cH^{-\infty}$ is such that the distribution 
$j_\eta(\alpha)$ vanishes on the subset $V \cdot \cO$, for an $e$-neighborhood $V$ 
in $G$, then $\alpha =0$. 
\end{lem}

\begin{prf} Let $(\delta_n)_{n \in \N}$ be a $\delta$-sequence in 
$C^\infty_c(G,\R)$ and 
$\phi \in C^\infty_c(\cO,\R)$. 
Then $\delta_n * \phi \in C^\infty_c(V \cO,\R)$ if 
$n$ is sufficiently large. We therefore have 
\[ 0 
= j_\eta(\alpha)(\delta_n * \phi) 
= \alpha(U^{-\infty}(\delta_n * \phi)\eta) 
= \alpha(U^{-\infty}(\delta_n) U^{-\infty}(\phi)\eta) 
= \la U^{-\infty}(\phi)\eta, U^{-\infty}(\delta_n^*)\alpha \ra.\] 
As $\sH_\eta(\cO)$ is total, 
$U^{-\infty}(\delta_n)\alpha = 0$ if $n$ is sufficiently large, 
and since this sequence converges to $\alpha$ in $\cH^{-\infty}$, 
we obtain $\alpha =0$. 
\end{prf}

\begin{prop} \mlabel{prop:distrubsupp} 
{\rm(Full support property)} 
Suppose that 
$\eta \in \cH^{-\infty}$ is such that 
$\sH_\eta(\cO)$ is total in  $\cH_\eta(G)$ for every 
non-empty open subset $\cO \subeq G$.
If $\alpha \in \cH^{-\infty}$ is such that $\supp(j_\eta(\alpha)) \not= G$, 
then $\alpha = 0$. 
\end{prop}

\begin{prf} If $\supp(j_\eta(\alpha)) \not= G$, then 
there exists a non-empty open subset $\cO_0 \subeq G$ on which 
$j_\eta(\alpha)$ vanishes. Let $\cO \subeq \cO_0$ be a non-empty 
open relatively compact subset. Then there exists an 
identity neighborhood $V \subeq G$ with $V \cO \subeq \cO_0$. 
As $\sH_\eta(\cO)$ is total in $\cH$, Lemma~\ref{lem:supp1} implies 
that $\alpha =0$.   
\end{prf}

The following corollary generalizes 
Lemma~\ref{lem:sw64-genb} to distributional matrix coefficients. 

\begin{cor} For any unitary representation $(U,\cH_D)$ on a subspace 
$\cH_D \subeq C^{-\infty}(G)$ for which $C_U^0 \not=\eset$, any 
non-zero distribution $D' \in \cH_D$ satisfies $\supp(D') = G$. 
\end{cor}

\begin{prf} This follows by combining Theorem~\ref{thm:sw64-gen-abs} 
with Proposition~\ref{prop:distrubsupp}. 
\end{prf}

\section{$C$-positive representation 
of $3$-graded Lie groups} 
\mlabel{sec:3}

After the preparations in Section~\ref{sec:2}, we now turn to standard 
subspaces $\sV = \sV_{(h,\tau_G,U)}$ specified by the BGL construction 
for an antiunitary representation $(U,\cH)$ of $G_\tau 
= G \rtimes \{\1,\tau_G\}$ as in \eqref{eq:bgl} by 
\[ J_\sV = U(\tau_G) \quad \mbox{ and } \quad 
\Delta_{\sV}^{-it/2\pi} = U(\exp th) \quad \mbox{ for } \quad t \in \R.\] 
In this section we assume that $\sV$ is {\it regular} 
in the sense that the Lie wedge $\L(S_\sV)$ has interior 
points. This  leads us to assumption (B4) below (\cite[Thm.~4.4]{Ne19}).

We now consider the following setting which is a 
non-abelian generalization of \cite{NOO20}: 
\begin{itemize}
\item[\rm(B1)] $G$ is a connected Lie group with Lie algebra~$\g$. 
\item[\rm(B2)] $\tau_G \in \Aut(G)$ is an involution and $h \in \g$ 
is fixed by $\tau = \L(\tau_G)\in \Aut(\g)$. 
\item[\rm(B3)] $C \subeq \g$ is a pointed 
closed convex cone invariant under $\Ad(G)$ and $-\tau$, and 
$\g = \g_C + \R h$, where $\g_C = C - C$ is the ideal generated by~$C$. 
We write $G_C \trile G$ for the connected normal 
subgroup corresponding to $\g_C$ and $A := \exp(\R h)$. 
If $h \in \g$, then $G = G_C$, and if $h \not\in\g$, then 
$G =  G_C A$. 
\item[\rm(B4)] $h$ is an {\it Euler element}, i.e., 
$\ad h$ is diagonalizable with eigenvalues $\{-1,0,1\}$,  
and $\tau = e^{\pi i \ad h}\res_{\g}$. 
\end{itemize}

Then $\g = \g^{-1}\oplus \g^0 \oplus \g^1$, 
where $\tau\res_{\g^j} = (-1)^j \id_{\g^j}$ and $h \in \g^0 = \g^\tau$ 
defines the grading by $\g^j = \ker(\ad h  - j \1)$. 
Further, (B3/4) imply that $\fq := \g^{-\tau}$ is generated by 
the cone
\begin{equation}
  \label{eq:cq}
C \cap \fq = C\cap (\g_1 \oplus \g_{-1}) = C_+ \oplus - C_-
\end{equation}
(cf.\ \cite[Lemma~3.1]{NOO20}). 

The group $G_\tau := G \rtimes \{\1,\tau_G\}$ has a natural grading 
and every  antiunitary representation $(U,\cH)$ of $G_\tau$ 
defines a standard subspace $\sV = \sV_{(h,\tau_G,U)}$ as in \eqref{eq:bgl}  
(\cite{BGL02}, \cite{NO17}). From 
\cite[Thm.~3.4]{Ne19b} we know that, for $C = C_U$, 
\begin{equation}
  \label{eq:sv2}
S_\sV = \exp(C_+) G_{\sV} \exp(C_-) 
= G_{\sV} \exp(C_+ + C_-), 
\end{equation}
where $\L(G_\sV) = \g^0$ and $C_+ + C_-$ is a pointed generating cone in $\fq$ 
(see (B3) and \eqref{eq:cq}).

\begin{rem} If $C$ has interior points, then 
$\Xi := G \exp(i C^0) = S_C^0 \subeq S_C$ is a complex 
manifold with the antiholomorphic involution 
\[ \tau_S(g \exp(ix)) = \tau_G(g) \exp(-i \tau(x)) \] 
and the fixed point set 
\[ S^c := G^{\tau_G} \exp(i (C^0 \cap \fq)).\] 
Since $C$ is invariant under $e^{\R \ad h}$, we obtain for each $t \in \R$ 
a unique automorphism $\beta_t$ of $S_C$
which induces the automorphism $e^{t\ad h}$ on the Lie wedge $\L(S_C)$. 

Now $S_\sV^0$ plays the role of a 
``wedge  domain'' in the ``Shilov boundary'' $G$ of $S_C^0$, 
on which we may ``realize'' the standard subspace~$\sV$ 
as $\sH_\sE(S_\sV^0) \subeq \cH_\sE(G)$ for some 
$\sE \subeq \cH^{-\infty}$. 

\nin (c) If $C$ is not generating in $\g$ and 
$\g = \g_C + \R h$, then $\Xi := G_C \exp(iC^0)$ 
(the open complex Olshanski semigroup associated to the pointed 
generating invariant cone $C \subeq \g_C$) 
is a complex manifold 
with an $\R^\times$-action and we shall realize the representation 
in holomorphic functions on~$\Xi$, with boundary value maps to~$G_C$. 
In this case $S_\sV = (S_\sV \cap G_C)A$ 
and we have the continuous inclusion 
\[ \cH^\infty(U) \into  \cH^\infty(U\res_{G_C}) \] 
which dualizes by Lemma~\ref{lem:charsmooth}(b) in the appendix 
to a linear injection 
\begin{equation}
  \label{eq:disvecincl}
 \cH^{-\infty}(U\res_{G_C})  = \dd U^{-\infty}(\cU(\g_C)) \cH \into 
\dd U^{-\infty}(\cU(\g)) \cH = \cH^{-\infty}(U),  
\end{equation}
and we have to work with elements 
$\eta \in \cH^{-\infty}(U\res_{G_C})$. 
\end{rem}

\subsection{Distribution vectors associated to standard subspaces} 
\mlabel{subsec:3.1}

In this subsection we develop some results concerning the connection between 
standard subspaces and distribution vectors. We shall assume (B1-4), that 
$(U,\cH)$ is an antiunitary representation with 
$C = C_U$ pointed and that $\sV \subeq \cH$ is the standard subspace 
specified by the triple $(h,\tau_G,U)$. 
For later reference, we note that \eqref{eq:SE} implies 
that, for the symplectic orthogonal space 
$\sV'$ (see \eqref{eq:sympdual}), 
which is the standard subspace specified by 
the triple $(-h,\tau_G, U)$, the endomorphism semigroup 
(cf.\ \eqref{eq:sv2}) is given by 
\begin{equation}
  \label{eq:sv'}
  S_{\sV'} = S_{\sV}^{-1}.
\end{equation}

\begin{defn} We write $\sV^\infty := \sV \cap \cH^\infty$ for the subspace of smooth 
vectors contained in $\sV$ and 
\[ \sV^{-\infty} := \{ \alpha \in \cH^{-\infty} \: 
\Im\alpha((\sV')^\infty) = \{0\}\},\] 
where $\sV' = \sV^{\bot_\omega} = \sV_{(-h,\tau_G, U)}$ 
is the symplectic orthogonal space of $\sV$ 
(see \eqref{eq:sympdual}).
\end{defn}

\begin{lem}
  \mlabel{lem:vinfty} 
The following assertions hold: 
\begin{itemize}
\item[\rm(a)] $\sV^\infty$ is dense in $\sV$ and 
$(\sV')^\infty$ is dense in $\sV'$. 
\item[\rm(b)] $\sV^{-\infty}$ is the weak-$*$-closure of the real subspace 
$\sV$ of $\cH^{-\infty}$. 
\item[\rm(c)] $\sV^{-\infty} = \{ \alpha \in \cH^{-\infty} \: (\forall 
\phi \in C^\infty_c(S_\sV^0,\R)) \, U^{-\infty}(\phi)\alpha \in \sV \}$. 
\item[\rm(d)] $\sV^{-\infty} \cap i\sV^{-\infty}= \{0\}$. 
\item[\rm(e)] $\sV^\infty$ is $\dd U(\g)$-invariant. 
\item[\rm(f)] $\sV^{-\infty}$ is $\dd U^{-\infty}(\g)$-invariant. 
\end{itemize}
\end{lem}

\begin{prf} (a) Let $(\delta_n)_{n \in \N}$ be a $\delta$-sequence in 
$C^\infty_c(S_\sV^0,\R)$. 
For $\xi \in \sV$, the sequence $U(\delta_n)\xi$ of smooth vectors is contained 
in $\sV$ and converges to~$\xi$. We likewise conclude that 
$(\sV')^\infty$ is dense in $\sV'$. 
 
\nin (b) An element $\xi \in \cH^\infty$ annihilates the subspace $\sV$ of $\cH^{-\infty}$ 
under the real bilinear pairing 
\[ \cH^{-\infty} \times \cH^\infty \to \R, \quad 
(\alpha, \xi) \mapsto \Re \alpha(\xi) \] 
if and only if $\Re \la \xi, \sV \ra = \{0\}$, which is equivalent to 
$\xi \in (i \sV') \cap \cH^\infty = i (\sV')^\infty$. 
Therefore the weak-$*$-closure of $\sV$ in $\cH^{-\infty}$ is the annihilator 
of $i (\sV')^\infty$, which consists of all elements $\alpha$ satisfying 
$\{0\} = \Re\alpha(i (\sV')^\infty)$, which is equivalent to 
$\{0\} = \Im\alpha((\sV')^\infty)$, i.e., to $\alpha \in \sV^{-\infty}$. 

\nin (c) If $\alpha \in \sV^{-\infty}$ and $\phi 
 \in C^\infty_c(S_\sV^0,\R)$, then any $\xi \in (\sV')^\infty$ satisfies 
\[ \Im (U^{-\infty}(\phi)\alpha)(\xi) 
= \int_G \phi(g) \Im (U^{-\infty}(g)\alpha)(\xi)\, dg  
= \int_G \phi(g) \Im \alpha(U(g^{-1})\xi)\, dg = 0 \] 
because, for $\phi(g) \not=0$, the relation $U(g^{-1})\xi \in (\sV')^\infty$ 
follows from $S_{\sV'} = S_{\sV}^{-1}$ (see \eqref{eq:sv'}).

Suppose, conversely, that $\alpha \in \cH^{-\infty}$ is such that, for any 
$\phi \in C^\infty_c(S_{\sV}^0,\R)$ the vector $U^{-\infty}(\phi)\alpha$ is contained in $\sV$. 
As $S_\sV$ has dense interior (\cite[Thm.~4.4]{Ne19}),  
$C^\infty_c(S_{\sV}^0,\R)$ contains a $\delta$-sequence $(\delta_n)_{n \in \N}$. 
For $\xi \in (\sV')^\infty$, we then obtain 
\[ \Im \alpha(\xi) 
= \lim_{n \to \infty} \Im (U^{-\infty}(\delta_n) \alpha)(\xi) 
= \lim_{n \to \infty} \Im \la \xi, U^{-\infty}(\delta_n) \alpha \ra = 0.\] 
This means that $\alpha \in \sV^{-\infty}$. 

\nin (d) A distribution vector $\alpha$ is contained in $\sV^{-\infty}  \cap i \sV^{-\infty}$ 
if and only if $(\sV')^\infty = \sV' \cap \cH^\infty\subeq \ker \alpha$. 
Let $\xi \in \sV'$ and $(\delta_n)_{n \in \N}$ be a $\delta$-sequence of test functions 
supported in the interior of $S_{\sV'} = \tau_G(S_\sV) = S_{\sV}^{-1}$. 
Then $U(\delta_n) \xi \in (\sV')^\infty \subeq \ker \alpha$ implies that 
$U(\delta_n)(\sV' + i \sV') \subeq \ker \alpha$, and since 
$U(\delta_n) \: \cH \to \cH^\infty$ is continuous and $\sV'$ is standard, it follows that 
$U(\delta_n) \cH \subeq \ker \alpha$. Now the assertion follows 
from $U(\delta_n) v \to v$ in $\cH^\infty$ for every smooth vector~$v$, 
a consequence of the continuity of the $G$-action on the Fr\'echet space 
$\cH^\infty$ (cf.\ \cite[Thm.~4.4]{Ne10}). 

\nin (e) Let $\xi \in \sV^\infty$ and $x \in \g$. If $(\delta_n)_{n \in \N}$ 
is a $\delta$-sequence in $C^\infty_c(S_{\sV}^0)$, then we have already observed 
under (d) that $U(\delta_n) \xi \to \xi$ in the topology of $\cH^\infty$. 
As $\dd U(x) U(\delta_n) \xi = U(x^R \delta_n) \xi \in \sV$ 
by \eqref{eq:derrep2} in Appendix~\ref{app:a}, 
passing to the limit $n \to \infty$ 
implies $\dd U(x) \xi \in \sV^\infty$. 

\nin (f) For $\alpha \in \sV^{-\infty}$, $\xi \in (\sV')^\infty$ and $x \in \g$, we have 
\[ (\dd U^{-\infty}(x)\alpha)(\xi) 
=  -\alpha(\dd U(x)\xi) \in \R \] 
because (e), applied to the standard subspace $\sV'$, implies that 
$\dd U(x) \xi \in(\sV')^\infty$. 
\end{prf}

Since we shall use it several times below, we recall the following 
important fact on standard subspaces 
from \cite[Prop.~3.10]{Lo08}: 

\begin{lem} \mlabel{lem:lo-3.10} 
Suppose that  $\sH_1 \subeq \sH_2$ are standard subspaces of $\cH$. 
If 
\begin{itemize}
\item[\rm(a)] $\Delta_{\sH_2}^{it} \sH_1 = \sH_1$ for every $t \in \R$, or 
\item[\rm(b)] $\Delta_{\sH_1}^{it} \sH_2 = \sH_2$ for every $t \in \R$, 
\end{itemize}
then $\sH_1 = \sH_2$. 
\end{lem}

\begin{prf} That (a) implies $\sH_1 = \sH_2$ follows from 
\cite[Prop.~3.10]{Lo08}). From (b) we obtain by dualization 
$\sH_2' \subeq \sH_1'$ with 
$\Delta_{\sH_1'}^{it} \sH_2' = \sH_2'$ for $t \in \R$, so that 
we obtain 
$\sH_1' = \sH_2'$ with (a), hence $\sH_1 = \sH_2$ also holds in this case. 
\end{prf}

The following theorem is an interesting tool 
to obtain nice descriptions of standard subspaces in concrete 
situations. Here a subtle point is that we assume 
$\sE \subeq \sV^{-\infty}$, but we 
shall see in Lemma~\ref{lem:key} below how this assumption can be verified 
in terms of the action of $A = \exp(\R h)$ on $\cH^{-\infty}$. 

\begin{thm} \mlabel{thm:3.6}
For a real subspace $\sE \subeq \cH^{-\infty}$ invariant 
under $A = \exp(\R h)$ and satisfying 
$\cH_\sE(G) = \cH$, the following assertions hold: 
\begin{itemize}
\item[\rm(a)] If $\eset\not=\cO\subeq G$ is open, 
then $\sH_\sE(\cO)$ is cyclic {\rm(Reeh--Schlieder property)}. 
\item[\rm(b)] If $\sE \subeq \sV^{-\infty}$, then $\sH_\sE(S_\sV^0) = \sV.$ 
\end{itemize}
\end{thm}

If $C= C_U$ has interior points, then $G = G_C$, and 
the Reeh--Schlieder property 
(Theorem~\ref{thm:sw64-gen-abs}) implies that $\sH_\sE(\cO)$ is cyclic 
in $\cH_\sE(G)$ for every non-empty open subset $\cO \subeq G$. 
Assertion (a) above shows that this remains true under the weaker assumption~(B3).

\begin{prf}  (a) For $\alpha \in \sE$ and $\phi \in C^\infty_c(\cO,\R)$, 
the support $\supp(\phi)$ is a compact subset of $\cO$. Hence 
there exists an $e$-neighborhood $V \subeq G_C$ with 
$V \supp(\phi) \subeq \cO$. This shows that 
\begin{equation}
  \label{eq:inc-abc}
 U(C^\infty_c(V,\R)) U^{-\infty}(\phi)\alpha 
\subeq U^{-\infty}(C^\infty_c(\cO,\R))\alpha \subeq \sH_\sE(\cO).
\end{equation}
Applying Theorem~\ref{thm:sw64-gen-abs} to 
$\sE = \R\eta$ for $\eta := U^{-\infty}(\phi)\alpha \in \cH \subeq \cH^{-\infty}$ 
and the open subset $V\subeq G_C$, we first see that 
$\sH_{\eta}(V)$ is total in $\cH_{\eta}(G_C)$, 
and thus 
\[ U(G_C)\eta \subeq \cH_{\eta}(G_C) =  \cH_\eta(V) 
{\buildrel{\eqref{eq:inc-abc}} \over \subeq} \cH_\sE(\cO).\] 
This implies that $\cH_\sE(\cO)$ is $G_C$-invariant. 

Lemma~\ref{lem:fragment}(a)(ii) now implies that 
$\cH_\sE(\cO) = \cH_\sE(G_C\cO).$ 
As $A$ is abelian and $G = G_CA$, the open subset $G_C\cO \subeq G$ 
is invariant under all inner automorphisms. 
Therefore the $A$-invariance of $\sE$ implies that 
$\cH_\sE(\cO)$ is also $A$-invariant, hence $G$-invariant, 
and thus $\cH_\sE(\cO) = \cH_\sE(G)$ (Lemma~\ref{lem:sw64-genb}). 

\nin (b) First, Lemma~\ref{lem:vinfty}(c) implies that 
$\sH_\sE(S_\sV^0) \subeq \sV$, hence that $\sH_\sE(S_\sV^0)$ is separating. 
As $\sH_\sE(S_\sV^0)$ is cyclic by (a), it is standard. 
The invariance of $\sE$ under $U^{-\infty}(A)$ and the invariance 
of $S_\sV^0$ under conjugation with $A$ entail that 
$\sH_\sE(S_\sV^0)$ is invariant under $\Delta_\sV^{i\R} = U(A)$ 
(see \eqref{eq:covl1} in the appendix). 
Now $\sH_\sE(S_\sV^0) = \sV$ follows from~Lemma~\ref{lem:lo-3.10}.   
\end{prf}

For $\sE = \sV$, we have $\cH_\sV(G) = \cH$, and 
Theorem~\ref{thm:3.6}(b) reduces to 
the tautology $\sH_\sV(S_\sV^0) = \sV$, so that this theorem 
is most interesting if $\sE$ is small. Actually it is 
finite-dimensional in many interesting situations (see Section~\ref{sec:5}). 
In particular, we would 
also like to have that $\sH_\sE((S_\sV^0)^{-1}) = \sH_{\sE}(S_{\sV'}^0) \subeq \sV'$, 
but this requires 
\begin{equation}
  \label{eq:seinintersect}
   \sE \subeq \sV^{-\infty} \cap (\sV')^{-\infty}
\end{equation}
by Lemma~\ref{lem:vinfty}(c). 
This is an interesting point because it may happen that the symplectic form 
is non-degenerate on $\sV$, i.e., $\sV \cap \sV' = \{0\}$, 
but that nevertheless 
the subspaces $\sV^{-\infty}$ and $(\sV')^{-\infty}$ have non-trivial 
intersection. 
As we shall see in Subsection~\ref{subsec:3.4}, 
the irreducible antiunitary positive energy representation of 
$\Aff(\R)$ provides an example where $\sV \cap \sV' = \{0\}$ 
follows from the fact that $\Delta-\1$ is injective, but in this 
case $\sV^{-\infty} \cap (\sV')^{-\infty}$ may contain $J_\sV$-fixed 
$\Delta$-eigenvectors. Note that 
\[ \sV \cap \sV' = \{ \xi \in \cH \: \Delta_\sV \xi = \xi = J_\sV \xi \} 
= \ker(\Delta_\sV - \1)^{J_\sV}\] 
is contained in the $1$-eigenspace of $\Delta_\sV$.

\subsection{Extending orbit maps of distribution vectors} 
\mlabel{subsec:3.2}

In \cite[Prop.~2.1]{NOO20} we have seen that an element 
$\xi \in \cH$ is contained in the standard subspace $\sV$ if and only 
if the orbit map 
\[ \alpha^\xi \:\R \to \cH, \quad 
\alpha^\xi(t) :=  \Delta_\sV^{-it/2\pi}\xi \] extends to a 
continuous map on the closed strip 
$\oline{\cS_\pi} \to \cH$, holomorphic on $\cS_\pi$,  
and satisfying $\alpha^\xi(\pi i) = J_\sV \xi$. In this subsection 
we consider a similar condition for distribution vectors. 
This condition specifies a linear 
subspace $\cH^{-\infty}_{{\rm ext},J}\subeq \cH^{-\infty}$. We then show that this 
space is invariant under 
$U^{-\infty}(S_\sV^0)$ and $U^{-\infty}(C^\infty(S_\sV^0))$, which in turn 
leads to the important result that it is contained in $V^{-\infty}$ 
(Lemma~\ref{lem:key}).

\begin{defn}
Let $\eta \in \cH^{-\infty}$ be a distribution vector. We say that 
\[ \alpha^\eta \:  \R \to \cH^{-\infty}, \quad 
\alpha^\eta(t) := U^{-\infty}(\exp th) \eta \] 
{\it extends} if there exists a weak-$*$-continuous extension 
$\alpha^\eta \: \oline{\cS_\pi} \to \cH^{-\infty}$ which is weak-$*$-holomorphic 
on $\cS_\pi$. We write $\cH^{-\infty}_{\rm ext}$ for the linear subspace 
of distribution vectors with this property and 
$\cH^{-\infty}_{{\rm ext},J}$ for the subspace 
of those $\eta \in \cH^{-\infty}_{\rm ext}$ 
for which the extension to $\oline{\cS_\pi}$ satisfies 
$\alpha^\eta(\pi i) = J_\sV \eta$.   
\end{defn}

Note that $\sV \subeq \cH^{-\infty}_{{\rm ext},J}$ follows from the continuity 
of the inclusion $\cH \into \cH^{-\infty}$ with respect to the weak-$*$-topology 
(\cite[Prop.~2.1]{NOO20}). To address the invariance properties 
of $\cH^{-\infty}_{\rm ext}$, we start with an abstract lemma.

\begin{lem} \mlabel{lem:cartes}
Let $X$ be a locally compact space and 
$f \: X \to \cH^{-\infty}$ be a weak-$*$-continuous map. 
Then the following assertions hold: 
\begin{itemize}
\item[\rm(a)] $f^\wedge \: X \times \cH^\infty \to \C, f^\wedge(x,\xi) := f(x)(\xi)$, 
is continuous. 
\item[\rm(b)] If, in addition, $X$ is a complex manifold and 
$f$ is antiholomorphic, then $\oline{f^\wedge}$ is holomorphic. 
\end{itemize}
\end{lem}

\begin{prf} (a) Since the assertion is local in the first argument, we may w.l.o.g.\ 
assume that $X$ is compact. Then $f(X)$ is pointwise bounded, 
hence equicontinuous because 
$\cH^\infty$ is a Fr\'echet space and therefore barrelled (\cite[Thm.~33.2]{Tr67}). 
To see that $f^\wedge$ is continuous in $(x,\xi)$, we note that 
\[ f(x')(\xi') - f(x)(\xi) 
= f(x')(\xi' - \xi) + (f(x')-f(x))(\xi). \] 
As $f(\cdot)(\xi)$ is continuous and $f(X)$ is equicontinuous, 
the continuity of $f^\wedge$ follows. 

\nin (b) If, in addition, $X$ is a complex manifold 
and $f$ is antiholomorphic, 
then $\oline{f^\wedge}$ is a continuous map on the product of two complex Fr\'echet 
manifolds which is holomorphic in each argument separately. 
Hence the assertion follows from Hartogs' Theorem. Alternatively, 
one may combine \cite[Prop.~1.2.8]{GN} with \cite[Thm.~2.1.12]{GN} 
to see that $\oline{f^\wedge}$ is holomorphic. 
\end{prf}

Since the semigroup $S_C$ is not a manifold because $C$ is a closed 
cone and $C$ may have empty interior, it is not at all clear what 
holomorphic  functions with values in $S_C$ should be. Here is a definition. 
\begin{defn}  \mlabel{def:holodisc} ($S_C$-valued holomorphic functions) 
Let 
\[ q_S \: \tilde S_C \to \tilde G_\C, \quad 
q_S(g \exp(ix)) = \eta_{\tilde G}(g) \exp_{\tilde G_\C}(ix), \quad  g\in \tilde G, 
x \in C,\] 
where $\eta_{\tilde G} \: \tilde G \to \tilde G_\C$ 
is the universal homomorphism for which 
$\L(\eta_{\tilde G}) \: \g \to \g_\C$ is the inclusion, and $\tilde G_\C$ is a 
$1$-connected Lie group with Lie algebra~$\g_\C$ 
(cf.\ Definition~\ref{def:olshanski}).
If $M$ is a complex manifold, then we call a continuous map 
$f \: M \to \tilde S_C$ {\it holomorphic} if the composition 
\break  $q_S \circ f \: M \to \tilde G_\C$ is holomorphic. 

A map $f \: M \to S_C$ is called {\it holomorphic}, if, for every 
$1$-connected open subset $U \subeq M$ a lift 
$\tilde f_U \: U \to \tilde S_C$ of $f\res_U \: U \to S_C$ 
(which exists if $U$ is $1$-connected) is holomorphic. 
\end{defn}

\begin{lem} \mlabel{lem:betaact} 
For $s \in S_C$, we write 
\[ \beta^s(t) = \exp(th)s\exp(-th) \] 
for the orbit map under the conjugation action of the 
modular one-parameter group. 
For every $s \in S_\sV$, this orbit map extends to a continuous map 
$\beta^s \: \oline{\cS_\pi}\to S_C$ on the closed strip 
$\oline{\cS_\pi}$ which is holomorphic on $\cS_\pi$ 
in the sense of {\rm Definition~\ref{def:holodisc}}. The so obtained map 
\begin{equation}
  \label{eq:contact}
 \oline{\cS_\pi} \times S_{\sV} \to S_C, \quad 
(z, s) \mapsto \beta^s(z)
\end{equation}
is continuous. 
In addition, 
$\beta^s(\cS_\pi) \subeq S_C^\circ := G \exp(i C^\circ)$ for $s \in S_\sV^0$, 
where $C^\circ$ denotes the interior of $C$ in its span $\g_C = C-C$.
\end{lem}

\begin{prf} For $s = g\exp(x_1 + x_{-1})\in S_\sV$ with $g \in G_\sV$, 
$x_{\pm 1} \in C_\pm$, we have 
\begin{equation}
  \label{eq:betarel1}
\beta^s(z) = g\exp(e^z x_1 + e^{-z} x_{-1}). 
\end{equation}
If $s \in S_\sV^0$, i.e., $x_{\pm 1} \in C_\pm^0$, and 
$z \in \cS_\pi$, then 
$\beta^s(z) \in S_C^\circ$ because, for $z = a + i b$, $0 < b <  \pi$, we have  
\[ \Im(e^z x_1 + e^{-z} x_{-1}) 
=\sin(b)(e^a x_1 - e^{-a} x_{-1}) \in C_\fq^0 \subeq C^\circ.\]
In particular, we see that $\beta^s(z) \in S_C$, and \eqref{eq:betarel1} shows that 
the map \eqref{eq:contact} is continuous. 
It also shows that all maps $\beta^s \: \cS_\pi \to S_C$, 
$s \in S_\sV$, are holomorphic in the sense of {\rm Definition~\ref{def:holodisc}}.
\end{prf}

\begin{lem} \mlabel{lem:hinftyext}
The subspaces $\cH^{-\infty}_{\rm ext}$ and $\cH^{-\infty}_{{\rm ext},J}$ 
are invariant under $U^{-\infty}(S_\sV^0)$ and the algebra 
$U^{-\infty}(C^\infty_c(S_\sV^0,\R))$. 
\end{lem}

\begin{prf} {\bf Step 1:} $U^{-\infty}(S_\sV^0)\cH^{-\infty}_{\rm ext}\subeq \cH^{-\infty}_{\rm ext}$. \\
Let $\eta \in \cH^{-\infty}_{\rm ext}$. 
For $t \in \R$ and $g \in S_\sV$, we have 
\[ \alpha^{U^{-\infty}(g)\eta}(t) 
= U^{-\infty}(\exp th) U^{-\infty}(g) \eta 
= U^{-\infty}(\beta^g(t)) U^{-\infty}(\exp th) \eta 
= U^{-\infty}(\beta^g(t)) \alpha^\eta(t).\] 
By Lemma~\ref{lem:betaact} and Proposition~\ref{prop:2.2}, 
this curve extends to the function 
\[\oline{\cS_\pi} \to \cH^{-\infty}, \quad 
z \mapsto F(z,g) := U^{-\infty}(\beta^g(z)) \alpha^\eta(z).\] 

\nin {\bf Claim 1:} 
The map $F \: \oline{\cS_\pi} \times S_\sV \to \cH^{-\infty}, 
F(z,g) = U^{-\infty}(\beta^g(z)) \alpha^\eta(z)$ is 
weak-$*$-continuous. \\ 
As $\alpha^\eta$ is weak-$*$-continuous on the 
locally compact space $\oline{\cS_\pi}$, the weak-$*$-continuity of 
$F$ follows from 
Lemma~\ref{lem:cartes}(a) and the continuity of the maps 
\begin{equation}
  \label{eq:16}
\oline{\cS_\pi} \times S_\sV \to \cH^\infty, \quad 
(z,g) \mapsto  U(\beta^g(z)^*)\xi,  \qquad \xi \in \cH^\infty,
\end{equation}
which in turn follows from 
Lemma~\ref{lem:2.1} and Lemma~\ref{lem:betaact}. 

\nin {\bf Claim 2:} For $g \in S_\sV^0$, the map 
$F(\cdot,g)$ is  weak-$*$-holomorphic on $\cS_\pi$. \\
As $\alpha^\eta$ is weak-$*$-holomorphic on $\cS_\pi$, 
Claim 2 follows from Lemma~\ref{lem:cartes}(b) and the antiholomorphy of the maps 
\[ \cS_\pi \to \cH^{\infty}, \quad z \mapsto U(\beta^g(z)^*)\xi 
= U(\beta^{g^{-1}}(\oline z))\xi,  \qquad \xi \in \cH^\infty,\] 
which in turn follows from Lemma~\ref{lem:2.1} 
and the holomorphy of $\beta^g$ on $\cS_\pi$. \\
\nin Claims 1 and 2 imply that, for $g \in S_\sV^0$, we have 
 $U^{-\infty}(g)\eta \in \cH^{-\infty}_{\rm ext}$ with 
$\alpha^{U^{-\infty}(g)\eta} = F(\cdot,g)$. 

\nin {\bf Step 2:} $U^{-\infty}(C^\infty_c(S_\sV^0,\R))\cH^{-\infty}_{\rm ext}\subeq \cH^{-\infty}_{\rm ext}$. \\
For $\psi \in C^\infty_c(S_\sV^0,\R)$, we consider the map 
\[ f\:  \oline{\cS_\pi} \to \cH^{-\infty}, 
\quad 
f(z) :=  \int_{\supp(\psi)} \psi(g) F(z,g)\, dg
= \int_{S_\sV} \psi(g) U^{-\infty}(\beta^g(z))\alpha^\eta(z)\, dg. \] 
That $f$ is weak-$*$-continuous follows from Claim 1 and 
\cite[Lemma~1.1.11]{GN}. That 
it is weak-$*$-holomorphic on~$\cS_\pi$ 
likewise follows from Claim 2 with~\cite[Lemma~1.3.15]{GN}. 
This shows that $U^{-\infty}(\psi)\eta \in \cH^{-\infty}_{\rm ext}$ 
because $f = \alpha^{U^{-\infty}(\psi)\eta}$. 

\nin {\bf Step 3:} Invariance of $\cH^{-\infty}_{{\rm ext},J}$. \\
We now assume that $\eta\in \cH^{-\infty}_{{\rm ext},J}$, so that 
$\alpha^\eta(\pi i) = J_\sV \eta$. Then 
\begin{equation}
  \label{eq:17}
\alpha^{U^{-\infty}(g)\eta}(\pi i) 
= F(\pi i,g) 
= U^{-\infty}(\beta^g(\pi i)) \alpha^\eta(\pi i)
= U^{-\infty}(\tau_G(g)) J_\sV \eta 
= J_\sV U^{-\infty}(g) \eta 
\end{equation}
shows that  $U^{-\infty}(g)\eta \in \cH^{-\infty}_{{\rm ext},J}$. 
We further obtain with \eqref{eq:17} 
\begin{align*}
 \alpha^{U^{-\infty}(\psi)\eta}(\pi i) 
&= f(\pi i) 
= \int_{S_\sV} \psi(g) U^{-\infty}(\beta^g(\pi i))\alpha^\eta(\pi i)\, dg\\
&= \int_{S_\sV} \psi(g) J_\sV U^{-\infty}(g) \eta\, dg 
= J_\sV \int_{S_\sV} \psi(g)  U^{-\infty}(g) \eta\, dg 
= J_\sV U^{-\infty}(\psi)\eta.
\qedhere\end{align*}
\end{prf}

The following technical lemma is of key importance in this section. It provides a 
sufficient condition for elements $\eta \in \cH$ to be contained 
in $\sV$ in terms of a rather weak holomorphic extension 
requirement on the orbit map~$\alpha^\eta$. 
Combined with Lemma~\ref{lem:hinftyext}, its helps us to construct 
elements $U^{-\infty}(\phi)\eta \in \sV$. 

\begin{lem} \mlabel{lem:key-abstract}
Let $\cD \subeq \cH$ be a dense complex subspace, so that 
$v \mapsto \la \cdot, v \ra$ injects $\cH$ into the space 
$\cD^\sharp$ of antilinear functionals $\cD \to \C$. 
Let $\sV \subeq \cH$ be a standard subspace for which $\cD$ is invariant 
under $J_\sV$ and $\Delta_\sV^{i\R}$. 
Suppose that $\eta \in \cH$ is such that 
the orbit map 
\[ \alpha^\eta \: \R\to \cD^\sharp, \quad t \mapsto 
\eta \circ \Delta_\sV^{it/2\pi} \] extends to a map 
$\alpha^\eta \: \oline{\cS_\pi} \to \cD^\sharp$ which is 
pointwise continuous,  pointwise holomorphic on $\cS_\pi$, and 
satisfies $\alpha^\eta(\pi i) = J_\sV \eta$, 
for $(J_\sV \eta)(\xi) := \oline{\eta(J_\sV \xi)}$. 
Then $\la \xi, \eta \ra \in \R$ for every $\xi \in \cD \cap \sV'$, 
and if $\cD \cap \sV'$ is dense in $\sV'$, then $\eta \in \sV$. 
\end{lem}

\begin{prf} Let $\xi \in \cD \cap \sV'$. 
By assumption, we have a continuous function 
\[ f \: \oline{\cS_\pi} \to \C, \quad 
f(z) = \alpha^\eta(z)(\xi) \] 
which is holomorphic on $\cS_\pi$. This function satisfies for $t \in \R$ 
\begin{equation}
  \label{eq:s1}
f(t + \pi i) 
= \alpha^\eta(t + \pi i)(\xi) 
= \alpha^\eta(\pi i)(\Delta_\sV^{it/2\pi}\xi) 
= (J_\sV\eta)(\Delta_\sV^{it/2\pi}\xi) 
= \oline{\eta(\Delta_\sV^{it/2\pi}J_\sV\xi)}.
\end{equation}
As $\xi \in \sV'$, we also have for $t \in \R$ 
\[ f(t) = \alpha^\eta(t)(\xi) 
= \la \Delta_{\sV}^{it/2\pi}\xi, \eta \ra,\] 
and this function extends to a continuous function on 
$-\oline{\cS_\pi}$, holomorphic on $-\cS_\pi$, given by 
\[ f(z) =  \la \Delta_\sV^{i \oline  z/2\pi} \xi, \eta \ra.\] 
We thus obtain a continuous function 
\[ f \:  \oline{\cS_{-\pi,\pi}} = \{ z \in \C \: -\pi \leq \Im z \leq \pi \} \to \C, \] 
holomorphic on the complement of the real line in the interior 
$\cS_{-\pi,\pi}$, so that Morera's Theorem implies that it is 
holomorphic on~$\cS_{-\pi,\pi}$. On the lower boundary we have
\[   f(t - \pi i) 
= \la \Delta_\sV^{it/2\pi} \Delta_\sV^{-1/2}\xi, \eta \ra 
= \la \Delta_\sV^{it/2\pi} J_\sV \xi, \eta \ra 
= \eta(\Delta_\sV^{it/2\pi} J_\sV \xi) = \oline{f(t + \pi i)}.\] 
Therefore the function 
\[ F \: \{ z \in \C \: -\pi \leq \Im z \leq \pi \} \to \C, \quad 
F(z) := f(z) - \oline{f(\oline z)} \] 
is continuous, holomorphic on the interior, and vanishes on the line 
$\R + \pi i$. This implies that $F = 0$, and evaluating on the real line shows that 
$f(\R) \subeq \R$. For $t = 0$, we obtain in particular 
$f(0) = \la \xi, \eta \ra \in \R$. 

If $\cD \cap \sV'$ is dense in $\sV'$, then this further leads to 
$\eta \in (\sV')' = \sV$. 
\end{prf}

\begin{lem} \mlabel{lem:key}
$\cH^{-\infty}_{{\rm ext},J} \subeq \sV^{-\infty}$. 
\end{lem}

\begin{prf} First we show that 
  \begin{equation}
    \label{eq:step1}
\cH^{-\infty}_{{\rm ext},J} \cap \cH \subeq \sV.
  \end{equation}
We apply Lemma~\ref{lem:key-abstract} 
with $\cD = \cH^\infty$ and $\eta \in \cH \cap \cH^{-\infty}_{{\rm ext},J}$. 
As $\sV' \cap \cH^\infty$ is dense in $\sV'$ by 
Lemma~\ref{lem:vinfty}(a), we obtain $\eta \in \sV$. 
This proves \eqref{eq:step1}. 

For $\eta \in \cH^{-\infty}_{{\rm ext},J}$ and 
$\psi \in C^\infty_c(S_\sV^0,\R)$, Lemma~\ref{lem:hinftyext} 
implies that 
$U^{-\infty}(\psi) \eta \in \cH^{-\infty}_{{\rm ext},J}$, 
and since this is actually an element of $\cH$, 
\eqref{eq:step1} shows that $U^{-\infty}(\psi) \eta \in \sV$. 
Now the assertion follows from Lemma~\ref{lem:vinfty}(c). 
\end{prf}

\begin{prop} Let $F \subeq \cH^{-\infty}_{{\rm ext},J}$ be a real linear subspace 
which is $G$-cyclic in the sense that $\cH_\sF(G) = \cH$, 
let $A = \exp(\R h)$, and 
\[ \sE := \spann_\R (U^{-\infty}(A)\sF) \subeq \cH^{-\infty}.\] 
Then 
\[ \sH_\sE(S_\sV^0) = \sV.\] 
\end{prop}

\begin{prf} Since the subspace $\cH^{-\infty}_{{\rm ext},J}$ 
is $A$-invariant, 
we have $\sE \subeq \cH^{-\infty}_{{\rm ext},J} \subeq \sV^{-\infty}$ 
(Lemma~\ref{lem:key}). 
As $\sF$ is $G$-cyclic, $\cH_\sE(G) \supeq \cH_\sF(G) = \cH$, 
so that Theorem~\ref{thm:3.6}(b) implies that ${\sH_\sE(S_\sV^0) = \sV}$. 
\end{prf}

\begin{rem} \mlabel{rem:3.12} 
(a) If $\eta \in \cH^{-\infty}$ is contained 
in a finite dimensional complex $A$-invariant subspace 
$\cK \subeq \cH^{-\infty}$, 
then the representation of $A$ on $\cK$ is continuous because 
it is continuous on its dual space $\cH^{\infty}/\cK^\bot$. 
Hence it extends to a holomorphic representation $\rho$ of $A_\C \cong \C$ 
on the finite dimensional complex vector space $\cK$, and therefore 
$\cK \subeq \cH^{-\infty}_{\rm ext}$. 

If, in addition,  $J\cK \subeq \cK$ and $J_\cK := J\res_{\cK}$, then 
the real subspace 
\[ \sE := \Fix(J_\cK \rho(\exp(\pi i h))) \subeq \cH_{{\rm ext},J} \subeq \sV^{-\infty}\]
of $\cK$ satisfies 
\[ \sH_\sE(S_\sV^0) = \sV.\]

\nin (b) Below we shall need the following more general fact. 
We assume that $(\rho,\cK)$ is a norm-continuous representation 
of $A\cong \R$ on the Hilbert space $\cK$ by symmetric operators 
and that we have 
a continuous $A$-equivariant map $\eta \: \cH^\infty \to \cK$ 
with dense range. 
The adjoint map $\eta^* \: \cK \to \cH^{-\infty},
\eta^*(\xi)(v) := \la \eta(v), \xi \ra$ defines a weak-$*$ continuous 
equivariant embedding. 
Since $\eta^*$ is $A$-equivariant and $\rho$ extends to a 
holomorphic representation of $A_\C$ on $\cK$, we then have 
$\eta^*(\cK) \subeq \cH^{-\infty}_{\rm ext}$. 
\end{rem}

\begin{rem}
We assume that $\ker(U)$ is discrete. 
Clearly, regularity of $\sV$, i.e., that $\L(S_\sV)$ spans $\g$, 
implies $e \in \oline{S_\sV^0}$, but the converse is not 
clear. In view of the Germ Theorem (\cite[Thm.~4.1]{Ne19}), 
$e \in \oline{S_\sV^0}$ is equivalent to $e \in \oline{S_{C,{\rm inv}}^0}$, 
where 
\[ S_{C, {\rm inv}} = \{ g \in G \:  
(\forall z \in \oline{\cS_\pi})\ \beta^g(z) \in G \exp(iC) \}.\] 
From \cite[Lemma~5.4(ii)]{Ne19} it follows that $e \in \oline{S_{C,{\rm inv}}^0}$ 
implies that $\tau = e^{\pi i \ad h}\res_\g$, hence 
$e^{2\pi i \ad h} = \id_{\g_\C}$, so that  $\ad h$ is diagonalizable 
with integral eigenvalues. Presently we do not know how to derive from 
 $e \in \oline{S_{C,{\rm inv}}^0}$ that $\L(S_\sV)$ has interior points. 
The main difficulty is to show that $\Spec(\ad h) \subeq \{-1,0,1\}$. 
\end{rem}

Although we will not need it in the following, we record the following invariance 
property: 

\begin{lem} The subspaces 
$\cH^{-\infty}_{{\rm ext},J} \subeq \cH^{-\infty}_{\rm ext}$ are both invariant under 
$\dd U^{-\infty}(\g)$. 
\end{lem}

\begin{prf} If suffices to prove invariance under $\dd U^{-\infty}(\g^j)$ 
for $j = 1,0,-1$. 
Let $\eta \in \cH^{-\infty}_{\rm ext}$ and 
$x \in \g^j$, so that $[h,x] = j x$. 
For $\eta_x := \dd U^{-\infty}(x)\eta$, we then we then have 
\[ \alpha^{\eta_x}(t) 
= \dd U^{-\infty}(e^{t \ad h} x)\alpha^\eta(t) 
= e^{tj} \dd U^{-\infty}(x)\alpha^\eta(t) \quad \mbox{ for } \quad t \in \R.\] 
Hence the map 
\[ \alpha^{\eta_x} \: \oline{\cS_\pi} \to \cH^{-\infty}, \qquad 
z \mapsto 
\dd U^{-\infty}(e^{z \ad h} x)\alpha^\eta(z) 
= e^{jz} \dd U^{-\infty}(x)\alpha^\eta(z) \] 
is weak-$*$-continuous and weak-$*$-holomorphic on the interior. 
This proves the invariance of $\cH^{-\infty}_{\rm ext}$. 
If, in addition, $\alpha^\eta(\pi i) = J \eta$, then we further get 
\[ \alpha^{\eta_x}(\pi i) 
= \dd U^{-\infty}(e^{\pi i \ad h} x)\alpha^\eta(\pi i) 
= \dd U^{-\infty}(\tau(x)) J\eta
= J \dd U^{-\infty}(x) \eta = J \eta_x.\qedhere\] 
\end{prf}

\subsection{An example: The $ax + b$-group} 
\mlabel{subsec:3.4}

We consider the affine group $G = \Aff(\R)_0$ of the real line with 
\[ G_\tau = \Aff(\R) \cong \R \rtimes \R^\times 
\quad \mbox{ and } \quad 
\tau_G(b,a) = (-b,a).\] 
The Lie algebra data is given by 
\[ C = [0,\infty), \quad h := (0,1), \quad q := (1,0), \quad \mbox{ and } \quad 
\g^0 = \R h,\quad \g^1 = \R q.\] 

\begin{prop} For the irreducible antiunitary $C$-positive representation 
$(U,\cH)$ of $\Aff(\R)$, given by 
\[ \cH = L^2\Big(\R^\times_+, \frac{dp}{p}\Big) 
\quad \mbox{ and } \quad 
(U(b,e^t)f)(p) = e^{ibp} f(e^t p), \quad 
U(0,-1)f = \oline f,\] 
the following assertions hold: 
\begin{itemize}
\item[\rm(a)] The power functions $(p^s)_{\Re s > 0}$ define distribution vectors 
\[ \eta_s(f) := \int_{\R^\times} \oline{f(p)} p^s\, \frac{dp}{p}\] 
for the restriction $U_N$ to the translation group $N = \R \times \{1\}$. 
\item[\rm(b)] These distribution vectors transform under the action of the dilation 
group $\R^\times$ by 
\begin{equation}
  \label{eq:trafos}
 U^{-\infty}(0,a) \eta_s = a^s \eta_s \quad \mbox{ for } \quad a > 0 
\quad \mbox{ and } \quad 
U^{-\infty}(0,-1) \eta_s = \eta_{\oline s}.
\end{equation}
\item[\rm(c)]  The distribution $D_s := j_{\eta_s}(\eta_s) \in \cS'(\R)$ 
{\rm(cf.\ \eqref{eq:dist-inc} in Appendix~\ref{app:a})} 
coincides with the Fourier transform of the measure 
$p^{2 \Re s-1}\, dp$ on $\R_+$. 
\end{itemize}
\end{prop}

\begin{prf} (a) As $\cH^{-\infty}(U_N) = \spann_\C \big(\dd U^{-\infty}(\cU(\R q)) \cH\big)$ 
(Lemma~\ref{lem:charsmooth}), it 
suffices to show that there exists a polynomial $F(p) \in \C[p]$ 
with $F(p)^{-1} p^s \in L^2(\R^\times_+, \frac{dp}{p})$. 
For $F(p) := (1 + p^2)^n$, we have 
\[ \|F(p)^{-1}p_s\|_2^2 
= \int_0^\infty \frac{p^{2 \Re s}}{(1 + p^2)^{2n}}\, \frac{dp}{p},\] 
and this integral is finite if and only if 
$\Re s > 0$ and $2n > \Re s$. 

\nin (b) Next we note that 
\[ \big(U^{-\infty}(0,e^t) \eta_s\big)(f) 
= \eta_s(U(0,e^{-t})f)
= \int_{\R^\times} \oline{f(e^{-t}p)} p^s\, \frac{dp}{p}
= \int_{\R^\times} \oline{f(p)} (e^tp)^s\, \frac{dp}{p}
= e^{ts} \eta_s(f)\] 
and 
\[ \big(U^{-\infty}(0,-1) \eta_s\big)(f) 
= \oline{\eta_s(\oline f)} 
= \int_{\R^\times} \oline{f(p)} p^{\oline s}\, \frac{dp}{p}
= \eta_{\oline s}(f),\] 
so that $\eta_s \in \cH^{-\infty}$ is an $\exp(\R h)$-eigenvector satisfying 
\eqref{eq:trafos}. 

\nin (c) For $\phi \in C^\infty_c(\R,\R)$, we have 
\[ (U^{-\infty}(\phi) \eta_s)(p) 
= \tilde \phi(p) p^s, \qquad 
\tilde\phi(p) = \int_\R e^{ipx} \phi(x)\, dx,   \] 
because 
\[ \eta_s(U(\phi^*)f) 
= \int_0^\infty p^s \oline{\tilde{\phi^*}(p)} \oline{f(p)}\, \frac{dp}{p}
= \int_0^\infty p^s \tilde \phi(p) \oline{f(p)}\, \frac{dp}{p}.\]
For the distribution $D_s = j_{\eta_s}(\eta_s) \in C^{-\infty}(\R)$, we thus 
obtain 
\[ D_s(\phi) = \eta_s(U^{-\infty}(\phi)\eta_s) 
= \int_0^\infty p^s \oline{\tilde\phi(p) p^s}\, \frac{dp}{p}
= \int_0^\infty p^{2 \Re s} \oline{\tilde\phi(p)}\, \frac{dp}{p}
= \int_0^\infty p^{2 \Re s} \hat{\oline\phi}(p)\, \frac{dp}{p}.\]
This is (c). 
\end{prf}

To identify corresponding standard subspaces, we note that 
$\alpha^{\eta_s}(z) = e^{sz} \eta_s$ implies that 
$\alpha^{\eta_s}(\pi i) = e^{\pi i s} \eta_s$ and 
$J \eta_s = \eta_{\oline s}$. For $s \in \R$, it follows that 
\[ \tilde \eta_s := e^{-\frac{s \pi i}{2}} \eta_s 
\in \cH^{-\infty}_{{\rm ext},J} \subeq \sV^{-\infty}.\] 
Accordingly, $\sV \subeq \cH$ is generated by the functions 
\[ \tilde\phi \tilde \eta_s = e^{-\frac{s \pi i}{2}} \tilde \phi \eta_s 
\quad \mbox{ for  }\quad \phi \in C^\infty_c((0,\infty),\R).\] 
For $s \not\in \R$, we put 
\[ \tilde \eta_s := 
\eta_s + e^{-\pi i \oline s} \eta_{\oline s}  \] 
and obtain 
$J \tilde \eta_s  = e^{\pi i s} \eta_s + \eta_{\oline s} 
=  \alpha^{\tilde\eta_s}(\pi i),$ so that 
\[ \tilde\eta_s \in \cH^{-\infty}_{{\rm ext},J} \subeq \sV^{-\infty}.\]

\section{Realizing $C$-positive representations 
on tubes} 
\mlabel{sec:4} 

In this section we show that, for semisimple Lie algebras 
$\g$, which are direct sums of simple hermitian ideals of tube type, 
and any pointed generating invariant cone $C \subeq \g$, 
the (irreducible) $C$-positive unitary representation can be realized 
in Hilbert spaces of holomorphic functions on a tube domain. This 
realization will be used in Section~\ref{sec:5} to exhibit a suitable 
real subspace $\sE \subeq \cH^{-\infty}_{{\rm ext},J}$ to which the results 
from Section~\ref{sec:3} apply.

Let $\g$ be a semisimple Lie algebra 
and $h \in \g$ be an Euler element for which 
$\fh := \g^0(h)$ contains no non-zero ideal of $\g$, i.e., the Lie algebra 
$\g$ is generated by  $\g^{\pm 1} = \g^{\pm 1}(h)$. 
We consider the Lie groups
\[ G := \Inn(\g) = \Aut(\g)_0 \quad \mbox{ and } \quad 
G_\tau := G  \rtimes \{ \1, \tau \} 
\quad \mbox{ for } \quad \tau = e^{\pi i \ad h}\res_{\g}. \]
Then $G_\tau$ is a graded Lie group and we put $H := \Inn_\g(\fh) = (G^\tau)_0$. 
We also fix a Cartan involution $\theta$ of $\g$ fixing the center 
pointwise and satisfying $\theta(h) = - h$. The corresponding 
Cartan decomposition is written $\g = \fk \oplus \fp$, 
and we write $K := \la \exp \fk \ra$ for the corresponding 
maximal compact subgroup of~$G$.

We further assume that all simple ideals of $\g$ are 
hermitian, so that the existence of an Euler element 
implies that they are tube type (cf.~\cite[Thm.~3.12]{MN20}; 
\cite[Thm.~5.6]{O91}) 
and that we may consider the eigenspace 
$E := \g^1(h)$ as a, not necessarily simple, 
unital euclidean Jordan algebra. We may further pick 
the unit element $e\in E$ in such a way that it is fixed by $H_K= H^\theta$. Then 
\[ H_K \cong \Aut(E,e)_0 \quad \mbox{ and } \quad 
H \cong \Str(E)_0 \subeq H_\C := \Str(E_\C)_0 \] 
are the identity components of the real and complex structure groups. 
Accordingly, we identify them with linear groups acting on $E_\C$. 
We then have the two associated complex domains: 
\begin{itemize}
\item the tube domain $\cT := \cT_{E_+} := E + i E_+ \subeq E_\C$, and 
\item the unit ball 
$\cD := \{ z\in E_\C \:  \|z\|_\infty < 1 \}$, 
with respect to the spectral norm $\|\cdot\|_\infty$ of $E_\C$, 
which is a bounded symmetric domain 
\end{itemize}
The {\it Cayley transform} 
\[ p \: \cT \to \cD, \quad 
p(z) := (z - ie)(z+ie)^{-1} \] 
maps the tube $\cT$ biholomorphically onto $\cD$. Its inverse is given 
by 
\[ c \: \cD \to \cT, \quad 
c(z) := i(e+z)(e-z)^{-1}\] 
(cf.~\cite[Thm.~X.4.3]{FK94}).
The differential of the Cayley 
transform is given by 
\begin{equation}
  \label{eq:diffcayley}
 \dd p(z) = 2i P(z + i e)^{-1} \quad \mbox{ for } \quad z \in \cT,
\end{equation} 
where 
\[ P \: E_\C \to \End(E_\C), \qquad P(z) := 2 L(z)^2 - L(z^2), \qquad 
L(z)w := zw, \] 
is the {\it quadratic representation} 
(\cite[p.~192]{FK94}). 
We have 
$P(E_\C^\times) \subeq \Str(E_\C)$ 
(\cite[Prop.~VIII.2.4]{FK94}), hence in particular 
\[ P(\cT) \subeq H_\C \quad \mbox{ because} \quad 
P(ie) = -P(e) = - \1 \in \C^\times\1 = \exp(\C h)\subeq H_\C\] 
(cf.\ \cite[p.~168]{FK94}). 

The group 
$G\cong \Conf(E)_0$ is generated by the subgroups 
$N^+ := \exp(\g^1(h))\cong (E,+)$ (the translation group of $E$), 
the subgroup $H$ 
and an element $\eta \in Z(K)$ acting on $E_\C^\times$ by 
the involution 
\begin{equation}
  \label{eq:dereta}
 \eta(z) = - z^{-1} \quad \mbox{ with differential } \quad 
\dd\eta(z) = P(z)^{-1}
\end{equation}
(\cite[Prop.~II.3.3(i)]{FK94}). 
This is the point symmetry in the point $ie$ of the Riemannian 
symmetric space~$\cT$.
The element $\eta \in G$ implements the 
Cartan involution $\theta$ of $G$ by 
\[ \theta(g) = \eta \circ g \circ \eta^{-1}= \eta \circ g \circ
\eta. \] 

We write $G_\C := \Aut(\g_\C)_0$ and observe that $H_\C = \la \exp \fh_\C \ra$. 
For $g \in H$, we put 
\[ g^* := \theta(g)^{-1},\] 
and if $\oline\theta$ denotes the antiholomorphic involution of 
$H_\C$ extending $\theta$, we likewise put 
\[ g^* := \oline\theta(g)^{-1} \quad \mbox{ for } \quad g \in H_\C.\] 
We also use this notation for elements of the simply connected covering 
group $q_{H_\C} \: \tilde H_\C \to H_\C$ and the lift of $\oline\theta$ to 
$\tilde H_\C$. We note that the kernel 
 of $q_{H_\C}$ is a discrete central subgroup.
As $\cT$ is simply connected, there exists a holomorphic lift 
\begin{equation}
  \label{eq:tildep}
\tilde P \: \cT \to \tilde H_\C. 
\end{equation}
As $P(ie) = -\1$, any such lift is determined by 
$\tilde P(ie) \in q_{H_\C}^{-1}(-\1)$, which is a discrete subgroup 
of the center of $\tilde H_\C$; a coset of $\ker(q_{H_\C})$. 
We also observe that 
$\tilde P(ie)$ is contained in the connected 
subgroup $(\tilde H_\C)^{\oline\theta}$ of unitary elements 
with respect to the involution $g^* = \oline\theta(g)^{-1}$. 

We write $\tau_{H_\C}(g) = \oline g = \tau g \tau$ 
for the antiholomorphic automorphism of $H_\C$ induced by 
complex conjugation on~$E_\C$. Its lift to $\tilde H_\C$ 
is denoted $\tau_{\tilde H_\C}$. We write 
$H^\sharp = (\tilde H_\C)^{\tau_{\tilde H_\C}}$ for its 
connected group of fixed points 
(\cite[Thm.~IV.3.4]{Lo69}). 

\subsection{Translating between tube and ball} 
\mlabel{subsec:4.1} 

Let $G(\cD) = \Aut(\cD)_0$ be the identity component of the 
group of holomorphic automorphisms of 
$\cD$ and $G(\cT) = \Aut(\cT)_0 \cong G$ be the corresponding group 
for the tube domain $\cT$. Both are subgroups 
of the conformal group $\Conf(E_\C)$, and in this group 
they are conjugate by the Cayley transform 
\begin{equation}
  \label{eq:cayconj}
G(\cD)=c^{-1} G(\cT)c \subeq G_\C.
\end{equation}
For the simply connected covering groups 
$q_{G(\cD)} \: \tilde G(\cD) \to G(\cD)$ and 
$q_{G(\cT)} \: \tilde G(\cT) \to G(\cT)$ we define action 
and $\cD$ and $\cT$, respectively, by 
$g.z := \alpha_{G(\cD)}(g)(z)$ and $g.z := \alpha_{G(\cT)}(g)(z)$. 

Let 
\begin{equation}
  \label{eq:liftalphac}
\tilde\alpha_c \: \tilde G(\cD) \to \tilde G(\cT) 
\end{equation}
be the isomorphism obtained by lifting the isomorphism 
\[ \alpha_c \: G(\cD) \to G(\cT),\quad  \alpha_c(g) := cgc^{-1}.\] 
We write 
\[ K(\cD) := G(\cD)^0 \quad \mbox{ and } \quad 
K(\tau) := G(\cT)^{ie} \] 
for the stabilizer groups of the base points $0 \in \cD$ and $ie \in \cT$. 

On the tube domain $\cT = E + i E_+$, we consider the 
{\it universal kernel} 
\[ Q \: \cT \times \cT \to H_\C, \quad 
Q(z,w) := P\Big(\frac{z-\oline w}{2i}\Big), \] 
which satisfies $Q(ie,ie) = P(e) = \1$. It is sesquiholomorphic 
in the sense that it is holomorphic in the first and antiholomorphic 
in the second argument. As $\cT$ is simply 
connected, this kernel lifts to a sesquiholomorphic kernel 
\[ \tilde Q \: \cT \times \cT \to \tilde H_\C 
\quad \mbox{ with }\quad 
\tilde Q(ie,ie) = \1.\] 
The action of $G(\cT)$ on $\cT$ defines a cocycle 
\[ J \: G(\cT) \times \cT \to H_\C, \qquad 
J(g,z) := J_g(z) := \dd \sigma_g(z) \] 
satisfying 
\begin{equation}
  \label{eq:cocyc1}
J(g_1 g_2,z) = J(g_1,g_2.z) J(g_2,z) 
\quad \mbox{ for } \quad 
g\in G, z \in \cT.  
\end{equation}
Since $\cT$ is simply connected, this cocycle lifts to a 
smooth cocycle 
\[ \tilde J \: \tilde G(\cT) \times \cT \to \tilde H_\C 
\quad \mbox{ determined  by } \quad \tilde J(e,ie) = \1, \] 
which is holomorphic in the second argument. The cocycle 
property of $\tilde J$ follows by the uniqueness of lifts, 
which immediately implies $\tilde J(e,z) = \1$ for every $z \in \cT$. 
We have 
\begin{itemize}
\item $\tilde J_g = 1$ for translations $g(x) = x + v$, $ \in E$, 
\item On the connected subgroup $\tilde H(\cT) = \la \exp \fh \ra 
\subeq \tilde G(\cT)$, the cocycle $\tilde J$ 
defines a morphism of Lie groups 
$\tilde J\res_{\tilde H(\cT)} \: \tilde H(\cT) \to H^\sharp \subeq \tilde H_\C$ 
integrating the inclusion $\fh \into \fh_\C$. 
\end{itemize}

\begin{lem}
  \mlabel{lem:III.2.1} 
  \begin{equation}
    \label{eq:kern-trafo-lift}
  \tilde Q(g.z, g.w) = 
\tilde J(g,z) \tilde Q(z,w) \tilde J(g,w)^*\quad \mbox{  for } \quad 
z \in \cT, g \in \tilde G(\cT).
 \end{equation}
\end{lem}

\begin{prf} (a) First we verify the corresponding relation 
for the $H_\C$-valued kernel~$Q$: 
\[ Q(g.z, g.w) = J(g,z) Q(z,w) J(g,w)^* 
\quad \mbox{  for } \quad z \in \cT, g \in \tilde G(\cT).\] 
Since $G = G(\cT)$ is generated by $H$, $\eta$ and translations 
and $J$ is a cocycle, we have to verify this relation only 
for these three cases. 

For $g \in H$ we have 
\[ Q(g.z, g.w) 
=   P({\textstyle{g.z - \oline {g.w} \over 2i }})
=   P(g.{\textstyle{z - \oline {w} \over 2i }})
 \ {\buildrel {!} \over =} \  g   P({\textstyle{z - \oline {w} \over 2i }}) g^* 
= J(g,z)  Q(z,w) J(g,w)^*,\] 
where we use \cite[p.~147]{FK94} for $!$. 
For $u \in E$ and the translation $g = \exp u \in N^+$, we obtain 
\[ Q(z+u,w+u) 
=   P\Big(\frac{z + u - \oline w - \oline u}{2i}\Big)
=   P\Big(\frac{z - \oline {w}}{2i}\Big)
=   Q(z,w) 
=   J(g,z)  Q(z,w)   J(g,w)^* \] 
because $J(g,z) = \1$. 
Finally we see with \cite[Lemma~X.4.4(i)]{FK94} that 
\begin{align*}
Q(-z^{-1}, -w^{-1}) 
&=   P({\textstyle{- z^{-1} + \oline w^{-1} \over 2i }}) 
= -{\textstyle{1\over 4}} P(- z^{-1} + {\oline w}^{-1}) \\ 
&= -{\textstyle{1\over 4}} P(-z)^{-1} P(- z + \oline w) P(\oline w)^{-1}  
= P(z)^{-1} P({\textstyle{z - \oline {w} \over 2i }}) P(\oline w)^{-1} \\ 
&= J(\eta, z) Q(z,w) J(\eta,w)^* 
\end{align*}
because 
$P(w)^* = \oline{P(w)} = P(\oline w)$  
and $J(\eta,z) = P(z)^{-1}$. 

\nin (b) Now we turn to the lifted kernel $\tilde Q$. 
Fix $z,w \in \cT$. Then both sides of \eqref{eq:kern-trafo-lift} 
take the same values in $g = e$. By (a), they are lifts of the 
same maps to $H_\C$. Hence the assertion follows from the uniqueness 
of continuous lifts. 
 \end{prf}

\begin{rem} The preceding lemma has some remarkable consequences. 

\nin (a) For $g \in H$, we have 
\[ Q(g.ie, g.ie) = gg^*, \] 
which is the quadratic representation of the Riemannian symmetric 
space $E_+ \cong i E_+ \subeq \cT$ with values in its isometry 
group~$(H,\theta)$.

\nin (b) For elements of the stabilizer group $\tilde K 
= \tilde K(\cT) = \tilde G(\cT)^{ie}$, 
we have 
\[ \tilde J(g,ie)  \tilde J(g,ie)^* = \1 \quad \mbox{ for } \quad 
g \in \tilde K, \] 
so that $\tilde J$ defines a homomorphism 
\[ \tilde J \: \tilde K \to (\tilde H_\C)^{\oline\theta} = \tilde K(\cD).\] 
\end{rem}
Using the Cayley transform, we obtain a kernel on $\cD$ by 
\begin{align} \label{eq:2kerneltrafo}
  Q^\cD(z, w) 
&:= 4\cdot \dd c(z)^{-1} Q(c(z), c(w)) (\dd c(w)^{-1})^* \notag \\ 
&= 4\cdot \dd p\big(c(z)\big) Q(c(z), c(w)) \big(\dd p(c(w))\big)^*\notag\\
&{\buildrel \eqref{eq:diffcayley} \over =}\ \  P\Big(\frac{c(z)+ie}{2}\Big)^{-1}  Q(c(z), c(w)) 
\Big(P\Big(\frac{c(z)+ie}{2}\Big)^{-1}\Big)^*. 
\end{align}
The kernel $Q^\cD$ also takes values in $H_\C$, 
and $c(0) = ie$ leads in particular to 
\[ Q^\cD(0,0) 
= P(ie)^{-1}  Q(ie, ie) (P(ie)^{-1})^* = \1,\] 
so that there exists a unique continuous lift
\begin{equation}
  \label{eq:qdlift}
 \tilde Q^\cD\colon \cD \times \cD \to \tilde H_\C
\quad \mbox{ with } \quad \tilde Q^\cD(0,0) = \1.
\end{equation}
As for the tube $\cT$, the cocycle 
\[ J^\cD\colon G(\cD)\times \cD\to H_\C,
\qquad J^\cD(g,z):=\dd g(z)\] 
lifts to a cocycle 
\[ \tilde J^{\cal D}\colon \tilde G(\cD)\times \cD \to \tilde H_\C.\] 
As 
\[ J^\cD(k,z)=k\quad \mbox{ for}\quad  k\in K(\cD)= G(\cD)^0, \]
the restriction of $J^\cD$ provides an inclusion 
\[  J^\cD \: K(\cD) = G(\cD)^0 \into H_\C \] 
(cf.~\cite[Thm.~X.5.3, Prop.~X.3.1]{FK94}). Its image is a maximal 
compact subgroup, so that $H_\C \cong K(\cD)_\C$. \\

\subsection{Unitary representations} 
\mlabel{subsec:4.2} 

Now we turn to unitary representations of the simply connected groups 
$\tilde G(\cD)$ and $\tilde G(\cT)$ on Hilbert spaces of holomorphic functions 
on $\cD$ and $\cT$, respectively.  
We start with a holomorphic involutive representation 
$(\rho, \cK)$ of the complex involutive Lie group $\tilde H_\C$ 
with $g^* = \oline\theta(g)^{-1}$ on the complex Hilbert space~$\cK$. 
This representation does not have to be finite dimensional, but 
holomorphy implies that the operators of the derived representations 
are all bounded and that $\rho$ is norm-continuous. Conversely, 
every norm-continuous $*$-representation of $\tilde H$ defines a 
holomorphic representation of its universal complexification $\tilde H_\C$. 

\begin{defn} A holomorphic $*$-representation $(\rho,\cK)$ of 
$\tilde H_\C$ is said to be {\it $\cT$-positive} if the kernel 
\[ Q_\rho := \rho \circ \tilde Q, \: \cT \times \cT \to B(\cK), \quad  
Q_\rho(z,w) = P_\rho\Big(\frac{z - \oline w}{2i}\Big) \] is 
positive definite. 
\end{defn}

Then we obtain a representation of $\tilde G(\cD)$ 
on the space $\Hol(\cD,\cK)$ by 
\begin{equation}
  \label{eq:piD}
 (U^\cD_\rho(g)f)(z) 
:= J^\cD_\rho(g, g^{-1}.z) f(g^{-1}.z) 
\quad \mbox{ for } \quad g \in \tilde G(\cD), z \in \cD, 
J^\cD_\rho := \rho \circ \tilde J^\cD,
\end{equation}
and a representation of $\tilde G(\cT)$ 
on the space $\Hol(\cT,\cK)$ by 
\begin{equation}
  \label{eq:pitau}
 (U_\rho(g)f)(z) := J_\rho(g, g^{-1}.z) f(g^{-1}.z)  
\quad \mbox{ for } \quad g \in \tilde G(\cT), z \in \cT, 
J_\rho := \rho \circ \tilde J.
\end{equation}
The space $\Hol(\cD,\cK)$ contains a unitary subrepresentation 
of $G(\cD)$ if and only if 
the kernel $Q_\rho^\cD := \rho \circ \tilde Q^\cD$ 
is positive definite (\cite[Prop.~XII.2.1]{Ne99}; see 
also \cite[Thm.~XII.2.6]{Ne99} for the link to highest weight representations). 
To see that the Cayley transform provides a natural intertwining operator, 
we use a holomorphic lift 
$\tilde P \: \cT \to \tilde H_\C$ of the quadratic representation~$P$ 
(cf.\ \eqref{eq:tildep}). 
As 
\begin{equation}
  \label{eq:jcrels}
 J(c,z) := J_c(z) := \dd c(z) = \dd p^{-1}(z) = \dd p(c(z))^{-1} 
= \frac{1}{2i} P(z + ie) 
= \frac{2}{i} P\Big(\frac{z + ie}{2}\Big),
\end{equation}
up to a central factor $c_0 \in Z(\tilde H_\C)$, 
\[ \tilde J_c(z) := c_0 \tilde P(z + ie) \] 
is a lift of $J_c$. 

For $g \in \tilde G(\cD)$ and $\tilde g := \tilde\alpha_c(g)\in \tilde G(\cT)$, we have 
$\tilde g.(c(z)) = c(g.z)$, and thus 
\[ J_{\tilde g}(c(z)) J_c(z) = J_c(g.z) J_g(z) \quad \mbox{ for } \quad 
z \in \cD.\] 
For any holomorphic lift $\tilde J_c$ of $J_c$, we thus have the relation 
\[ \tilde J_{\tilde g}(c(z)) \tilde J_c(z) \tilde J_g(z)^{-1} 
\tilde J_c(g.z)^{-1} \in \ker(q_{H_\C}).\] 
For $g = e$, this element is the identity of $\tilde H_\C$, and 
since $\ker(q_{H_\C})$ is discrete,  
\[ \tilde J_{\tilde g}(c(z)) \tilde J_c(z) \tilde J_g(z)^{-1} 
\tilde J_c(g.z)^{-1} = e \quad \mbox{ for } \quad 
g \in \tilde G(\cD), z \in \cD.\] 
This shows that 
\begin{equation}
  \label{eq:lift-intertwice}
\tilde J_{\tilde g}(c(z)) \tilde J_c(z) = 
\tilde J_c(g.z)\tilde J_g(z) \quad \mbox{ for } \quad 
g \in \tilde G(\cD), z \in \cD.
\end{equation}

From \eqref{eq:2kerneltrafo} we first derive the relation 
\begin{equation}
  \label{eq:2liftrel}
  \tilde Q^\cD(z, w) 
= \tilde P\Big(\frac{c(z)+ie}{2}\Big)^{-1}  \tilde Q(c(z), c(w)) 
\Big(\tilde P\Big(\frac{c(w)+ie}{2}\Big)^{-1}\Big)^*. 
\end{equation}
Composing with the $*$-representation $\rho$ 
and putting $P_\rho := \rho \circ \tilde P$, 
now leads to\begin{footnote}
{There is a certain freedom in normalizing the kernel on the tube domain. 
We have chosen a normalization for which $Q_\rho(ie,ie) = \id_\cK$, and 
this is why we use the $P_\rho$-factors to transform the 
kernels and not the factors $J_\rho(c,c^{-1}(z))$ obtained from 
the differential of the Cayley transform $c \: \cD \to \cT$. The source 
of the trouble is the fact that $dc(0) = dp(ie)^{-1} = 2i \1$ is not unitary.
}\end{footnote}

\begin{equation}
  \label{eq:posdefrel}
  Q_\rho^\cD(z, w) 
= P_\rho\Big(\frac{c(z)+ie}{2}\Big)^{-1}   Q_\rho(c(z), c(w)) 
P_\rho\Big(\frac{c(w)+ie}{2}\Big)^{-*}
\end{equation}
This relation implies in particular that the kernel 
$\cQ_\rho^\cD$ on $\cD \times \cD$ is positive definite if and 
only if the kernel $\cQ_\rho$ on $\cT \times \cT$ 
is positive definite, i.e., if the representation 
$\rho$ is $\cT$-positive 
(cf.~\cite[Thm.~I.1.4]{Ne99}). From now one we assume that this is the case. 
Let $\cH^\cD_\rho \subeq \Hol(\cD,\cK)$ and 
$\cH_\rho \subeq \Hol(\cT,\cK)$ denote the reproducing kernel 
Hilbert spaces with kernel $Q_\rho^\cD$ and 
$Q_\rho$, respectively.

\begin{lem} \mlabel{lem:gamma-intertwine} 
The map 
\[ \Gamma \: \Hol(\cT,\cK) \to \Hol(\cD,\cK), \qquad 
\Gamma(f)(z) := P_\rho\Big(\frac{z + i e}{2}\Big)^{-1} f(c(z)), \]
intertwines the representation of the groups 
$G(\cT)$ and $G(\cD)$ in the sense that 
\[  \Gamma \circ U_\rho(\tilde\alpha_c(g)) =   
U_\rho^\cD(g) \circ \Gamma \quad \mbox{ for } \quad g \in \tilde G(\cD).\] 
If the kernel $Q_\rho$ is positive definite, then it restricts to a unitary operator 
$\Gamma \: \cH_\rho \to \cH_\rho^\cD$. 
\end{lem}

\begin{prf} First we show that the slightly modified map 
$\Gamma_0(f)(z) := \tilde J_\rho(c,z)^{-1} f(c(z))$ 
intertwines the two representation. 
For $\tilde g:= \tilde\alpha_c(g)  \in \tilde G(\cT)$ and 
$f \in \Hol(\cD,\cK)$ we have 
  \begin{align*}
\Gamma_0(U_\rho(\tilde g)f)(z) 
&= \tilde J_\rho(c,z)^{-1} \tilde J_\rho(\tilde g^{-1},c(z))^{-1} f(\tilde g^{-1}.c(z)) \\ 
&= \big(\tilde J_\rho(\tilde g^{-1},c(z))\tilde J_\rho(c,z)\big)^{-1} f(c(g^{-1}.z)) \\ 
& {\buildrel {\eqref{eq:lift-intertwice}} \over =}\ \ 
 \big(\tilde J_\rho(c,g^{-1}.z)\tilde J_\rho(g^{-1},z)\big)^{-1} f(c(g^{-1}.z)) \\ 
&= \tilde J_\rho(g^{-1},z)^{-1} \tilde J_\rho(c,g^{-1}.z)^{-1} f(c(g^{-1}.z))\\ 
&= \big(U_\rho(g) \Gamma_0(f)\big)(z).
  \end{align*}
Here we have used the relation 
\[   \tilde J_\rho(\tilde g^{-1},c(z))\tilde J_\rho(c,z) 
= \tilde J_\rho(c,g^{-1}.z)\tilde J_\rho(g^{-1},z)\]
which follows from \eqref{eq:lift-intertwice}. 
Next we observe that, by \eqref{eq:jcrels}, 
there exists a central element $z \in \tilde H_\C$ such that 
$\Gamma(f) = \rho(z) \Gamma_0(f)$. As composition with $\rho(z)$ 
commutes with the representation of $\tilde G(\cD)$, this 
implies the intertwining property of~$\Gamma$.

To see that $\Gamma$ restrict to a 
unitary operator $\cH_\rho \to \cH^\cD_\rho$, we show that 
it maps $\cH_\rho$ isometrically onto the subspace with reproducing
kernel $Q^\cD_\rho$. Writing, for simplicity, $\Gamma(f)(z) = \beta(z) f(c(z))$ 
with $\beta(z) := P_\rho\big(\frac{z + i e}{2}\big)^{-1}$, 
the injective map $\Gamma$ maps $\cH_\rho$ unitarily onto a reproducing 
kernel Hilbert space $\im(\Gamma) \subeq \Hol(\cD,\cK)$ on which the point evaluations are given by 
\[ \ev_z = \beta(z) \circ \ev_{c(z)} \circ \Gamma^{-1} 
= \beta(z) Q_{\rho, c(z)} \Gamma^*.\]
Therefore the reproducing kernel of $\im(\Gamma)$ is given by 
\begin{align*}
 \ev_z \ev_w^* 
&=  \beta(z) Q_{\rho, c(z)} \Gamma^* \Gamma Q_{\rho, c(w)}^* \beta(w)^*
=  \beta(z) Q_{\rho, c(z)} Q_{\rho, c(w)}^* \beta(w)^*\\
&=  \beta(z) Q_\rho(c(z),c(w)) \beta(w)^* 
\ \ {\buildrel \eqref{eq:posdefrel} \over = } \ \  Q_\rho^\cD(z,w).
\end{align*}
This implies that $\Gamma(\cH_\rho) = \cH_\rho^\cD$ and that 
$\Gamma\res_{\cH_\rho}$ is unitary (\cite[Lemma~I.1.5]{Ne99}). 
\end{prf}

\subsection{Antiunitary extensions} 
\mlabel{subsec:4.3} 

In this subsection we discuss antiunitary extensions of the unitary 
representations of $\tilde G(\cD)$ and $\tilde G(\cT)$ on 
$\cH_\rho^\cD$ and $\cH_\rho$ respectively. We refer to \cite{NO17} for 
more background on antiunitary representations. 
We are mainly interested in antiunitary  extensions 
to the simply connected group $\tilde G_\tau \cong \tilde G(\cT)_\tau$. 
The involution 
$\tau$ acts on $\cT$ and $\cD$ by 
\[ \tau_\cT(z) = - \oline z \quad \mbox{ and }\quad \tau_\cD(z) =  \oline z.\] 
From 
\begin{equation}
  \label{eq:cayley-tau}
\tau_\cT(c(z)) = i(e +\oline z)(e - \oline z)^{-1}   
=   c(\oline z)=   c(\tau_\cD(z))
\end{equation}
we derive that the Cayley transform intertwines 
$\tau_\cT$ with $\tau_\cD$. 
We define 
\[ \tau_G(g) := \tau_\cD \circ g \circ \tau_\cD 
\quad \mbox{ for } \quad g \in G(\cD),
 \quad \mbox{ and } \quad 
\tau_G(g) := \tau_\cT \circ g \circ \tau_\cT \quad \mbox{ for } \quad 
g \in G(\cT).\] 
Then \eqref{eq:cayley-tau} implies that 
\begin{equation}
  \label{eq:cayley-g-inter}
c \tau_G(g) c^{-1} = \tau_G(cgc^{-1}) \quad \mbox{ for } \quad  
g \in \cG(\cD).
\end{equation}

\begin{defn} \mlabel{def:4.4}
Let $\tilde G(\cD)$ denote the universal covering 
of the group $G(\cD) = \Aut(\cD)_0$ and 
$\tau_G$ denote the involution of $\tilde G(\cD)$ induced 
by conjugating with $\tau_\cD$. 
An antiunitary representation $(U, \cH)$ of 
\[ \tilde G(\cD)_{\tau} := \tilde G(\cD) \rtimes \{\1,\tau_G\} \] 
is called a {\it positive energy representation} if 
\[ -i\dd U(\bd)  \geq 0 \quad \mbox{ holds for } \quad \bd := -ih. \] 
\end{defn}

Positive energy representations $(U,\cH)$ of 
$\tilde G(\cD)_{\tau}$  are direct sums 
of representations which are holomorphically induced from 
unitary representations of $\tilde K(\cD)$ 
(see \cite{Ne13} for details and definitions), i.e., 
realized in reproducing kernel spaces 
$\cH_\rho^\cD \subeq \Hol(\cD,\cK)$ 
for norm-continuous unitary representations $(\rho, \cK)$ of $\tilde K(\cD)$. 

For any such representation, the unitary one-parameter group 
$U_t := U(\exp t \bd)$ acts on $f \in \cH_\rho^\cD$ by 
\[ (U_t f)(z) 
= \rho(\exp t\bd) f(\exp(-t\bd).z) 
= \rho(\exp t\bd) f(e^{it}z),\] 
so that 
\[ - i \partial U(\bd) f 
= -i\partial \rho(\bd) f + \dd f(z)z,\] 
where $(Ef)(z) := \dd f(z)z$ is the Euler operator on holomorphic functions. 
Therefore 
\begin{equation}
  \label{eq:specinf}
\Spec(-i\partial U(\bd)) \subeq 
\Spec(-i\partial \rho(\bd)) + \N_0.
\end{equation}
As $\rho$ is norm-continuous, the operator $\partial \rho(\bd)$ is bounded, 
and thus $-i\partial U(\bd)$ is bounded from below. 
If, in addition, $(\rho,\cK)$ is irreducible, then 
$-i \dd \rho(\bd) = \zeta \1$ for some $\zeta \in \R$ and 
\[ - i \partial U(\bd) f 
= \zeta f + \dd f(z)z\] 
shows that $\cK = \ker(i\partial U(\bd) + \zeta \1)$ is the minimal 
eigenspace of $-i\partial U(\bd)$. 
Here we use that, if $f \: \cD \to \cK$ is holomorphic and 
$ f= \sum_{n = 0}^\infty f_n$ is its Taylor expansion, then 
\[ \dd f(z) z = \sum_{n = 0}^\infty n f_n(z).\]  

We recall that the positive definiteness of the kernel 
$Q_\rho^\cD$ implies that, for the 
corresponding unitary representation $U_\rho^\cD$, the operator 
$-i \partial U_\rho^\cD(\bd)$ is bounded from below. As 
$\bd$ belongs to a minimal invariant cone in $\g$, it even follows that 
$-i \partial U_\rho^\cD(\bd) \geq 0$ and hence that 
 $-i\partial \rho(\bd) \geq 0$ (see \cite[Thms.~X.4.1, XIV.1.3]{Ne99}). 
Actually the positive cone $C_\rho \subeq \fk(\cD)$ contains 
the intersection $C_{U^\cD_\rho} \cap \fk(\cD)$, which  
has interior points.

\begin{prop} \mlabel{prop:c.5}
Suppose that the representation $(\rho, \cK)$ of $\tilde H_\C$ is irreducible.  
For the representations  
$(U_\rho^\cD, \cH^\cD_\rho)$ of $\tilde G(\cD)$ and 
$(U_\rho, \cH_\rho)$ of $\tilde G(\cT)$,  the following are equivalent: 
\begin{itemize}
\item[\rm(a)] $(U_\rho^\cD, \cH^\cD_\rho)$ 
extends to an antiunitary representation of 
$\tilde G(\cD)_\tau$. 
\item[\rm(b)] $(\rho,\cK)$ extends to an 
antilinear representation of $\tilde H_{\C,\tau}$, 
i.e., there exists a conjugation $J_\cK$ on $\cK$ satisfying 
$J_\cK \rho(g) J_\cK = \rho(\tau_{\tilde H_\C}(g))$ for 
$g \in \tilde H_\C$.
\item[\rm(c)] $(U_\rho, \cH_\rho)$ 
extends to an antiunitary representation of 
$\tilde G(\cT)_\tau$.
\end{itemize}
\end{prop}

\begin{prf} (a) $\Rarrow$ (b): Suppose that $(U_\rho^\cD, \cH^\cD_\rho)$ 
extends to an antiunitary representation of 
$\tilde G(\cD)_\tau$ by $J := U_\rho^\cD(\tau_G)$. Then 
\[ J (-i\partial U_\rho^\cD(\bd)) J 
= i\partial U_\rho^\cD( \tau(\bd)) 
= i\partial U_\rho^\cD(-\bd) 
= -i\partial U_\rho^\cD(\bd) \] 
implies that $J$ commutes with $i\dd U_\rho^\cD(\bd)$, hence preserves its eigenspaces 
and in particular the minimal eigenspace $\cK$ of constant functions. 
Therefore  the restriction $J_\cK := J\res_{\cK}$ defines an 
antiunitary extension of the unitary representation 
$(\rho,\cK)$ of $\tilde K(\cD)$ to the group 
$\tilde K(\cD)_\tau  = \tilde K(\cD) \rtimes \{\1,\tau\}$. 
As $\tilde K(\cD)$ and $\tilde H_\C$ are simply connected, 
and  the homomorphism $\tilde J \: \tilde K(\cD) \to \tilde H_\C$ induces  
on the Lie algebra level an inclusion $\fk(\cD) \to \fh_\C$ 
of a real form, $\tilde J$ defines the universal complexification 
of $\tilde K(\cD)$. Hence 
the finite dimensional representation $\rho$ extends to a holomorphic representation 
of $\tilde H_{\C}$, so that we obtain a representation 
of $\tilde H_{\C,\tau}$. 

\nin (b) $\Rarrow$ (c): Let $J_\cK$ be a conjugation on $\cK$ extending 
the representation $\rho$ to $\tilde H_{\C,\tau}$. We define 
\[(Jf)(z) := J_\cK f(\tau(z))= J_\cK f(-\oline z).\] 
It is clear that $J$ is an antilinear involution on 
$\Hol(\cT,\cK)$. To see that it is antiunitary on $\cH_\rho$, 
we have to verify that 
\[ Q_\rho(-\oline z, - \oline w) 
= J_\cK Q_\rho(w,z) J_\cK \quad \mbox{ for } \quad z,w \in \cD.\] 
The left hand side is given by 
\[ Q_\rho(-\oline z, - \oline w) 
= \rho(\tilde Q(-\oline z, -\oline w)),\] 
and the right hand side by  
\[ J_\cK Q_\rho(w,z) J_\cK 
= \rho\big(\tau \tilde Q(w,z) \tau)
= \rho\big(\tau_{\tilde H_\C}(\tilde Q(w,z))).\] 
By the uniqueness of lifts, it therefore suffices to verify that 
\begin{equation}
  \label{eq:qrela}
\tau_{H_\C}(Q(w,z)) 
= Q(\tau(z), \tau(w)) \quad \mbox{ for } \quad z,w \in \cD.
\end{equation}

We start with the observation that 
\begin{equation}
  \label{eq:pconj1}
P(-\oline z) 
= P(\oline z) = \oline{P(z)} = \tau_{H_\C}(P(z)) \quad \mbox{ for }\quad 
z \in E_\C^\times.
\end{equation}
We further have 
\begin{equation}
  \label{eq:pconj2}
Q(\tau(z), \tau(w)) = Q(-\oline z,-\oline w) 
= P\Big(\frac{-\oline z+ w}{2i}\Big) = Q(w,z), 
\end{equation}
which leads to 
\begin{align}
  \label{eq:pconj3}
Q(\tau(z), \tau(w)) 
&= Q(w,z) 
= P\Big(\frac{w-\oline z}{2i}\Big) 
= \tau_{H_\C}\Big(P\Big(\frac{-\oline w+z}{-2i}\Big)\Big)\notag \\
&= \tau_{H_\C}\Big(P\Big(\frac{z-\oline w}{2i}\Big)\Big)
= \tau_{H_\C}(Q(z,w)).
\end{align}
This proves \eqref{eq:qrela}. 
Lifting \eqref{eq:pconj2} to $\tilde H_\C$ and composing with 
the extension of $\rho$ to $\tilde H_{\C,\tau}$ by 
$\rho(\tau) := J_\cK$, we get 
\begin{equation}
  \label{eq:pconj4}
Q_\rho(\tau(z), \tau(w)) 
= J_\cK Q_\rho(z,w) J_\cK.
\end{equation}
This implies that $(Jf)(z) = J_\cK f(-\oline z)$ defines an 
antiunitary involution on $\cH_\rho \subeq \Hol(\cT,\cK)$.  

Next we show that $U_\rho(\tau) := J$ defines an antiunitary extension 
of $U_\rho$ to $\tilde G(\cT)_\tau$. As 
$\tau(g) = \tau g \tau$ holds in $\Diff(\cT)$ for any $g \in G(\cT)$, 
we obtain for $g \in \tilde G(\cT)$ the relation 
\[ (J U_\rho(g)f)(z) 
= J_\cK J_\rho(g,g^{-1}.\tau(z)) f(g^{-1}.\tau(z)) 
\ {\buildrel{!}\over =}\ 
J_\rho(\tau(g),\tau(g)^{-1}.z) J_\cK f(\tau(\tau(g)^{-1}.z)) \] 
follows from 
\begin{align*}
 \tau_{H_\C}(J(g,g^{-1}.\tau(z))) 
&= \tau_{H_\C}(J(g^{-1},\tau(z)))^{-1} 
= \tau_{H_\C}(\dd(g^{-1})(-\oline z))^{-1} \\
&= \dd(\tau(g)^{-1})(z)^{-1} 
= J(\tau(g), \tau(g)^{-1}.z).  
\end{align*}

\nin (c) $\Rarrow$ (a): Since we have a 
unitary intertwining operator $\Gamma \: \cH_\rho \to \cH_\rho^\cD$ 
(Lemma~\ref{lem:gamma-intertwine}), 
it suffices to show that $\Gamma$ is also compatible with the 
involutions $\tau_G$ on the groups $\tilde G(\cD)$ and $\tilde G(\cT)$ in 
the sense that 
\[\Gamma \circ U_\rho(\tau_G(g)) = U^\cD_\rho(\tau_G(g)) \circ \Gamma 
\quad \mbox{ for } \quad g \in \tilde G(\cT).\] 
As 
\[  \Gamma \circ U_\rho(\tilde\alpha_c(g)))
 =   U_\rho^\cD(g) \circ \Gamma  \] 
by Lemma~\ref{lem:gamma-intertwine}, the assertion 
follows from \eqref{eq:cayley-g-inter}, which, in view of the uniqueness
of lifts, implies that $\tilde\alpha_c(\tau_G(g)) 
= \tau_G(\tilde\alpha_c(g))$ for 
$g \in \tilde G(\cD)$. 
\end{prf}

\begin{lem} The homomorphism 
\[ \tilde J \: \tilde G^{\tau_G} \to \tilde H_\C \] 
is injective and defines the universal complexification of $\tilde G^{\tau_G}$. 
In particular, it defines an isomorphism $\tilde G^{\tau_G} \to H^\sharp 
= \la \exp_{\tilde H_\C} \fh \ra$. 
\end{lem}

\begin{prf} Since the image of $\tilde J$ is the connected Lie subgroup 
with Lie algebra $\fh$, it suffices to show that it is injective. 

\nin (a) Recall that $\tilde J^\cD \: \tilde K(\cD) \to \tilde H_\C$ is the 
universal complexification of the simply connected group 
$\tilde K(\cD)$. We know from the Gelfand--Raikov Theorem 
(\cite[Thm.~XI.5.2]{Ne99})  that the contractive holomorphic representations 
of $S_C$ separate the points of $S_C$. 
This implies that the $\cT$-positive representations of $\tilde H_\C$ 
separate the points of $\tilde H_\C$ because composition with 
$\tilde J^\cD$ separates the points of $\tilde K(\cD)$. 

\nin (b) Now let $g \in \tilde G^{\tau_G}$ with $\tilde J(g) = e$. 
As the action of $\tilde G^{\tau_G}$ on $E_\C$ factors through the action 
of $\tilde H_\C$, defined by the covering map into $H_\C$, 
$g$ acts trivially on $\cT$. As $\tilde J(g) = e$ also acts trivially 
on $\cK$ for every $\cT$-positive representation $(\rho,\cK)$, 
it follows that $U_\rho(g) = \1$. From (a) we now derive that $g = e$. 
\end{prf}

\begin{rem}  As any $\cT$-positive norm-continuous representation 
$(\rho,\cK)$ of $\tilde H_\tau$ defines a holomorphic representation 
of its universal complexification, we have $\ker \rho \supeq \ker(\eta_{\tilde H})$, 
i.e., $\rho$ factors through a representation of the subgroup 
$H^\sharp \subeq \tilde H_\C$. In the preceding lemma we have seen that 
this subgroup can be identified with $\tilde G^{\tau_G}$ via the isomorphism~$\tilde J$.
\end{rem}

\begin{ex} For the Jordan algebra 
$E = \Sym_n(\R)$, $\g =\sp_{2n}(\R)$ and 
$\fh = \gl_n(\R)$, we have 
\[ \tilde H_\C\cong \tilde \GL_n(\C) \cong \C \times \SL_n(\C).\] 
The real subgroup corresponding to $\fh$ is the non-simply connected group
$H^\sharp \cong \R \times \SL_n(\R)$. The natural map 
$H^\sharp \to \Sp_{2n}(\R)$ lifts to a homomorphism 
$H^\sharp \to \tilde\Sp_{2n}(\R)$ because the inclusion 
$\SO_n(\R) \into \U_n(\C)\subeq \Sp_{2n}(\R)$ induces the trivial homomorphism 
$\pi_1(\SO_n(\R)) \to \pi_1(\Sp_{2n}(\R))$. 
\end{ex}

\begin{rem} \mlabel{rem:4.9}
(a) In the proof of Proposition~\ref{prop:c.5}, the implications 
(b) $\Rarrow$ (c) $\Rarrow$ (a) hold for any $\cT$-positive 
holomorphic $*$-representation 
$(\rho,\cK)$ of $\tilde H_\C$. Only the implication (a) $\Rarrow$ (b) requires 
additional assumptions; such as irreducibility.

\nin (b) If the conditions of Proposition~\ref{prop:c.5} 
are satisfied, then 
the unitary intertwining operator $\Gamma \: \cH_\rho \to \cH_\rho^\cD$ 
intertwines the conjugation $J$ on $\cH_\rho \subeq \Hol(\cT,\cK)$ defined 
by $(Jf)(z) := J_\cK f(-\oline z)$ with the conjugation $J^\cD$ 
on $\cH_\rho^\cD$ defined by 
$(J^\cD f)(z) := J_\cK f(-\oline z)$. 
In fact, we have 
\begin{align*}
(J^\cD \Gamma(f))(z) 
&= J_\cK P_\rho\Big(\frac{-\oline z + ie}{2}\Big) f(c(-\oline z)) 
= P_\rho\Big(\frac{z + ie}{2}\Big) J_\cK f(- \oline{c(z)}) 
= \Gamma(J f)(z).
\end{align*}

\nin (c) For any unitary representation 
$(U,\cH)$ of $\tilde G$, the direct sum 
$U \oplus (U^*\circ \tau_G)$  on $\cH \oplus \cH^*$ 
extends to an antiunitary representation of $\tilde G_\tau$, 
so that the situations of Proposition~\ref{prop:c.5} can always 
be achieved by this doubling process 
(cf.\ \cite[Lemma~2.10]{NO17}). 
The same argument applies to a holomorphic $*$-representation 
$(\rho, \cK)$ of the symmetric Lie group 
$(\tilde H_\C, \oline\theta)$, for which the representation 
$\rho\oplus (\rho^*\circ \oline\theta)$ on $\cK \oplus \cK^*$ always 
extends to $\tilde H_{\C,\tau}$. 

\nin (d) Uniqueness of extensions: From \cite[Thm.~2.11]{NO17} we 
know that any two extensions of a unitary representation 
$(U,\cH)$ of $\tilde G$ to an antiunitary representation of 
$\tilde G_\tau$ are unitarily equivalent, regardless of whether the 
representation $U$ is irreducible or not. 

\nin (e) Irreducible holomorphic representations 
$(\rho, \cK)$ of $(\tilde H_\C, \oline\theta)$ are finite dimensional 
and every finite dimensional irreducible  complex representation 
$(\beta,\cK)$ of the complex Lie algebra $\fh_\C$ on a complex vector space $\cK$ 
integrates to a holomorphic representation 
$(\rho,\cK)$ of $\tilde H_\C$. Moreover, $\cK$ carries a Hilbert space 
structure for which $\rho$ is a $*$-representation if and only if, 
for each $x \in \fz(\fh)$, the operator 
$\beta(x)$ is diagonalizable with real eigenvalues. 
Here we use that $\theta(x) = x$ for $x \in \fz(\fh)$.  
Then the subgroup $\tilde H_\C^{\oline\theta}$ is represented 
by a compact group and the scalar product can be obtained by averaging any 
given scalar product on $\cK$. 

\nin (f) A representation of $\tilde H_\C$ extends to 
$\tilde H_{\C,\tau}$ if and only if there exists a conjugation $J_\cK$ 
on $\cK$ commuting with $\beta(\fh)$. This is equivalent to the representation 
$(\beta, \cK)$ of the Lie algebra $\fh$ being {\it real} 
in the sense that $\cK_\R := \Fix(J_\cK)$ is a real $\fh$-invariant subspace 
and the representation on $\cK$ arising by complex linear extension. 
Conversely, every real $*$-representation 
$(\rho, \cK_\R)$ of $\fh$, resp., the group $H^\sharp \subeq \tilde H_\C$, 
extends to a complex $*$-representation 
$(\rho, (\cK_\R)_\C)$ of $H^\sharp_\tau$ by $\rho(\tau)(x + iy) 
:= x - iy$ and this representation extends to a holomorphic representation 
of the universal complexification $\tilde H_\C$. 
Therefore the irreducible extendable representations of $\tilde H_{\C,\tau}$ 
correspond to the irreducible real $*$-representations of $\fh$. 

\nin (g) The classification of real irreducible representations can easily 
be reduced to the classification of the complex ones by the following observations 
(\cite[Ch.~8]{On20}, \cite{Iw59}). If $(\beta,V)$ is a real simple 
$\fh$-module, then we say it is of real, complex of quaternionic 
type if $\End_\fh(V) \cong \R,\C$ or $\H$. 
If $V$ is of real type, then $V_\C \cong \oline{V_\C}$ is simple. 
If $V$ is of complex type, then $V$ 
itself is complex simple with $\oline V \not\cong V$ 
(there is no antilinear automorphism), and if 
$V$ is of quaternionic type, then $V \cong \oline V$ is simple complex, 
and there exists an antilinear endomorphism $J$ with $J^2 = - \1$. 
We thus obtain for each simple real $\fh$-module a simple complex 
one, and the three types are reflected in the (non-)existence of 
antilinear endomorphisms $J$ with $J^2 = \pm \1$.

Thus every simple complex module $V$ with $\oline V \not\cong V$ 
defines by restriction of scalars a real simple module 
$V^\R$ of complex type, and $\oline V^\R \cong V^\R$. 
If $\oline V \cong V$, then  $V$ there exists an 
antilinear endomorphism $J$ with $J^2 = \pm\1$. 
If $J^2 = \1$, then $V^J = \Fix(J)$ is a real simple module of real type, 
and if $J^2 =- \1$, then $V^\R$ is a real simple module of 
quaternionic type. 

The three types of complex simple modules can be identified in 
terms of their highest weights. 
Let $\fb \subeq \fh^{\theta}$ be maximal abelian 
and $\fa \subeq \fh^{-\theta} \cap \fz_\fh(\fb)$ be maximal abelian, 
so that $\fc := \fa \oplus \fb$ is a Cartan subalgebra 
of $\fh$ which is ``maximally compact''. In particular $\fb$ contains 
regular elements of $\fh$. We parametrize irreducible complex representations 
$(\beta_\lambda, V_\lambda)$ of $\fh$ 
by their highest weights $\lambda \: \fc \to  \C$ which are real on $\fa$,  
purely imaginary on $\fb$ and dominant with respect to  a Weyl chamber 
intersecting $i\fb$, so that the Weyl chamber can be chosen invariant 
under $\theta$. 
As the weights of the complex conjugate representation are the complex 
conjugates of the weights and $-\lambda \circ \theta = \oline\lambda$ on $\fc$, 
the highest weight of the complex conjugate representation is
 $\oline{V_\lambda}$ is $-w.(\lambda \circ \theta) = w.\oline\lambda$, where 
$w \in \cW$ is the longest element in the Weyl group. 
Therefore  $\oline{V_\lambda}\cong V_\lambda$ is equivalent to 
$\lambda = w.\oline\lambda$. 
If this is the case, then 
there exists an antilinear endomorphism $J$ with $J^2 = (-1)^{\eps_\lambda} \1$. 
The number $\eps_\lambda$ can be determined from $\lambda$ and the 
root system; see \cite[Thm.~8.3]{On20} and the subsequent discussion 
for details. 
\end{rem}

\subsection{$L^2$-realization} 
\mlabel{subsec:4.4} 

The representations $(\rho, \tilde H_\C)$ that are of interest for 
us are the $\cT$-positive ones. This condition 
is equivalent to the positive definiteness 
of the function 
$P_\rho \: \cT \to B(\cK)$ on the $*$-semigroup $\cT$ with involution 
$z^* := - \oline z$, which in turn is equivalent to the existence 
of a unique $\Herm^+(\cK)$-valued measure $\mu$ on $E_+^\star$ with 
\[ P_\rho = \cL(\mu), \qquad 
\cL(\mu)(z) = \int_{E^*} e^{-\lambda(z)}\, d\mu(\lambda)\] 
(cf.\ \cite[Thm.~V.12]{HN01})

\begin{lem} The measure $\mu$ has the following transformation properties:
  \begin{itemize}
  \item[\rm(a)] $J_\cK \mu J_\cK = \mu$, i.e., $\mu$ takes 
values in the subspace $\Sym^+(\cK^{J_\cK})$ of positive symmetric 
operators on real Hilbert space $\cK^{J_\cK}$. 
  \item[\rm(b)] $(g^{-1})_*\mu = \rho(g) \mu \rho(g)^*$ for $g \in 
\tilde H$; i.e., for every compactly supported continuous function we have 
\[ \int_{E^*} \phi(g^{-1}.\lambda)\, d\mu(\lambda) 
 = \int_{E^*} \phi(\lambda)\, \rho(g)\, d\mu(\lambda) \, \rho(g)^*.\]
  \end{itemize}
\end{lem}

\begin{prf} (a)  The relation 
$\tau_{H_\C}(P(z)) = \oline{P(z)} = P(\oline z) = P(-\oline z)$ 
from \eqref{eq:pconj1} implies that 
\[\cL(\mu)(\oline z) =  P_\rho(\oline z) 
= J_\cK P_\rho(z) J_\cK = J_\cK \cL(\mu)(z)  J_\cK 
= \cL(J_\cK \mu J_\cK)(\oline z)
 \quad \mbox{ for } \quad z \in E_+ + i E = -i \cT.\]  
As any tempered measure on $E_+^\star$ is uniquely determined 
by its Laplace transform, the assertion follows.

\nin (b) Starting from the relation 
\[ P_\rho(g.z) = \rho(g) P_\rho(z) \rho(g)^* \quad \mbox{ for } \quad 
g \in \tilde H, z \in \cT,\] 
we obtain 
\[ \cL((g^{-1})_*\mu)(z) 
=  \cL(\mu)(g.z) 
= \rho(g) \cL(\mu)(z) \rho(g)^*
= \cL(\rho(g) \mu \rho(g)^*)(z).\] 
As in (a), this implies (b). 
\end{prf}

With a slight modification of the normalizations 
in \cite[\S 3.2]{NOO20} (a factor $2$ in the exponent), it follows that 
\begin{equation}
  \label{eq:Phi}
\Phi \: L^2(E^*,\mu,\cK) \to \Hol(\cT, \cK), \quad 
\Phi(f)(z) = \int_{E^*} e^{\frac{i\lambda(z)}{2}} d\mu(\lambda) f(\lambda) 
\end{equation}
maps $L^2(E^*,\mu,\cK)$ injectively 
onto the reproducing kernel Hilbert space $\cH_Q$ with $B(\cK)$-valued kernel 
\[ Q_\rho(z,w) = \cL(\mu)\Big(\frac{z- \oline w}{2i}\Big) 
= \int_{E^*} e^{i\lambda(z-\oline w)/2}\, d\mu(\lambda).\] 
Note that 
\begin{equation}
  \label{eq:eval1}
\la \xi, \Phi(f)(z) \ra  = \la e_{-i \oline z/2}\, \xi, \mu \cdot f \ra 
\quad \mbox{ for } \quad v \in \cK, z \in \cT.
\end{equation}

The map $\Phi$ is a unitary intertwining operator for the 
antiunitary representation of the group 
$E \rtimes \tilde H_\tau$ on $L^2(E^*,\mu,\cK)$, defined by 
\begin{equation}
  \label{eq:Urep}
 (U(x,g)f)(\lambda) := e^{-\frac{i\lambda(x)}{2}} \rho^\theta(g) f(g^{-1}.\lambda), 
\qquad U(0,\tau)f := J_\cK \circ f, 
\end{equation}
where $\rho^\theta :=\rho \circ \theta$ is the $\theta$-twist of $\rho$. 
Here we use that the representation of $E \rtimes \tilde H_\tau$ on 
$\cH_\rho \subeq \Hol(\cT,\cK)$ has the form 
\[  (U_\rho(x,g)f)(z) 
:= \rho(g) f(g^{-1}.(z - x)) 
\quad \mbox{ for } \quad g \in \tilde H, x \in E, z \in \cT,\] 
and 
\[ (U_\rho(\tau)f)(z) := J_\cK f(-\oline z).\] 

\begin{ex} The case where $\cK = \C$ is one-dimensional is particularly 
interesting and easily described. Then the representation $\rho \: \tilde H 
\to \GL_1(\R) \cong \R^\times$ is a real character. We now take a closer 
look at these representations for the case where $E$ is simple. 

Let $E$ be a simple euclidean Jordan algebra or rank $r$ 
whose Pierce subspaces are of dimension~$d$. For 
\begin{equation}
  \label{eq:walachset}
s \in \Big\{0, \cdots, (r-1)\frac{d}{2}\Big\} \cup 
\Big((r-1)\frac{d}{2}, \infty\Big), 
\end{equation} 
we consider the corresponding Riesz measure $\mu_s$ on 
$E_+^\star$ whose Fourier (Laplace) transform satisfies 
\[ \cL(\mu_s)(z)= \tilde\mu_s(iz) = \Delta_E(z)^{-s} \quad \mbox{ for } \quad 
z \in E_+ + i E = -i \cT,\] where 
$\Delta_E$ is the Jordan determinant (\cite{FK94}). 
Recall the relations 
\[ \Delta_E(g.x) = {\det}_E(g)^{r/n} \Delta_E(x)\quad \mbox{ and } \quad 
\Delta_E(P(y)x) =\Delta_E(y)^2 \Delta_E(x) \] 
(\cite[Prop.~III.4.2]{FK94}). 
For the corresponding one-dimensional representation $(\rho_s, \C)$ 
of $\tilde H$, the relations 
\[ \Delta_E(z)^{-s} = P_{\rho_s}(z) = \rho_s(\tilde P(z)) 
\quad \mbox{ and } \quad 
\tilde P(g.z) = g \tilde P(z) g^* \] 
now lead to $\rho_s(gg^*) = {\det}_E(g)^{-rs/n}$, which in turn shows that 
\begin{equation}
  \label{eq:det-s-rel}
 \rho(g)  = {\det}_E(g)^{-rs/2n}.
\end{equation}
It follows in particular that, as a representation of $\tilde H$, 
$\rho$ actually factors through the group $H$ itself. 
\end{ex}

\subsection{Distribution vectors} 
\mlabel{subsec:4.5} 

The main purpose of this section is to realize $C$-positive 
unitary representations of $\tilde G$ in such a way that 
we get a nice picture of a sufficiently large space of distribution 
vectors. This is finally done in this subsection, where we formulate the 
precise results on distribution vectors that we shall use in 
Section~\ref{sec:5} to construct local nets of standard subspaces. 
The distribution vectors are most easily identified in the $L^2$-realization. 

\begin{prop} \mlabel{prop:c.10} For $\xi \in \cK$, the corresponding constant 
function on $E^*$ defines a distribution vector $\eta_\xi$ of the 
unitary representation of $(E,+)$ on $L^2(E^*,\mu;\cK)$ by 
\[ \eta_\xi(f) = \int_{E^*} \la d\mu(\lambda) f, \xi \ra.\] 
These distribution vectors have the following properties: 
\begin{itemize}
\item[\rm(a)] The subspace $\{ \eta_\xi \: \xi \in \cK\}$ 
in $L^2(E^*,\mu;\cK)^{-\infty}$ 
is cyclic for the group $(E,+)$. 
\item[\rm(b)] $U^{-\infty}(0,g)\eta_\xi = \eta_{\rho^\theta(g)\xi}$ 
for $g \in \tilde H$ and $\xi \in \cK$. 
\item[\rm(c)]  $U^{-\infty}(0,\tau)\eta_\xi = \eta_{J_\cK \xi}$. 
\end{itemize}
\end{prop}

\begin{prf} Let $\|\cdot\|$ denote the euclidean norm on $E^*$. 
Since the measure $\mu$ is tempered 
(\cite[Thm.~V.12]{HN01}), there 
exists an $m \in \N$ such that, for every $\xi \in \cK$, the integral 
\[ \int_{E^*} \frac{1}{(1 + \|\lambda\|^2)^{2m}}\, 
\la \xi, \dd \mu(\lambda) \xi \ra \] 
is finite.  This implies that 
$\xi \in (1 + \|\lambda\|^2)^m L^2(E^*,\mu;\cK) 
\subeq L^2(E^*,\mu;\cK)^{-\infty}$, because 
the enveloping algebra $S(E)$ of the 
Lie algebra of the abelian group $(E,+)$ acts on $L^2(E^*,\mu;\cK)$ 
by multiplication with polynomial functions on~$E^*$.

\nin (a) We first observe that, for 
$\phi \in C^\infty_c(E)$, we have 
\[ \big(U(\phi)f\big)(\lambda) 
= \int_E \phi(x) e^{-i\lambda(x)/2} f(\lambda)\, d\mu_E(x) 
= \hat\phi\Big(\frac{\lambda}{2}\Big) f(\lambda), 
\quad \mbox{ where } \quad 
\hat\phi(\lambda) :=  \int_E \phi(x) e^{-i\lambda(x)}\, d\mu_E(x)\] 
is the Fourier transform of $\phi$. As 
$\cS(E^*) \cdot \cK \subeq \cS(E^*,\cK)$ 
is dense and $\cS(E^*,\cK)$ maps to a dense subspace of $L^2(E^*,\mu;\cK)$, 
the assertion follows. 

\nin (b) For $g \in \tilde H$, we have 
\begin{align*}
\eta_{\rho(g)^*\xi}(f) 
&= \la f, \rho(g)^*\xi \ra 
= \int_{E^*} \la d\mu(\lambda) f(\lambda), \rho(g)^*\xi \ra
= \int_{E^*} \la f(\lambda), d\mu(\lambda) \rho(g)^*\xi \ra\\
&= \int_{E^*} \la \rho(g^{-1})^* f(\lambda), 
\rho(g) d\mu(\lambda) \rho(g)^*\xi \ra\\
&= \int_{E^*} \la \rho(g^{-1})^* f(g^{-1}.\lambda), d\mu(\lambda) \xi \ra\\
&= \int_{E^*} \la \rho^\theta(g) f(g^{-1}.\lambda), d\mu(\lambda) \xi \ra
= \int_{E^*} \la (U(0,g)f)(\lambda), d\mu(\lambda) \xi \ra \\
&= \eta_\xi(U(0,g)f)= (U^{-\infty}(0,g^{-1})\eta_\xi)(f).
\end{align*}
This implies~(b).

\nin (c) This assertion follows from 
\begin{align*}
\eta_{J_\cK\xi}(f) 
&= \int_{E^*} \la d\mu(\lambda) f(\lambda), J_\cK\xi \ra
= \int_{E^*} \la f(\lambda), d\mu(\lambda) J_\cK \xi \ra\\
&= \int_{E^*} \la f(\lambda), J_\cK d\mu(\lambda) \xi \ra
= \int_{E^*} \la d\mu(\lambda) \xi, J_\cK f(\lambda)  \ra\\
&= \oline{\eta_\xi(U(0,\tau)f)}
= (U^{-\infty}(0,\tau)\eta_\xi)(f).\qedhere 
\end{align*}
\end{prf}

For later use, we collect some information on the boundary values of the 
holomorphic functions in $\cH_\rho \subeq \Hol(\cT,\cK)$. 

\begin{lem} \mlabel{lem:C.11} 
For $\phi \in C^\infty_c(E)$ and $F \in \cH_\rho$, we consider the 
holomorphic function 
\[F_\phi = U(\phi^*)F \: \cT \to \cK, \quad 
 F_\phi(z) := \int_E \oline{\phi(x)} F(z + x)\, d\mu_E(x), \] 
where $dx = d\mu_E(x)$ denotes a Haar measure on~$E$.
This function extends to a continuous function on 
$\oline{\cT} = E + i \oline{E_+}$ which is smooth on~$E$.
\end{lem}

\begin{prf} We write $F = \Phi(f)$ for some 
$f \in L^2(E^*,\mu;\cK)$ as in \eqref{eq:Phi}. Then  the integrated 
representation of 
the convolution algebra $C^\infty_c(E)$ on $L^2(E^*,\mu;\cK)$ is given 
by 
\[ \big(U(\phi)f\big)(\lambda) 
= \int_E \phi(x) e^{-i\lambda(x)/2} f(\lambda)\, d\mu_E(x) 
= \hat\phi(\lambda/2) f(\lambda).\] 
We thus obtain 
\begin{align*}
F_\phi(z) 
&= \int_E \phi^*(x) F(z-x)\, d\mu_E(x) 
 = (U_\rho(\phi^*)F))(z) 
 = (U_\rho(\phi^*)\Phi(f))(z) \\
& = (\Phi(U(\phi^*)f))(z) 
= \int_{E^*} e^{i\lambda(z)/2} d\mu(\lambda)\, \oline{\hat\phi(\lambda/2)} f(\lambda).
\end{align*}
Next we observe that the function 
$\oline{\hat\phi(\lambda/2)} f(\lambda)$ on $E^*$ 
is $\mu$-integrable because $\hat\phi$ 
is a Schwartz function, hence $L^2$. As 
\[ |e^{i\lambda(z)/2}|
= e^{-\Im \lambda(z)/2}
= e^{-\lambda(\Im z)/2} \leq 1 \quad \mbox{ for } \quad 
z \in \oline\cT,\] 
it follows from the Dominated Convergence Theorem that 
$F_\phi$ extends to a continuous function on~$\oline\cT$. 
Its restriction to $E$ is the Fourier transform 
of the measure 
$d\mu(\lambda) \oline{\hat\phi(\lambda/2)} f(\lambda)$, 
which remains finite when multiplied by any polynomial. Therefore its 
Fourier transform is smooth. 
\end{prf}

\begin{rem} (a) The preceding lemma implies in particular that 
$\bo(F)(\phi) := F_\phi(0)$ 
is defined for every $F \in \cH_\rho$ and $\phi \in C^\infty_c(E)$.
To see that this defines a tempered distribution on~$E$, we note 
that, for $F = \Phi(f)$, $f \in L^2(E^*,\mu;\cK)$, we have 
\[ F_\phi(0) 
= \int_{E^*} \oline{\hat\phi(\lambda/2)}\, d\mu(\lambda)  f(\lambda),\] 
and that the measure $\mu \cdot f$ on $E^*$ is tempered. 
Therefore the continuity of the Fourier transform 
$\cS(E) \to \cS(E^*)$ implies that 
$\bo(F)$ defines a $\cK$-valued tempered distribution on $E$. 
This leads to an injective linear map 
\begin{equation}
  \label{eq:boundval}
 \bo \: \cH_\rho \to \cS'(E,\cK), \quad 
\bo(F)(\phi) = F_\phi(0),  
\end{equation}
whose image is a Hilbert space of tempered distributions 
on which the scalar product is defined by the requirement that $\bo$ is isometric. 

\nin (b) To identify this Hilbert space 
as a reproducing kernel Hilbert space, 
we write 
\[ \ev_\phi \: \bo(\cH_\rho) \to \cK \] 
for evaluation in $\phi$, which is a linear map, but 
antilinear in $\phi$. 
For $F \in \cH_\rho$, we then have 
\begin{align*}
\la \bo^* \ev_\phi^*\xi, F \ra 
&= \la \xi, F_\phi(0) \ra 
= \int_{E^*} \oline{\hat\phi(\lambda/2)}\, 
\la \xi, d\mu(\lambda) \Phi^{-1}(F)(\lambda) \ra \\ 
&= \la \hat\phi(\cdot/2)\xi, \Phi^{-1}(F) \ra_{L^2} 
= \la \Phi({\hat\phi}(\cdot/2) \xi), F \ra_{\cH_\rho}.
\end{align*}
We conclude that 
\[ \bo^* \ev_\phi^*\xi 
= \Phi({\hat\phi}(\cdot/2) \xi) \in \cH_\rho.\] 
The corresponding reproducing kernel is therefore given by 
\[  \la \xi, \ev_\phi \ev_\psi^*\eta \ra 
= \la \bo^*\ev_\phi^* \xi, \bo^*\ev_\psi^*\eta \ra 
= \la {\hat\phi}(\cdot/2) \xi, {\hat\psi}(\cdot/2) \eta \ra
_{L^2} 
= \int_{E^*} \oline{\hat\phi(\lambda/2)} 
{\hat\psi}(\lambda/2) 
\la \xi, d\mu(\lambda) \eta \ra.\] 

Next we note that the holomorphic function 
\[ D \: \cT \to B(\cK), \quad 
D(z) = \int_{E^*} e^{i\lambda(z)/2}\ d\mu(\lambda)  \] 
has distributional boundary values $D \in \cS'(E,B(\cK))$, given by 
\[ D(\phi) 
= \int_{E^*} \tilde{\oline\phi}(\lambda/2)\, d\mu(\lambda) 
= \int_{E^*} \oline{\hat\phi(\lambda/2)}\, d\mu(\lambda)\]  
(\cite[Lemma 3.22]{NOO20}). 
This distribution satisfies 
\[ D(\psi^* * \phi) 
= \int_{E^*} \oline{\hat\phi(\lambda/2)}
{\hat\psi}(\lambda/2)\, d\mu(\lambda) 
= \ev_\phi \ev_\psi^*,\]  
so that it represents the reproducing kernel 
of the subspace $\bo(\cH_\rho) \subeq \cS'(E,\cK)$ 
(cf.\ \cite[Def.~7.1.5]{NO18}, which does not have the factor $\frac{1}{2}$). 
\end{rem}

\section{Nets of standard subspaces} 
\mlabel{sec:5} 

Let $\g$ be a direct sum of simple hermitian Lie algebras of tube type 
(see the table in the introduction for a list of these 
Lie algebras) 
and $(\rho, \cK)$ be a $\cT$-positive norm-continuous representation  
of $\tilde H_{\C,\tau}$ (cf.~Remark~\ref{rem:4.9}), 
so that we obtain an antiunitary representation 
$(U_\rho, \cH_\rho)$ of $\tilde G_\tau$ on the Hilbert subspace 
$\cH_\rho \subeq \Hol(\cT,\cK)$. 
In this section we show that, for the standard subspace 
$\sV \subeq \cH_\rho$ specified by the pair $(h,\tau_G)$, there exists a 
real subspace $\sE \subeq \cK$ contained in $\sV^{-\infty}$,  
to which we can apply Section~\ref{sec:3}. This leads to the identity 
$\sV = \sH_\sE(S_\sV^0)$ and to a net $\sV(\cO) := \sH_\sE(\cO)$ 
of cyclic subspaces associated to non-empty open subsets $\cO \subeq G$. 
As a consequence of the construction in Section~\ref{sec:4}, 
this covers all irreducible 
$C$-positive antiunitary representations $(U,\cH)$ of $\tilde G_\tau$ 
(Theorem~\ref{thm:5.1}). 

\subsection{Nets of standard subspaces on Lie groups}

We briefly recall our setting. In this section $\g$ denotes a semisimple Lie algebra. 
\begin{itemize}
\item We assume that $\g$ contains a closed pointed generating invariant 
convex cone $C \subeq \g$. 
\item We fix an {\it Euler element} $0 \not=h \in [\g,\g]$, 
i.e., $\ad h$ is diagonalizable with $\Spec(\ad h) \subeq  \{-1,0,1\}$. 
All these elements are conjugate under inner automorphisms 
(cf.~\cite[Ex.~6.1]{MN20}), so that it suffices to consider one 
such element $h$. Together with the preceding item, the 
existence of such an element is equivalent 
to all simple non-compact ideals of $\g$ being hermitian of tube type 
(see the proof of~\cite[Thm~3.12]{MN20} and \cite[Thm.~5.6]{O91}). 
\item We assume that $\g^0(h) = \ker(\ad h)$ contains no proper ideal, 
i.e., that $\g$ is generated as a Lie algebra by the eigenspaces $\g^{\pm 1}(h)$. 
This excludes compact ideals in $\g$. 
\item We consider on $\g$ the involution 
$\tau_h := e^{\pi i \ad h}$ fixing~$h$. 
\item Then the eigenspace $E := \g^1 = \g^1(h)$ 
is a unital euclidean Jordan algebra, where the unit element $e \in E$ 
can be chose such that the open cone $E_+$ of positive squares 
satisfies 
\[ \oline{E_+} = E \cap C = \g^1(h) \cap C =: C_+.\] 
\end{itemize}

As in Section~\ref{sec:4}, we consider the connected Lie group 
$G := \Aut(\g)_0$ and write 
$q_G \: \tilde G \to G$ for its universal covering group. 
Let $\tau_G \in \Aut(\tilde G)$ be the involutive automorphism 
integrating $\tau_h \in \Aut(\g)$. Then 
\[ \tilde G_\tau := \tilde G \rtimes \{\id_G, \tau\} \] 
is a graded Lie group with two connected components. 
We write 
\[ \fh = \g^0 = \ker(\ad h), \qquad 
H := \la \exp_G \fh \ra \subeq G,  \quad \mbox{ and } \quad 
H_\tau := H \times \{\id, \tau\} \subeq G_\tau.\] 
We also note that 
\[ \tilde G^{\tau_G} = \{ g\in \tilde G \: \tau_G(g)= g \} \] 
is connected because $\tilde G$ is simply connected 
(\cite[Thm.~IV.3.4]{Lo69}), so that 
$q_G$ restricts to a covering map 
\[ q_H \: \tilde G^{\tau_G} \to H.\]

\begin{thm} {\rm(Realization Theorem)} \mlabel{thm:5.1} 
For each irreducible
antiunitary representation $(U,\cH)$ of $\tilde G_\tau$ 
for which $C_U$ is pointed and generating, 
there exists an irreducible involutive finite dimensional 
representation $(\rho, \cK)$ of the product group 
\[ \tilde H_\tau := \tilde H \times \{\1,\tau\}\] 
with the following properties: 
\begin{itemize}
\item[\rm(a)] The holomorphic extension of $\rho$ to 
the universal complexification $\tilde H_\C$ is $\cT$-positive, i.e., 
the $B(\cK)$-valued kernel 
\[ Q_\rho(z,w) := P_{\rho}\Big(\frac{z - \oline w}{2i}\Big) \] 
on the tube domain $\cT = E + i E_+$ is positive definite. 
\item[\rm(b)] $(U,\cH)$ is equivalent 
to the representation of $\tilde G_\tau$ on the corresponding 
reproducing kernel Hilbert space 
$\cH_\rho  \subeq \Hol(\cT,\cK),$  given by 
\begin{equation}
  \label{eq:j1}
(U_\rho(g)f)(z) = 
J_{\rho}(g, g^{-1}.z) f(g^{-1}.z) \quad \mbox{ 
for } \quad g \in \tilde G, z \in \cT,
\end{equation} 
and 
\begin{equation}
  \label{eq:j2}
(U_\rho(\tau_G)f)(z) := J_\cK f(-\oline z),
\end{equation}
where  $J_\cK = \rho(\tau)$, 
$\tilde J \: \tilde G \times \cT \to \tilde H_\C$ is the lift 
of the cocycle $J \: G \times \cT \to H_\C, J(g,z) := \dd g(z)$, and 
$J_\rho := \rho \circ \tilde J$. 
\end{itemize}
\end{thm}

\begin{prf} Recall the isomorphism $\tilde\alpha_c \: 
\tilde G(\cD)_\tau \to \tilde G_\tau = \tilde G(\cT)_\tau$ from \eqref{eq:liftalphac}. 
Composing with $\tilde\alpha_c$, 
any irreducible antiunitary representation $(U,\cH)$ of $\tilde G_\tau$ for which 
$C_U$ is pointed and generating defines a representation 
$(U^c,\cH)$ of $\tilde G(\cD)_\tau$ with the same property. 
From \cite[Thm.~X.3.9]{Ne99} we know that this is a highest weight 
representation for a suitably chosen positive system, hence 
can be realized in the space $\Hol(\cD,\cK)$ of 
holomorphic functions on $\cD$, where 
$\cK$ carries an irreducible 
antiunitary representation $(\rho^c,\cK)$ of the simply 
connected group $\tilde K(\cD)_\tau$ 
(Proposition~\ref{prop:c.5}, \cite[Thm.~XII.2.6]{Ne99}). 
We conclude that the kernel $Q_\rho^\cD := \rho \circ \tilde Q^\cD$ 
is positive definite (\cite[Prop.~XII.2.1]{Ne99}), 
Now Lemma~\ref{lem:gamma-intertwine} transports all this 
structure from $\cD$ to $\cT$, so that 
we obtain the required realization 
in a Hilbert space $\cH_\rho \subeq \Hol(\cT,\cK)$, 
where $(\rho, \cK)$ is a finite-dimensional $*$-representation 
of $\tilde H_\tau$, and the representation is given by 
\eqref{eq:j1} and \eqref{eq:j2}. 
\end{prf}

We now turn to the standard subspace $\sV = \sV_{(h,\tau_G,U)}$ specified by the 
multiplicative one-parameter group of $G_\tau$, 
corresponding to the pair $(h,\tau_G)$ via 
\begin{equation}
J_\sV = U(\tau_G) \quad \mbox{ and } \quad 
\Delta_{\sV}^{-it/2\pi} = U(\exp th) \quad \mbox{ for } \quad t \in \R.
\end{equation}
In the following proposition, we shall see in particular 
that the distribution vectors 
$\Phi(\eta_\xi) \in \cH_\rho^{-\infty}$ are invariant under the subgroup 
$N^- = \exp(\g^{-1}(h)) \subeq \tilde G(\cT)$. 
As the action of this subgroup is not so easily  accessible 
in the $L^2$-picture, we study this problem in the holomorphic picture on $\cH_\rho$. 

\begin{prop} \mlabel{prop:c.12} 
Let $(\rho,\cK)$ be a holomorphic $\cT$-positive $*$-representation 
of $\tilde H_\C$ and let $\ev_0 \: \cH_\rho^\infty \to \cK, F \mapsto F(0)$ be the 
evaluation in $0 \in \oline{\cT}$. 
Then the following assertions hold: 
\begin{itemize}
\item[\rm(a)] 
For each $\xi \in \cK$, we obtain a distribution vector  
$\ev_0^\xi \: \cH_\rho^\infty \to \C, \ev_0^\xi(F) := \la F(0), \xi \ra$. 
\item[\rm(b)] $U_\rho(g) \ev_0^\xi = \ev_0^{\rho(\oline\theta(\tilde J(g)))\xi}$ 
for $g \in \tilde G^{\tau_G}\subeq \tilde G$ and 
$\tilde J(g) \in \tilde H_\C$. 
\item[\rm(c)] $\ev_0^\xi$ is invariant under the subgroup 
$N^- = \exp(\g^{-1}(h)) \subeq \tilde G = \tilde G(\cT)$. 
\item[\rm(d)] 
The real subspace 
\begin{equation}
  \label{eq:defE}
 \sE := \{ \ev_0^\xi \: e^{-\pi i \cdot \dd \rho(h)} \xi = J_\cK \xi\} 
\end{equation}
is contained in $\cH^{-\infty}_{\rho, {\rm ext}, J}$ 
and invariant under the subgroup $P^- := N^- \tilde G^{\tau_G}$. 
\end{itemize}
\end{prop}

\begin{prf} (a) Let $F\in \cH_\rho$ be a smooth vector. 
Then it is in particular smooth for the representation 
of $(E,+) \cong \exp(\g^1(h))$. 
By the Dixmier--Malliavin Theorem (\cite{DM78}), it is of the form 
$F = U_\rho(\phi)F_0$ for some 
$\phi \in C^\infty_c(E)$ and $F_0 \in \cH_\rho$. 
Then 
\[ F(z) = \int_E \phi(x) F_0(z-x)\, d\mu_E(x),\] 
so that Lemma~\ref{lem:C.11} implies that $F$ extends to a 
continuous function on $\oline{\cT}$ which is smooth on~$E$. 
Therefore $\ev_0$ is defined.

Next we note that, for $F = \Phi(f)$ and $f \in L^2(E^*,\mu;\cK)^\infty$,  
\begin{align*}
\Phi(\eta_\xi)(F) 
&=  \Phi(\eta_\xi)(\Phi(f)) 
=  \eta_\xi(f) 
= \int_{E^*} \la f(\lambda), d\mu(\lambda) \xi \ra\\
&= \int_{E^*} \la d\mu(\lambda) f(\lambda),  \xi \ra
= \la \Phi(f)(0), \xi \ra 
= \la F(0), \xi \ra =\ev_0^\xi(F).
\end{align*}
As $\eta_\xi$ is a distribution vector for the representation 
of $(E,_+)$ on $L^2(E^*,\mu;\cK)$,  the same holds for 
$\ev_0^\xi$ and the representation of the translation group $(E,+)$ 
on $\cH_\rho$. 

\nin (b) follows from 
\begin{align*}
(U^{-\infty}_\rho(g) \ev_0^\xi)(F)
&=  \ev_0^\xi(U^{\infty}_\rho(g^{-1}) F)
= \la (U^{\infty}_\rho(g^{-1}) F)(0), \xi \ra \\
&= \la \rho(g^{-1}) F(0), \xi \ra 
= \la F(0), \rho(g^{-1})^* \xi \ra 
= \la F(0), \rho(\theta(g)) \xi \ra.
\end{align*}

\nin (c) For $g \in N^-$, we have $U_\rho(g)F \in \cH_\rho^\infty$ and 
\[ (U_\rho(g)F)(0) 
= \lim_{z \to 0} (U_\rho(g)F)(z)
= \lim_{z \to 0} J_\rho(g,g^{-1}.z) F(g^{-1}.z) 
=  J_\rho(g,g^{-1}.0) F(g^{-1}.0) = F(0) \] 
because $\dd g(0) = \1$ for every $g \in N^-$. 
This implies that the distribution 
vectors $\ev_0^\xi$ are $N^-$-invariant. 

\nin (d) On $\cT \subeq E_\C$, the one-parameter group $\exp(\R h)$ acts by $\exp(th)z = e^t z$. 
Then 
\[ U^{-\infty}(\exp th) \ev_0^\xi 
= \ev_0^{\rho(\exp -th)\xi}\quad \mbox{ for every } \quad 
\xi \in \cK\] 
follows from (b). 
This shows that 
$\ev_0^\xi \in \cH^{-\infty}_{\rho, {\rm ext},J}$ 
if and only if $e^{-\pi i \dd\rho(h)} \xi = J_\cK \xi.$
As $J_\cK$ and $\dd \rho(h)$ commute with $\rho(\tilde H)$, the 
real subspace $\sE$ is $U(\tilde G^{\tau_G})$-invariant. 
From (c) we have the invariance of each $\ev_0^\xi$ under the subgroup 
$N^-$, so that $\sE$ is invariant under $P^- = N^- \tilde G^{\tau_G}$. 
\end{prf}

We are now ready to prove the main theorem of this paper. 

\begin{thm} \mlabel{thm:5.4} 
For a holomorphic $\cT$-positive $*$-representation $(\rho, \cK)$ of 
$\tilde H_\C$, we consider the irreducible antiunitary representation 
$(U_\rho, \cH_\rho)$ of $\tilde G_\tau$ from \eqref{eq:j1} and \eqref{eq:j2}
 and the real subspace 
$\sE \subeq \cH^{-\infty}_\rho$ 
from \eqref{eq:defE}. 
Then the following assertions hold: 
\begin{itemize}
\item[\rm(a)] The standard subspace $\sV$ of $\cH_\rho$, specified by the 
conjugation $J_\sV = U_\rho(\tau_G)$ and the modular group 
$\Delta_\sV^{-it/2\pi} = U_\rho(\exp th)$, coincides with 
\[ \sH_\sE(S) = 
\oline{\spann_\R \big(U^{-\infty}(C^\infty_c(S,\R))\sE\big)},
\quad \mbox{ where } \quad S := \tilde G^{\tau_G} \exp(C_+^0 \oplus C_-^0) \] 
is an open connected subsemigroup of $G$. 
\item[\rm(b)] The open subsemigroup $S_\sV^0\supeq S$ also satisfies 
$\sH_\sE(S_\sV^0) = \sV.$ 
\item[\rm(c)] The prescription $\sV(\cO) := \sH_\sE(\cO)$ assigns to 
every non-empty 
open subset $\cO \subeq G$ a real cyclic subspace, and this net 
has the following properties: 
\begin{itemize}
\item[\rm(I)] {\rm Isotony:} $\cO_1 \subeq \cO_2$ implies $\sV(\cO_1) \subeq \sV(\cO_2)$. 
\item[\rm(Cov)] {\rm Covariance:} $\sV(g\cO) = U(g)\sV(\cO)$ for $g \in G$. 
\item[\rm(BW)] {\rm Bisognano--Wichmann property:} For 
$g \in G$ and $\cW := gS$, the subspace $\sV(\cW)$ is standard with 
\[ \Delta_{\sV(\cW)}^{-it/2\pi} = U(\exp t \Ad(g)h)\quad \mbox{  for  } \quad 
t \in \R, g \in G
\quad \mbox{ and } \quad J_{\sV(\cW)} = U(g \tau_G(g)^{-1}) J_\sV.\]
\item[\rm(Inv)] $\sV(gS)' = \sV(gS^{-1}) = \sV(g\tau_G(S))$ for $g \in G$. 
\end{itemize}
\item[\rm(d)] If $\eset \not=\cO \subeq G$ is open and there exists a 
$g \in G$ with $\cO \subeq gS$, then $\sV(\cO)$ is standard. 
\item[\rm(e)] If $A\cO S \subeq \cO \supeq S$ and $\sV(\cO)$ is standard, then 
$\cO \subeq S_\sV^0$. In particular
\begin{itemize}
\item[$\bullet$] If $g \in S^{-1}\setminus S_\sV$, 
then $\sV(AgS)$ is not standard. 
\item[$\bullet$] $\sV(G)$ is not standard. 
\end{itemize}
\end{itemize}
\end{thm}

\begin{prf} (a) From Proposition~\ref{prop:c.12}(d) and Lemma~\ref{lem:key} 
it follows that $\sE \subeq \cH_{\rho,{\rm ext},J}^{-\infty} \subeq \sV^{-\infty}$. 

As $S_\sV^0 = G_\sV \exp(C_+^0 \oplus C_-^0)$ by 
\eqref{eq:SE} in the introduction, and $\tilde G^{\tau_G} = (G_\sV)_0$, we have 
$S_\sV^0 = G_\sV S$ and in particular $S \subeq S_\sV$. For 
$\phi \in C^\infty_c(S,\R)$, 
Lemma~\ref{lem:vinfty}(c) implies $U^{-\infty}(\phi)\sE \subeq  \sV$, 
so that $\sH_E(S) \subeq \sV$. 
As $\sV$ is standard, $\sH_\sE(S)$ is also separating, hence standard. 
The invariance of $\sE$ under $U^{-\infty}(A)$ and the invariance 
of $S$ under conjugation with $A$ show that 
$\sH_\sE(S)$ is invariant under $\Delta_\sV^{i\R} = U(A)$. 
Now $\sH_\sE(S) = \sV$ follows from 
Lemma~\ref{lem:lo-3.10}. 

\nin (b) From (a) we recall that $S_\sV^0 = G_\sV S$. 
By (a), $\sV = \sH_\sE(S)$, and this subspace is invariant under $G_\sV$, 
so that Lemma~\ref{lem:sw64-genb} shows that $\sH_\sE(S) 
= \sH_\sE(G_\sV S)= \sH(S_\sV^0)$. 

\nin (c) Since $\sE \subeq \cH^{-\infty}$ is $G$-cyclic in the sense that 
$\cH_\sE(G) = \cH$ 
(Proposition~\ref{prop:c.10}(a)), Theorem~\ref{thm:3.6}
implies that each subspace $\sH_\sE(\cO)$, $\cO \not=\eset$, is cyclic. 
Properties (I) and (Cov) are trivial, as we have already observed in 
Definition~\ref{def:HE}. Property (BW) follows from (a) and the covariance 
of the BGL net. 

To verify (Inv), we first note that, 
replacing $h$ by $-h$ corresponds to replacing $\sV$ by 
$\sV'$ (\cite[Prop.~3.3]{Lo08}) and $S = \tilde G^{\tau_G} \exp(C_+^0 + C_-^-)$ by 
$S^{-1} = \tilde G^{\tau_G}  \exp(-C_+^0 - C_-^0)$. Therefore (a) implies 
$\sV' = \sH_\sE(S^{-1}) = \sV(S^{-1})$. For $g \in G$, we thus obtain 
\[ \sV(gS)' = (U(g)\sV)' = U(g)\sV' = U(g)\sV(S^{-1}) = \sV(g S^{-1}).\] 
Finally, we note that $\L(\tau_G)(C_\pm) = - C_\pm$ and $\tau_G(\tilde G^{\tau_G}) = \tilde G^{\tau_G}$ 
implies $\tau_G(S) = S^{-1}$. 

\nin (d) By (Cov) we have $\sV(\cO) \subeq U(g)\sV(S)$, and the right hand side 
is standard. Hence the cyclic subspace $\sV(\cO)$ is also separating and 
therefore standard. 

\nin (e) As $A\cO = \cO\supeq S$, 
(Cov) implies that the real subspace, $\sV(\cO)$ is invariant 
under the modular group $U(A)$ of $\sV = \sV(S) \subeq \sV(\cO)$. 
If $\sV(\cO)$ is standard, then Lemma~\ref{lem:lo-3.10} 
implies that $\sV(\cO) = \sV$. This in turn shows that, 
for $g \in \cO$, we have  
\[ U(g)\sV =U(g)\sV(S) = \sV(g S) 
\subeq \sV(\cO) = \sV.\]
We conclude that $\cO \subeq S_\sV$, and since $\cO$ is open, 
that $\cO \subeq S_\sV^0$. 

As $S_\sV \not=G$, it follows in particular that $\sH_\sE(G)$ is not standard. 
For $g \in S^{-1}\setminus S_\sV$, the relation
 $g^{-1} S \subeq S$ implies that 
$S \subeq gS \subeq A g S =: \cO$. Hence $\cO$ satisfies the assumptions from above. 
If $\cO \subeq S_\sV^0$, then $e \in \oline{S}$ leads to the contradiction 
$g \in S_\sV$. Therefore $\sV(\cO)$ is not standard. 
\end{prf} 

The preceding theorem provides for every 
$\cT$-positive holomorphic representation 
$(\rho,\cK)$ of $\tilde H_{\C,\tau}$, or, equivalently, 
any norm-continuous real representation 
$(\rho, \cK_\R)$ of $\tilde G^{\tau_G}$, a net of cyclic real subspaces, 
some of which are standard. 
Applying second quantization functors, we thus 
obtain free quantum fields in the sense of Haag--Kastler on the 
group $\tilde G$, where the left translates $\cW = gS$ of the semigroup 
$S$ play the role of wedge domains. 
Here the invariant cone $C \subeq \g$ defines a biinvariant causal structure 
in the sense of a cone field $(gC)_{g \in \tilde G}$ on the group $\tilde G$. 
For $\cW = S$, the modular involution corresponds to the involution 
$\tau_G$ on $\tilde G$ which exchanges $S$ and $S^{-1}$. The modular 
one-parameter group acts on $S$ by the left translations 
$\alpha_t(s) = (\exp th)s$. By left translation, we obtain the geometric 
actions of the modular objects on the domains $gS$.

\begin{rem} The results of this section extend easily 
to quasihermitian Lie groups, i.e., where the Lie algebra $\g$ 
is of the form 
$\g = \g_1 \oplus \g_2$, where
$\g_1$ is a sum of simple hermitian ideals of tube type and $\g_2$ is a compact 
Lie algebra. Then $\fh = \fh_1 \oplus \g_2$, 
and irreducible representations 
$(\rho, \cK)$ of $\tilde H \cong \tilde H_1 \times \tilde G_2$ 
are tensor products 
$\rho \cong \rho_1 \otimes \rho_2$, where $\rho$ is 
$\cT$-positive if and only if $\rho_1$ has this property. 
\end{rem}

\subsection{Jordan space-times and causal symmetric spaces}

The real subspace $\sE$ in Theorem~\ref{thm:5.4} 
that we used above is also invariant 
under the closed connected subgroups $P_1 := \tilde G^{\tau_G}$ and $P_2 := N^- 
\tilde G^{\tau_G}$ (Proposition~\ref{prop:c.12}). 
We thus obtain on the simply connected 
homogeneous spaces $M_j := \tilde G/P_j$ 
covariant nets of cyclic/standard subspaces as follows. 
Let  $q_j \: \tilde G \to M_j$ denote the canonical projection and put 
\[ \sV_{M_j}(\cO) := \sV(q_j^{-1}(\cO)) \quad \mbox{ for } \quad 
\eset\not= \cO \subeq M_j.\] 
For these nets isotony and $G$-covariance 
\[ \sV_{M_j}(g\cO) = U(g)\sV_{M_j}(\cO) \] 
are clear. For $\cW_j := q_j(S)$ it follows from 
Lemma~\ref{lem:sw64-genb} that 
\[ \sV_{M_j}(\cW_j) = \sV(q_j^{-1}(\cW_j)) = \sV(S P_j) = \sV(S).\] 
Hence the subspace $\sV_{M_j}(\cW_j)$ is standard and 
has the Bisognano--Wichmann property. 
Its modular conjugation is implemented on $M_j$ by 
\[ \tau_{M_j}(gP_j) = \tau_G(g) P_j,\] 
and the modular group by 
\[ \alpha_t(g P_j) = \exp(th) g P_j 
= \exp(th) g \exp(-th) P_j \quad \mbox{ for } \quad g \in G, t \in \R.\]

Here the manifold $M_1 = \tilde G/P_1$ is a symmetric space 
of Cayley type (cf.~\cite{HO97})  
and the manifolds $M_2 = \tilde G/P_2$ 
coincide with the Jordan space-times 
described by G\"unaydin in \cite{Gu93} 
as a natural class of causal manifolds with conformal symmetries,
represented by the group~$G$. 
These manifolds also coincide with the 
{\it simple space-time manifolds} 
classified in  \cite[Thm.~4]{MdR07} by Mack and de Riese.

\subsection{Affine groups} 

The subgroup $P^+ := N^+ \tilde G^{\tau_G} \subeq \tilde G$ acts 
by affine maps on the Jordan algebra $E \cong \g^1(h)$, where 
$N^+ \cong (E,+)$ acts by translations, and $G^{\tau_G}$ by linear maps 
leaving the cone $E_+$ invariant. For the special case, where 
$E \cong \R^{1,n-1}$ is $n$-dimensional Minkowski 
space,  $P^+ \cong E \rtimes \R^\times_+  \Spin_{1,n-1}(\R)$ 
is the Poincar\'e group extended by dilations. Restricting everything from 
$\tilde G$ to $P^+$, we immediately obtain 
nets of cyclic/standard subspaces on $E$, which we may consider as an open 
subset of the homogeneous space $G/P^-$. In this case it is of some 
interest to consider an even smaller group, such as the 
Lorentz group $\SO_{1,n-1}(\R)_0$ on $E = \R^{1,n-1}$, which does not 
necessarily contain the dilations. 

A natural setting for this is explored in \cite{NOO20}, and 
we now explain briefly how it connects to the present paper. 
Let $L \subeq H$ be a connected subgroup invariant under the 
Cartan involution $\theta$, and 
$q_L\: \tilde L \to L$ be its universal covering group. 
We consider a norm-continuous $*$-representation 
$(\rho, \cK)$ of $\tilde L_\tau = \tilde L \times \{\1,\tau\}$ 
and a tempered $\Herm^+(\cK)$-valued measure $\mu$ on $E_+^\star$ such that 
\[ (g^{-1})_* \mu = \rho(g) \mu \rho(g)^* \quad \mbox{ for } \quad 
g \in \tilde L.\] 
Then the Laplace transform $\cL(\mu)$ defines a holomorphic function on 
$\cT$, and 
\[ Q_\rho(z,w) := \cL(\mu)\Big(\frac{z - \oline w}{2i}\Big) \] 
is a positive definite $B(\cK)$-valued kernel on $\cT$. 
We thus obtain a 
unitary representation of $E \rtimes \tilde L_\tau$ on the 
corresponding Hilbert subspace $\cH_\rho \subeq \Hol(\cT,\cK)$, given by 
\begin{equation}
  \label{eq:pitau2}
 (U_\rho(x,g)f)(z) := \rho(g) f(g^{-1}.(z-x))
\quad \mbox{ for } \quad x \in E, z \in \cT, g \in \tilde L_\tau.
\end{equation}

Now we consider a element $h' \in \fl$, the Lie algebra of $L$, 
for which $\ad h'$ defines a $3$-grading on $\g$, and the involution 
$\tau'$ on $E \rtimes \tilde L$ integrating the Lie algebra involution 
$e^{\pi i \ad h'}$ on $E \rtimes \fl$. For 
$J_\cK := \rho(\tau')$, the real subspace 
\begin{equation}
  \label{eq:defE2}
 \sE := \{ \ev_0^\xi \: e^{-\pi i \cdot \dd \rho(h')} \xi = J_\cK \xi\} 
\end{equation}
is then contained in $\cH_\rho^{-\infty}(U_\rho\res_E)$ and invariant under 
$U_\rho(\tilde L^0)$, 
where $\tilde L^0 \subeq \tilde L$ is the centralizer of~$h'$. 
Now the standard subspace 
$\sV = \sV_{(h',\tau',U_\rho)}$ coincides with 
$\sH_\sE^E(\cW)$, where $\cW$ is the wedge domain 
\[ \cW = (E_+ \cap E^1(h')) \oplus E^0(h') \oplus (-E_+ \cap E^{-1}(h')). \] 
Using Lemma~\ref{lem:semidir}, we even get 
\[ \sH_\sE^E(\cW) = \sH_\sE^{E \rtimes \tilde L^0}(\cW \rtimes \tilde L^0).\] 
In the special case where $\ad h\res_{E} = \id_E$, 
we have $\tilde L^0 = \tilde L$. 
For more details on this situation, we refer to \cite{NOO20}.

\section{Perspectives} 
\mlabel{sec:6}

\subsection{Extension to more general groups} 
\mlabel{subsec:6.1} 

The results of Sections~\ref{sec:2} and \ref{sec:3} work for 
non-reductive Lie groups satisfying the conditions (B1)-(B4). There are various 
types of non-reductive groups with this property, such as 
the extended Jacobi group, discussed in some detail in \cite[Ex.~3.7]{Ne19b}. 
We briefly recall some observations from \cite{Ne19b}. 
Its Lie algebra is 
\[ \g = \hcsp(\R^{2n},\omega) = \heis(\R^{2n},\omega) \rtimes \csp_{2n}(\R), \] 
where $\omega$ is the canonical symplectic form on $\R^{2n}$, 
$\heis(\R^{2n},\omega) = \R \oplus \R^{2n}$ is the corresponding Heisenberg algebra 
with the bracket $[(z,v),(z',v')] = (\omega(v,v'),0)$, and 
\[ \csp_{2n}(\R) := \sp_{2n}(\R) \oplus \R \1 \] 
is the {\it conformal symplectic Lie algebra} of $(\R^{2n},\omega)$. 
The hyperplane ideal 
\[ \fj := \heis(\R^{2n},\omega) \rtimes \sp_{2n}(\R) \] 
(the {\it Jacobi algebra})  
contains a pointed generating invariant cone~$C$, 
corresponding to the non-negative polynomials of degree $\leq 2$ on $\R^{2n}$. 
The involution  $\tau(q,p)= (-q,p)$ is antisymplectic, and 
the operator $h := \shalf(\1 + \tau)\in \csp_{2n}(\R)$  defines 
a $3$-grading of $\g$ for which the cones $C_\pm$ generate $\g^{\pm 1}(h)$. 
The corresponding simply connected group $G$ has an 
irreducible unitary representation on 
\[ \cH 
:= L^2\Big(\R^\times_+, \frac{d\lambda}{\lambda}; L^2(\R^n)\Big) \cong 
 L^2\Big(\R^\times_+ \times \R^n, \frac{d\lambda}{\lambda} \otimes dx\Big),\]
where $\lambda$ parametrizes a family of mutually 
inequivalent irreducible representations of the Jacobi group on $L^2(\R^n)$.
 It would be very interesting to identify nets of standard subspaces 
in this representation if it is extended to an antiunitary representation 
of $G_\tau$ by 
\[ (Jf)(\lambda,x) := \oline{f(\lambda,-x)}.\]

\subsection{Relation to causal structures} 
\mlabel{subsec:6.2}

In Section~\ref{sec:5} we  constructed nets of cyclic subspaces 
on three levels: 
\begin{itemize}
\item On the group level, where the invariant cone 
$C \subeq \g$ defines a biinvariant causal structure on the group manifold 
$G$ and the semigroup $S^0 = \exp(C_+^0) (G^{\tau_G})_0 \exp(C_-^0)$ 
plays the role of a wedge domain. 
\item On the symmetric space $M = G/G^{\tau_G}$, whose tangent space in the origin 
is $\fq = \g^1 \oplus \g^{-1}$, in which $C_\fq := C \cap \fq = C_+ - C_-$ is a pointed 
generating $\Ad(G^{\tau_G})$-invariant cone that defines an invariant causal 
structure on $M$. Here the image of $S$ in $M$, which is 
the set $\Exp(C_+ + C_-)$, where $\Exp \: \fq \to M$ is the exponential function of 
the symmetric space~$M$, plays the role of a wedge domain in $M$. Actually 
it also is the causal future of the base point for another 
causal structure on $M$ corresponding to the cone $C_+ + C_-$. 
Note that the first causal structure turns $M$ into a compactly  causal  symmetric 
space in the sense of \cite{HO97} and the second one into a 
non-compactly causal symmetric space. 
\item For the Jordan space-time 
$M = \tilde G/\tilde P^-$, the tangent space in the base point is 
$\g^1$, which we considered also as a euclidean Jordan algebra $E$, 
and the causal structure is given by the cone $C_+ = \oline{E_+}$. 
Here the image of $S$ coincides with the image of $\exp(C_+)$ in $M$. 
It plays the role of a wedge domain and the future of the base point for the 
causal structure at the same time. 
\end{itemize}

These observations show that, to proceed beyond the class of spaces 
considered here, we need a better theory of wedge domains in 
ordered homogeneous spaces. Some first steps in this general program 
will be  carried out in \cite{NO20} for the classes of compactly causal and 
non-compactly causal symmetric spaces, such as de Sitter spaces.

\appendix 

\section{Distribution vectors for 
unitary representations} 
\mlabel{app:a} 

In this appendix we collect some material on distribution vectors 
of unitary representations that we use in this paper. 

\begin{defn} Let $G$ be a Lie group. 
We fix a left invariant Haar measure $\mu_G$ on $G$ 
and we often write $dg$ for $d\mu_G(g)$. 
This measure defines on $L^1(G) := L^1(G,\mu_G)$ the structure of a 
Banach-$*$-algebra by the {\it convolution product} and  \index{convolution product} 
\begin{equation}
  \label{eq:convol}
(\phi*\psi)(x) =\int_G \phi(g)\psi(g^{-1}x)\, d\mu_G (g),  
\quad \mbox{ and } \quad \phi^*(g) = \overline{\phi (g^{-1})}\Delta_G (g)^{-1}
\end{equation}
is the involution, where 
$\Delta_G : G\to \R_+$ is the {\it modular function} determined by \index{modular 
function}
\begin{align*}
  \label{eq:modfunc}
 \int_G \phi(y)\, d\mu_G(y) 
&=\int_G \phi(y^{-1})\Delta_G(y)^{-1}\, d\mu_G(y) \quad \mbox{ and } \\ 
\Delta_G(x)\int_G \phi(yx)\, d\mu_G(y)&=\int_G \phi(y)\, d\mu_G(y) \quad 
\mbox{ for } \quad \phi\in C_c (G).
\end{align*}
We put 
\begin{equation}
  \label{eq:check}
\varphi^\vee(g)=\varphi (g^{-1})\cdot\Delta_G(g)^{-1}
\quad \mbox{ so that } \quad 
\int_G \phi(g)\, d\mu_G(g) 
= \int_G \phi^\vee(g)\, d\mu_G(g).
\end{equation}
The formulas above show that we have two isometric actions of $G$ on 
$L^1(G)$, given by 
\begin{equation}
  \label{eq:left-right-action}
(\lambda_g f)(x) = f(g^{-1}x) \quad \mbox{ and }\quad 
(\rho_g f)(x) = f(xg) \Delta_G(g).
\end{equation}
Note that 
\begin{equation}
  \label{eq:veecov}
(\lambda_g f)^* = \rho_g f^*
\quad \mbox{ and } \quad (\lambda_g f)^\vee = \rho_g f^\vee.   
\end{equation}
\end{defn}

Now let $(U,\cH)$ be a continuous unitary representation of the Lie group~$G$, 
i.e., a homomorphism $U : G\to \U(\cH), g \mapsto U(g)$ 
such that, for each $\eta \in\cH$, the  orbit map
$U^\eta (g)=U(g)\eta$ is continuous. 
For $\phi\in L^1(G)$ the operator-valued integral 
\[ U(\phi) := \int_G \phi(g) U(g)\, dg \] 
exists and is uniquely determined by 
\begin{equation}
  \label{eq:l1-est}
\la \eta, U(\phi) \zeta \ra =\int_G \phi(g)\la \eta, U(g)\zeta\ra\, dg \quad 
\mbox{ for } \quad \eta,\zeta \in \cH.
\end{equation}
Then $\|U(\phi)\|\le \|\phi\|_1$ and the 
so-obtained continuous linear map $L^1(G) \to B(\cH)$ 
is a representation of the Banach-$*$-algebra $L^1(G)$, i.e., 
$U(\phi*\psi)=U(\phi)U(\psi)$ and $U(\phi^*)=U(\phi)^*.$ 
We also note that 
\begin{equation}
  \label{eq:covl1}
U(g) U(\phi) = U(\lambda_g \phi) \quad \mbox{ and } \quad 
U(\phi)U(g)  = U(\rho_g^{-1} \phi)\quad \mbox{ for } \quad 
g\in G, \phi \in L^1(G).
\end{equation}
For $\phi_g(x) := \phi(xg)$, we then have 
$\phi_g = \Delta_G(g)^{-1} \rho_g \phi$ by \eqref{eq:left-right-action}, 
and thus by \eqref{eq:covl1} 
\begin{equation}
  \label{eq:rightrel}
U(\phi_g) = \Delta_G(g)^{-1} U(\phi) U(g^{-1}) \quad \mbox{ for } \quad 
 g\in G.
\end{equation}

\subsection{The topology on $\cH^\infty$ and the 
space $\cH^{-\infty}$} 
\mlabel{subsec:app1}

A {\it smooth vector} is an element $\eta\in\cH$ for which the orbit map 
$U^\eta : G\to \cH, g \mapsto U(g)\eta$ 
is smooth. We write~$\cH^{\infty}$ for the space 
of smooth vectors. It carries the {\it derived representation} 
$\dd U $ of the Lie algebra $\fg$ given by
\begin{equation}
  \label{eq:derrep}
\dd U(x)\eta =\lim_{t\to 0}\frac{U(\exp t x)\eta -\eta}{t}.
\end{equation}
We extend this representation to a homomorphism 
$\dd U \:  \cU(\g) \to \End(\cH^\infty),$ 
where $\cU(\g)$ is the complex enveloping algebra of $\g$. This algebra 
carries an involution $D \mapsto D^*$ determined uniquely $x^* = -x$ for $x \in \g$.
For $D \in \cU(\g)$, we obtain a seminorm on $\cH^\infty$ by 
\[p_D(\eta )=\|\dd U(D)\eta\|\quad \mbox{ for } \quad \eta \in \cH^\infty.\] 
These seminorms define a topology on $\cH^\infty$ which turn the injection 
\begin{equation}
  \label{eq:tophinfty}
 \eta \: {\cal H}^\infty \to {\cal H}^{{\cal U}(\g_\C)}, \quad 
\xi \mapsto (\dd U(D)\xi)_{D \in {\cal U}(\g_\C)}
\end{equation}
into a topological embedding, where the right hand side carries the product 
topology (cf.\ \cite[3.19]{Mag92}).  It turns $\cH^\infty$ 
into a complete locally convex space 
for which the linear operators $\dd U(D)$, $D \in \cU(\g)$, are continuous. 
Since $\cU(\g)$ has a countable basis, 
countably many such seminorms already determine the topology, so that 
$\cH^\infty$ is metrizable. As it is also complete, it is a Fr\'echet space.
We also observe that 
the inclusion $\cH^\infty\hookrightarrow \cH$ is continuous.

The space $\cH^\infty$ of smooth vectors is $G$-invariant 
and we denote the corresponding representation by~$U^\infty$. We have 
the intertwining relation 
\[ \dd U(\Ad (g)x)= U(g) \dd U(x) U(g)^{-1} \quad \mbox{ for } \quad 
g \in G, x \in \g.\] 
If $\varphi \in C_c^\infty (G)$ and $\xi \in\cH$, then 
$U(\varphi) \xi \in \cH^\infty$ and differentiation under the integral sign 
shows that  
\begin{equation}
  \label{eq:derrep2}
\dd U(x) U(\varphi) \xi :=U(-x^R \varphi) \xi,  
\quad \mbox{ where } \quad 
(x^R\varphi)(g) =\frac{d}{dt}\Big|_{t=0} \phi((\exp tx) g). 
\end{equation}

A sequence $(\varphi_n)_{n \in \N}$ in $C^\infty_c(G)$ is called a 
{\it $\delta$-sequence} if $\int_G \varphi_n(g)\, dg = 1$ for every 
$n \in \N$ and, for every $e$-neighborhood $U \subeq G$, we have 
$\supp(\varphi_n) \subeq U$ if $n$ is sufficiently large. 
If $(\varphi_n)_{n \in \N}$ is a 
{$\delta$-sequence}, then $U(\varphi_n)\xi \to \xi$, so that $\cH^\infty$
is dense in~$\cH$.

We write $\cH^{-\infty}$ for the space 
of continuous anti-linear functionals on $\cH^\infty$. 
Its elements are called \textit{distribution vectors}. 
The group $G$, $\cU(\g)$ and $C^\infty_c(G)$ act on $\eta \in \cH^{-\infty}$ by
\begin{itemize}
\item $(U^{-\infty}(g)\eta ) (\xi ):= \eta  (U(g^{-1})\xi)$, $g \in G, 
\xi \in \cH^\infty$. If $U \: G \to \AU(\cH)$ is an antiunitary 
representation and $U(g)$ is antiunitary, then we have to modify 
this definition slightly by 
$(U^{-\infty}(g)\eta) (\xi ):= \oline{\eta(U(g^{-1})\xi)}$. 
\item $(\dd U^{-\infty}(D) \eta ) (\xi ):= \eta(\dd U(D^*) \xi)$, $D \in \cU(\g), 
\xi \in \cH^\infty$. 
\item $U^{-\infty}(\varphi) \eta =\eta \circ U^\infty(\varphi^*)$, 
$\varphi \in C_c^\infty (G).$ 
\end{itemize} 
We have natural $G$-equivariant linear embeddings 
\begin{equation}
  \label{eq:embindistr}
\cH^\infty \into \cH
\mapright{\xi \mapsto  \la \cdot, \xi \ra} \cH^{-\infty}. 
\end{equation}

It is an important feature of \eqref{eq:embindistr} 
that the representation of $\cU(\g)$ on 
$\cH^{-\infty}$ provides an embedding of the whole Hilbert space $\cH$ 
into a larger space on which the Lie algebra acts. The following 
lemma shows that, $\cH^\infty$ is the maximal $\g$-invariant subspace 
of $\cH \subeq \cH^{-\infty}$ and that the subspace $\cH$ generates 
$\cH^{-\infty}$ as a $\g$-module. 

\begin{lem} \mlabel{lem:charsmooth} 
The following assertions hold: 
  \begin{itemize}
  \item[\rm(a)] $\cH^\infty = \{ \xi \in \cH \subeq \cH^{-\infty} \: (\forall D \in \cU(\g)) 
\ \dd U^{-\infty}(D)\xi \in \cH\}.$ 
  \item[\rm(b)] ${\cal H}^{-\infty} = \Spann \big(\dd U^{-\infty}({\cal U}(\g)){\cal H}\big)$. 
  \end{itemize}
\end{lem}

\begin{prf} (a) This follows by combining 
\cite[Prop.~A.1]{Oeh18}, asserting that 
\[ \cD(\partial U(x)) = \{ \xi \in \cH \: \dd U^{-\infty}(x) \xi \in \cH \}, \] 
with the fact that 
\[ \cH^\infty = \bigcap \{ \cD(\partial U(x_1) \cdots \partial U(x_n))  \: 
n \in \N, x_1, \ldots, x_n\in \g\} \] 
(\cite[Lemma~3.4]{Ne10}). 

\nin (b) Let $\eta \in {\cal H}^{-\infty}$ and  
consider $\cH^\infty$ as a subspace of the topological product ${\cal H}^{{\cal U}(\g)}$. 
By the Hahn-Banach extension theorem, 
$\eta$ extends to a continuous antilinear functional 
$\tilde\eta$ on ${\cal H}^{{\cal U}(\g)}$. Since the dual of a direct product 
is the direct sum of the dual spaces, there exist 
$D_1, \ldots, D_n \in\cU(\g)$ and $\xi_1, \ldots, \xi_n \in \cH$, such that 
\[ \eta(\xi) 
= \sum_{j = 1}^n \la  \dd U(D_j) \xi, \xi_j \ra 
= \sum_{j = 1}^n \la  \xi, \dd U^{-\infty}(D_j^*) \xi_j \ra 
\quad \mbox{ for } \quad \xi \in \cH^\infty,\] 
which means that 
$\eta =  \sum_{j = 1}^n \dd U^{-\infty}(D_j^*) \xi_j.$ 
\end{prf}

For each $\phi \in C^\infty_c(G)$, the map 
$U(\phi) \: \cH \to \cH^{\infty}$ 
is continuous, so that its adjoint defines a weak-$*$-continuous maps 
$U^{-\infty}(\phi^*) \: \cH^{-\infty} \to \cH$. We actually have 
$U^{-\infty}(\phi)\cH^{-\infty} \subeq \cH^\infty$ 
as a consequence of the 
Dixmier--Malliavin Theorem \cite[Thm.~3.1]{DM78}, 
which asserts that every $\phi \in C^\infty_c(G)$ factors as 
$\phi = \phi_1 * \phi_2$ with $\phi_j \in C^\infty_c(G)$

\subsection{Equivariant embeddings into distributions on $G$} 

Let $G$ be a Lie group and $(U,\cH)$ be a unitary representation of $G$. 
We call a distribution vector $\eta \in \cH^{-\infty}$ 
{\it cyclic} if $U^{-\infty}(C^\infty_c(G))\eta$ is dense in $\cH$. 
Then \cite[Prop.~7.1.6]{NO18} yields a $G$-equivariant injection 
\begin{equation}
  \label{eq:dist-inc}
 j_\eta \: \cH^{-\infty} \to C^{-\infty}(G), \quad 
j_\eta(\alpha)(\phi) := \alpha(U^{-\infty}(\phi)\eta)
\quad \mbox{ for } \quad 
\phi \in C^\infty_c(G).
\end{equation}
In particular, 
\[ D := j_\eta(\eta) \in C^{-\infty}(G) \] 
is positive definite and 
\begin{equation}
  \label{eq:cHD}
\cH_D  := j_\eta(\cH) \subeq C^{-\infty}(G) 
\end{equation}
is a reproducing kernel Hilbert space of distributions, on which 
$G$ acts by left translations. For $\phi, \psi \in C^\infty_c(G)$, we have 
\begin{equation}
  \label{eq:j11}
j_\eta \circ U^{-\infty}(g) = \lambda_g \circ j_\eta \quad \mbox{ for } \quad 
g \in G,
\end{equation}
\begin{equation}
  \label{eq:j22}
j_\eta(U^{-\infty}(\phi)\alpha) = \phi * j_\eta(\alpha) 
\quad \mbox{ for } \quad \phi \in C^\infty_c(G),
\end{equation}
and 
\begin{equation}
  \label{eq:scalproHd}
\la U^{-\infty}(\phi)\eta, U^{-\infty}(\psi)\eta \ra_\cH 
= \la \phi * D, \psi * D \ra_{\cH_D} 
= D(\psi^* * \phi).
\end{equation}

\begin{rem}
To obtain embeddings into $C^{-\infty}(G)$ that are equivariant with respect to 
the action by right translations, we first 
extend the involution $\vee$ from \eqref{eq:check} to distributions by 
\begin{equation}
  \label{eq:distvee}
 D^\vee(\phi) := D(\phi^\vee) 
\end{equation}
to obtain by duality and \eqref{eq:veecov}
\begin{equation}
  \label{eq:vee-inter2}
(\lambda_g D)^\vee = \rho_g D^\vee \quad \mbox{ for }\quad g \in G.
\end{equation} 
Therefore the map 
\begin{equation}
  \label{eq:dist-inc2}
 j_\eta^\vee \: \cH^{-\infty} \to C^{-\infty}(G), \quad 
j_\eta^\vee(\alpha)(\phi) := \alpha(U^{-\infty}(\phi^\vee)\eta)
\end{equation}
is equivariant with respect to the action of $G$ on $C^{-\infty}(G)$ by 
right translations. 
\end{rem}

\nin Department of Mathematics, Friedrich-Alexander-University of Erlangen-N\"urnberg, Cauerstrasse 11, 91058 Erlangen, Germany; neeb@math.fau.de

\nin Department of Mathematics, Louisiana State University, Baton Rouge, LA 70803, U.S.A.; olafsson@math.lsu.edu


\begin{thebibliography}{aaaaaaaa}

\bibitem[Ar63]{Ar63} Araki, H., {\it 
A lattice of von Neumann algebras associated with the 
quantum theory of a free Bose field}, 
J. Math. Phys. {\bf 4} (1963), 1343--1362 
 
\bibitem[Ar64]{Ar64} Araki, H.,  {\it von Neumann algebras of local observables 
for free scalar field}, J. Mathematical Phys. {\bf 5} (1964),  1--13

\bibitem[AW63]{AW63} Araki, H., and E. J. Woods, {\it 
Representations of the canonical commutation relations describing 
a nonrelativistic infinite free Bose gas}, 
J. Math. Phys. {\bf 4} (1963), 637--662 

\bibitem[AW68]{AW68} Araki, H., and E. J. Woods,  {\it A classification of factors}, 
Publ. RIMS, Kyoto Univ. Ser. A {\bf 3} (1968), 51--130

\bibitem[BJL02]{BJL02} Baumg\"artel, H., Jurke, M., and F. Lledo, 
{\it Twisted duality of the CAR-algebra}, J.~Math. Physics {\bf 43:8} (2002), 
4158--4179

\bibitem[Bo92]{Bo92} Borchers, H.-J., {\it The CPT-Theorem in two-dimensional 
theories of local observables}, Comm. Math. Phys. {\bf 143} (1992), 315--332 

\bibitem[BR87]{BR87} Bratteli, O., and D.~W.~Robinson, ``Operator Algebras and Quantum Statistical
Mechanics I,'' 2nd ed.,
Texts and Monographs in Physics, Springer-Verlag, 1987 

\bibitem[BM96]{BM96} Bros, J., and U. Moschella, {\it Two-point functions and quantum 
fields in de Sitter universe}, Rev. Math. Phys. {\bf 8} (1996), 327--391

\bibitem[BGL93]{BGL93} Brunetti, R., Guido, D., and R. Longo, {\it 
Modular structure and duality in conformal quantum field theory}, 
Comm. Math. Phys. {\bf 156} (1993), 210--219 

\bibitem[BGL02]{BGL02} Brunetti, R., Guido, D., and R. Longo, {\it 
Modular localization and Wigner particles}, Rev. Math. Phys. {\bf  14} (2002), 759--785

\bibitem[BDFS00]{BDFS00} Buchholz, D., Dreyer, O., Florig, M., 
and S. J. Summers, 
{\it Geometric modular action and spacetime symmetry groups}, Rev. Math. Phys. {\bf 12:4} (2000), 475--560 

\bibitem[BLS11]{BLS11} Buchholz, D., Lechner, G., and S. J. Summers, {\it 
Warped convolutions, Rieffel deformations and the construction of quantum 
field theories}, Comm. Math. Phys. {\bf 304:1} (2011), 95--123 

\bibitem[DM78]{DM78} Dixmier, J., and P. Malliavin,
{\it Factorisations de fonctions et de vecteurs ind\'efiniment
diff\'erentiables}, Bull. Soc. math., 2e s\'erie {\bf 102} (1978),
305--330

\bibitem[EO73]{EO73} Eckmann, J.-P., and K.~Osterwalder, {\it An application of Tomita's theory 
of modular Hilbert algebras: duality for free Bose fields}, J. Funct. Analysis {\bf 13:1} (1973), 1--12 

\bibitem[EHW83]{EHW83} Enright, T.J., R. Howe, and N. Wallach, {\it A classification of 
unitary highest weight modules}, in ``Representation Theory of Reductive 
Groups,'' Progress in Math. {\bf 40}, Birkh\"auser Verlag, 1983, 97--143

\bibitem[FK94]{FK94} Faraut, J., and A. Koranyi, ``Analysis on Symmetric Cones,'' 
Oxford Math.\ Monographs, Oxford University Press, 1994 

\bibitem[GN]{GN} Gl\"ockner, H.,  and K.-H. Neeb, ``Infinite Dimensional 
Lie Groups, Vol. I, Basic Theory and Main Examples,'' book in preparation 

\bibitem[Gu93]{Gu93} G\"unaydin, M., {\it Generalized conformal and superconformal 
group actions and Jordan algebras}, Modern Phys. Letters A {\bf 8:15} (1993), 1407--1416

\bibitem[Ha96]{Ha96} Haag, R., ``Local Quantum Physics. Fields, Particles, Algebras,'' 
Second edition, Texts and Monographs in Physics,  Springer-Verlag, Berlin, 1996

\bibitem[HN01]{HN01} Hilgert, J., and K.-H. Neeb, {\it Vector-valued Riesz 
distributions on euclidian Jordan algebras}, J. Geom. Analysis 
{\bf 11:1} (2001), 43--75

\bibitem[HN12]{HN12}
Hilgert, J.,  and K.-H.\ Neeb,
``Structure and Geometry of Lie Groups,'' 
{Springer Monographs in Mathematics}, {Springer}, {New York}, {2012}

\bibitem[H{\'O}97]{HO97} Hilgert, J., and G. {\'O}lafsson, {\it Causal Symmetric Spaces, 
Geometry and Harmonic Analysis}, Perspectives in Mathematics {\bf 18}, Academic Press, 1997

\bibitem[Iw59]{Iw59} Iwahori, N., {\it On real irreducible representations of 
Lie algebra}, Nagoya Math. J. {\bf 14} (1959), 59--83 


\bibitem[Le15]{Le15} Lechner, G., {\it Algebraic Constructive Quantum 
Field Theory: Integrable Models and Deformation Techniques}, 
in ``Advances in Algebraic Quantum Field Theory,'' Eds. R.~Brunetti et al; 
Math. Phys. Stud., Springer, 2015, 397--449, arXiv:math.ph:1503.03822
 
\bibitem[LL15]{LL15} Lechner, G., and R.~Longo, 
\textit{Localization in Nets of Standard Spaces}, 
 Comm. Math. Phys. {\bf 336:1} (2015), 27--61 

\bibitem[Lo08]{Lo08}  Longo, R., {\it Real Hilbert subspaces, modular theory, 
$\SL(2, \R)$ and CFT} 
in ``Von Neumann Algebras in Sibiu'', 33-91, Theta Ser. Adv. Math. {\bf 10}, 
Theta, Bucharest, 2008  

\bibitem[Lo69]{Lo69} Loos, O., ``Symmetric Spaces I: General Theory,'' 
W. A. Benjamin, Inc., New York, Amsterdam, 1969


\bibitem[MdR07]{MdR07} Mack, G., and M. de Riese, {\it 
Simple space-time symmetries: generalizing conformal field theory}, 
J. Math. Phys. {\bf 48:5} (2007), 052304, 21 pp.

\bibitem[Mag92]{Mag92} Magyar, M., ``Continuous Linear Representations,'' North-Holland, 
Mathematical Studies 168, 1992

\bibitem[MN20]{MN20} Morinelli, V., and K.-H. Neeb, {\it 
Covariant homogeneous nets of standard subspaces}, in preparation 

\bibitem[Ne98]{Ne98} Neeb, K.-H., {\it Operator valued
positive definite kernels on tubes},
Monatshefte f\"ur Math. {\bf 126} (1998), 125--160

\bibitem[Ne99]{Ne99} Neeb, K.-H., ``Holomorphy and Convexity in Lie Theory,'' 
Expositions in Mathematics {\bf 28}, de Gruyter Verlag, Berlin, 1999

\bibitem[Ne10]{Ne10} Neeb, K.-H.,  {\it 
On differentiable vectors for representations of infinite dimensional 
Lie groups}, J. Funct. Anal. {\bf 259} (2010), 2814--2855

\bibitem[Ne13]{Ne13} Neeb, K.-H., {\it Holomorphic realization of unitary 
representations of Banach--Lie groups}, 
in ``Lie Groups: Structure, Actions, and Representations---In 
Honor of Joseph A. Wolf on the Occasion of his 75th Birthday,''
Huckleberry, A., Penkov, I., Zuckerman, G. (Eds.), 
Progress in Mathematics {\bf 306}, 2013; 185--223

\bibitem[Ne19]{Ne19} Neeb, K.-H., 
{\it Finite dimensional semigroups of unitary endomorphisms of standard 
subspaces}, arXiv:math.OA.1902.02266 

\bibitem[Ne19b]{Ne19b} Neeb, K.-H., 
{\it Semigroups in 3-graded Lie groups}, 
arXiv:math.OA.1912.13367 

\bibitem[N\'O17]{NO17} Neeb, K.-H., and G.\, \'Olafsson,  {\it 
Antiunitary representations and modular theory}, 
in ``50th Sophus Lie Seminar'', Eds. K. Grabowska et al, 
Banach Center Publications {\bf 113}; pp.~291--362; 
arXiv:math-RT:1704.01336 

\bibitem[N\'O18]{NO18} Neeb, K.-H., and G.\, \'Olafsson, 
``Reflection Positivity. A Representation Theoretic Perspective,'' 
Springer Briefs in Mathematical Physics {\bf 32}, 2018 



\bibitem[N\'O20]{NO20} Neeb, K.-H., and G.\, \'Olafsson, 
{\it Standard subspaces of Hilbert spaces of distributions 
on noncompactly causal symmetric spaces}, in preparation 

\bibitem[N\'OO20]{NOO20} Neeb, K.-H., G.\, \'Olafsson, and 
B. \O{}rsted, {\it Standard subspaces of Hilbert spaces of holomorphic 
functions on tube domains}, in preparation 

\bibitem[Oeh18]{Oeh18} Oeh, D., {\it Analytic extensions of 
representations of $\star$-semigroups 
without polar decomposition}, 
Internat. Math. Res. Notices, to appear; arXiv:math-RT:1812.10751 

\bibitem[\'O91]{O91} \'Olafsson, G., {\it Symmetric spaces of Hermitian type},
Differential Geom. Appl. {\bf 1} (1991), 195--233

\bibitem[On20]{On20} Onishchik, A.L., 
``Lectures on Real Semisimple Lie Algebras and Their Representations,'' 
ESI Lectures in Mathematics and Physics,  European Math. Soc. 
Publishing House, 2020

\bibitem[PEGW19]{PEGW19} Pejhan, H., Enayati, M., Gazeau, J.-P., and A. Wang, 
{\it ``Massive'' Rarita--Schwinger fields in de Sitter space}, 
arXiv:1909.12350v1 [gr-qc]

\bibitem[Schr97]{Schr97} Schroer, B., {\it Wigner representation theory of the Poincar\'e group, 
localization, statistics and the S-matrix},  Nuclear Phys. {\bf B 499-3} (1997), 519--546

\bibitem[Si74]{Si74} Simon, B., ``The $P(\Phi)_2$ Euclidean (Quantum) Field Theory'', 
Princeton Univ. Press, 1974 

\bibitem[TS16]{TS16} Thiemann, Th., and A. Stottmeister, 
{\it The microlocal spectrum condition, 
initial value formulations and background independence}, 
Journal of Mathematical Physics {\bf 57} (2016), 022303 

\bibitem[Tr67]{Tr67} Treves, F., ``Topological Vector Spaces, Distributions, and
Kernels,'' Academic Press, New York, 1967 

\bibitem[Wi93]{Wi93} Wiesbrock, H.-W., 
 {\it Half-sided modular inclusions of von 
Neumann algebras}, Commun. Math. Phys. {\bf 157} (1993), 83--92 


\end{thebibliography}
\end{document}